\documentclass[twocolumn, twocolappendix]{aastex631}

\usepackage{CJK}

\usepackage{graphicx}
\usepackage{natbib}
\usepackage{gensymb}
\hypersetup{linkcolor=blue,citecolor=blue,filecolor=blue,urlcolor=blue}

\usepackage{footmisc}
\usepackage{amsmath}
\usepackage{verbatim}
\newcommand{\HI}{H{\sc\ i}}
\newcommand{\HII}{H{\sc\ ii}}
\newcommand{\OI}{O{\sc\ i}}
\newcommand{\AlII}{Al{\sc\ ii}}
\newcommand{\SiII}{Si{\sc\ ii}}
\newcommand{\SiIII}{Si{\sc\ iii}}
\newcommand{\SiIV}{Si{\sc\ iv}}
\newcommand{\SII}{S{\sc\ ii}}

\newcommand{\CII}{C{\sc\ ii}}

\newcommand{\CIV}{C{\sc\ iv}}

\newcommand{\OVI}{O{\sc\ vi}}

\newcommand{\FeII}{Fe{\sc\ ii}}

\newcommand{\NII}{N{\sc\ ii}}
\newcommand{\NaI}{Na{\sc\ i}}
\newcommand{\Halpha}{H$\alpha$}

\newcommand{\MgII}{Mg{\sc\ ii}}
\newcommand{\CIII}{C{\sc iii}}

\newcommand{\zsun}{\rm Z_\odot}
\newcommand{\kms}{\rm km~s^{-1}}
\newcommand{\sigsfr}{\Sigma_{\rm SFR}}
\newcommand{\msun}{M_\odot}
\newcommand{\msunyr}{M_\odot~{\rm yr}^{-1}}
\newcommand{\msunyrkpc}{M_\odot~{\rm yr}^{-1}~{\rm kpc}^{-2}}
\newcommand{\vcen}{v_{\rm cen}}
\newcommand{\vout}{v_{\rm out, bulk}}

\newcommand{\vlsr}{v_{\rm LSR}}
\newcommand{\vhelio}{v_{\rm helio}}
\newcommand{\mstar}{M_*}
\newcommand{\dotmout}{\dot{M}_{\rm out}}
\newcommand{\mout}{M_{\rm out}}

\newcommand{\ulldr}{DR5}

\shorttitle{Outflows in the LMC}
\shortauthors{Zheng et~al.}
\begin{document}
\defcitealias{heckman15}{H15}
\defcitealias{chisholm15}{C15}
\defcitealias{xu22_classy3}{X22}

\begin{CJK*}{UTF8}{gbsn} % to add Asian style names 

\title{Braving the Storm: Quantifying Disk-wide Ionized Outflows in the Large Magellanic Cloud with ULLYSES}

%https://orcid.org/0000-0003-4158-5116
%\correspondingauthor{Yong Zheng}
%\email{zhengy14@rpi.edu}

\author[0000-0003-4158-5116]{Yong Zheng (郑永)}
\affiliation{Department of Physics, Applied Physics and Astronomy, Rensselaer Polytechnic Institute, Troy, NY 12180, {\color{blue}zhengy14@rpi.edu}}

\author[0000-0003-0789-9939]{Kirill Tchernyshyov}
\affiliation{Department of Astronomy, University of Washington, Seattle, WA 98195, USA}

\author[0000-0002-7134-8296]{Knut Olsen}
\affiliation{National Optical Astronomy Observatory, Tucson, AZ 85719, USA}

\author[0000-0003-1680-1884]{Yumi Choi}
\affiliation{National Optical Astronomy Observatory, Tucson, AZ 85719, USA}

\author[0000-0002-8366-2143]{Chad Bustard}
\affiliation{Kavli Institute for Theoretical Physics, University of California—Santa Barbara, Kohn Hall, Santa Barbara, CA 93107, USA}

\author[0000-0001-6326-7069]{Julia Roman-Duval}
\affiliation{Space Telescope Science Institute, 3700 San Martin Drive, Baltimore, MD 21218, USA}

\author{Robert Zhu}
\affiliation{Department of Astronomy, University of California, Berkeley, CA 94720, USA}

\author[0000-0003-4019-0673]{Enrico M. Di Teodoro}
\affiliation{Dipartimento di Fisica e Astronomia, Universit\`{a} degli Studi di Firenze, I-50019 Sesto Fiorentino, Italy}

\author[0000-0002-0355-0134]{Jessica Werk}
\affiliation{Department of Astronomy, University of Washington, Seattle, WA 98195, USA}

\author[0000-0002-1129-1873]{Mary Putman}
\affiliation{Department of Astronomy, Columbia University, New York, NY 10027, USA}

\author[0000-0002-5456-523X]{Anna F. McLeod}
\affiliation{Centre for Extragalactic Astronomy, Department of Physics, Durham University, South Road, Durham DH1 3LE, UK}
\affiliation{Institute for Computational Cosmology, Department of Physics, University of Durham, South Road, Durham DH1 3LE, UK}

\author[0000-0003-3520-6503]{Yakov Faerman}
\affiliation{Department of Astronomy, University of Washington, Seattle, WA 98195, USA}

\author[0000-0002-6386-7299]{Raymond C. Simons}
\affiliation{Department of Physics, University of Connecticut, 196A Auditorium Road Unit 3046, Storrs, CT 06269 USA}

\author[0000-0003-4797-7030]{Joshua Peek}
\affiliation{Space Telescope Science Institute, 3700 San Martin Drive, Baltimore, MD 21218, USA}

\begin{abstract}

The Large Magellanic Cloud (LMC) is home to many H{\sc\ ii} regions, which may lead to significant outflows. We examine the LMC's multiphase gas ($T\sim10^{4-5}$ K) in H{\sc\ i}, S{\sc\ ii}, Si{\sc\ iv}, and C{\sc\ iv} using 110 stellar sight lines from the HST's Ultraviolet Legacy Library of Young Stars as Essential Standards (ULLYSES) program. 
We develop a continuum fitting algorithm based on the concept of Gaussian Process regression and identify reliable LMC interstellar absorption over $v_{\rm helio}=175-375$ km s$^{-1}$. Our analyses show disk-wide ionized outflows in Si{\sc\ iv} and C{\sc\ iv} across the LMC with bulk velocities of $|v_{\rm out, bulk}|\sim20-60$ km s$^{-1}$, which indicates that most of the outflowing mass is gravitationally bound. 
The outflows' column densities correlate with the LMC's star formation rate surface densities ($\Sigma_{\rm SFR}$), and the outflows with higher $\sigsfr$ tend to be more ionized. 
Considering outflows from both sides of the LMC as traced by C{\sc\ iv}, we conservatively estimate a total outflow rate of $\dot{M}_{\rm out}\gtrsim 0.03~M_\odot {\rm yr}^{-1}$ and a mass loading factor of $\eta\gtrsim 0.15$.
We compare the LMC's outflows with those detected in starburst galaxies and simulation predictions, and find a universal scaling relation of $|v_{\rm out, bulk}|\propto \Sigma_{\rm SFR}^{0.23}$ over a wide range of star-forming conditions ($\Sigma_{\rm SFR}\sim10^{-4.5}-10^{2}~M_\odot {\rm yr}^{-1} {\rm kpc}^{-2}$).
Lastly, we find that the outflows are co-rotating with the LMC's young stellar disk and the velocity field does not seem to be significantly impacted by external forces; we thus speculate on the existence of a bow shock leading the LMC, which may have shielded the outflows from ram pressure as the LMC orbits the Milky Way. 

\end{abstract}


\section{Introduction}
\label{sec:intro}

Stellar feedback is a multiscale process. It requires a detailed understanding of small-scale star-forming regions, large-scale structures such as the interstellar medium (ISM) and the circumgalactic medium (CGM), as well as the delicate interplay among these gaseous structures \citep{mckee77}. 
Feedback-driven outflows enrich the CGM with metals, momentum, and energy; and, theoretical studies find that presence of stellar feedback is key to producing a realistic galaxy and a gaseous CGM with multiphase properties consistent with observations \citep[e.g.][]{vogelsberger14, hopkins14, schaye15, peeples19}. 

\begin{figure*}[t]
    \centering
    \includegraphics[width=\textwidth]{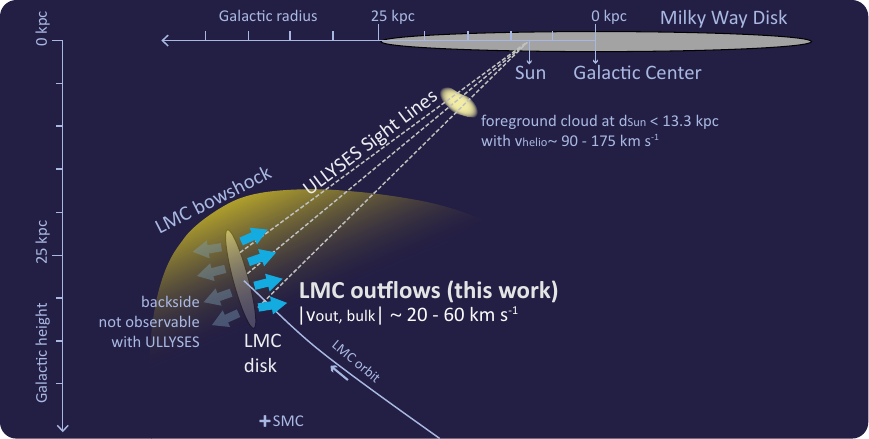}
    \caption{A schematic illustration of the present-day LMC with respect to the Milky Way disk. The locations, orientations, and sizes of the LMC (star-forming) disk and a potential bow shock are based on a hydrodynamic simulation of the LMC orbiting in the MW halo \citep[][see their Figure 4]{setton23}; the simulation assumes the LMC to be on its first infall \citep{besla07}. The location of the Small Magellanic Cloud (SMC) is indicated as a cross. We highlight in blue arrows the LMC's outflows on its near side, which we study in this work using 110 sight lines from the ULLYSES DR5. We also show the location of a foreground high-velocity cloud at $d_\odot<13.3$ kpc \citep{richter15, werner15}, which contaminates potential LMC outflow absorption over $\vhelio\sim90-175~\kms$. Our work focuses on the LMC gas at $\vhelio=175-375~\kms$ to minimize foreground contamination. }
    \label{fig:lmc_sketch}
\end{figure*}

Galactic outflows have been ubiquitously observed in star-forming galaxies \citep{veilleux20}. For example, the \NaI\ 5890/5896 \AA\ doublet probes dusty outflows with velocities up to $\sim1000~\kms$ in infrared (ultra)luminous starburst galaxies \citep[e.g.][]{heckman00, martin05, rupke05, chen10}. These outflows tend to be neutral; otherwise \NaI\ would not exist with its ionization potential at 5.1 eV. \cite{rubin14} studied cool outflows ($T\sim10^{4}$ K) in 
%105 
star-forming galaxies at $0.3<z<1.4$ using \MgII\ and \FeII\ doublets and found an outflow detection rate of $\sim66\%$ (see also \citealt{weiner09, erb12, davis23}). Warmer ionized outflows ($T\sim10^{4-5.5}$ K) in star-forming or starburst galaxies can be traced with numerous ions in the ultraviolet (UV) such as \SiII, \SiIII, \SiIV, \CIV, and \OVI\ with velocities up to a few hundreds of $\kms$ \citep[e.g.][]{heckman15, chisholm15, xu22_classy3, sirressi24}. Though with different tracers, a common finding among the above studies is that the velocities of outflows, regardless of their phases, correlate significantly with host galaxies' star formation activities, stellar masses, and circular velocities. 

%Observations of outflows in the literature: 
%\cite{rupke05} NaI infrared luminuous 78 starburst galaxies at $z<0.5$. 
%\cite{chen10} outflows detected in NaI lines, requires low ionizing flux, so only useful in dusty galaxies where the gas are shielded from photoionization.
%\cite{rubin14} study MgII and FeII outflows in 105 star-forming galaxies at $0.3<z<1.4$. 
%\cite{weiner09} MgII outflows of 1406 star forming galaxies at $z\sim1.4$. 
%\cite{erb12} outflows in FeII.  

%Star formation in dwarf galaxies has been observed to be enhanced in situations where the dwarf is likely undergoing some type of interaction.   For instance, \cite{stierwalt15} finds star formation is enhanced in dwarf galaxy pairs with the smallest projected separations.   The Fornax Cluster has a population of star-formation enhance dwarf galaxies for which the kinematics indicate they are falling into the cluster \citep{drinkwater01}.

For starburst galaxies, the bulk velocities of outflows  correlate with the galaxies' star formation rates (SFRs) as a power-law, $\vout\propto{\rm SFR}^{\alpha}$, with $\alpha\sim0.2-0.35$ \citep[e.g.][]{martin05, chisholm15, rupke18, xu22_classy3}. The power-law index $\alpha$ is shallower when considering the correlation between $\vout$ and SFR per surface area $\sigsfr$ ($\alpha\sim0.1-0.2$; e.g. \citealt{chen10, xu22_classy3, ReichardtChu24}). \cite{chu22} argue that these power-law indexes are indicative of energy-driven outflows, where the energy is mostly conserved as outflows break out of the ISM and propagate into the CGM \citep[see also][]{chen10, li17, kim20}.

Observationally, outflows have been measured largely based on stacking spectra from galaxies with similar physical properties to maximize spectral signal-to-noise ratios \citep[e.g.][]{chen10}, or collecting a sufficient sample with one sight line per galaxy to cover a wide parameter space \citep[e.g.][]{xu22_classy3}. While these approaches provide invaluable information on outflows over galactic scales, it remains unclear how outflows interact with their ambient environments on smaller scales. From theoretical perspectives, how outflows are generated and propagated in realistic environments such as the Solar neighborhood and varying star formation conditions have been an active area of research \citep[e.g.][]{li17, kim18, kim20, andersson23, tan23}; however, these simulations remain largely unconstrained because of the scarcity of observational details on sub-kpc scales. 

In this work, we examine how varying star-forming conditions impact the physical properties of ionized outflows in the Large Magellanic Cloud (LMC). The LMC is the closest galaxy that hosts many bright \HII\ regions, which makes it an ideal site to study how star formation drives outflows. Active star-forming regions are found across the LMC, such as 30 Doradus (30 Dor), N11, N44, N55, and N206 \citep{ambrocio-cruz16, mcleod19}. Table \ref{tb:lmc_info} lists the key physical parameters of the LMC, and Figure \ref{fig:lmc_sketch} illustrates the location and movement of the present-day LMC with respect to the Milky Way (MW). At a distance of 50.1 kpc \citep{freedman01} and moving in the MW halo at a Galactocentric velocity of $321~\kms$ \citep{kallivayalil13}, the LMC experiences strong headwinds due to ram pressure, which results in a truncated \HI\ disk \citep{salem15} and a potential bow shock leading the LMC \citep{setton23}. 

\begin{table}
\footnotesize
\centering
\caption{Key Physical Parameters of the LMC}
\begin{tabular}{ccl}
\hline
Parameter & Value & Reference(s)  \\ 
\hline 
%\hline  
$d$ & 50.1 & \citeauthor{freedman01} \\ 
(distance) & (kpc) & (\citeyear{freedman01})\\ 
%\hline 
$v_{\rm LMC, LSR}$ & $264.0\pm0.4$ & \citeauthor{choi22} \\ 
(systemic velocity) & ($\kms$) & (\citeyear{choi22})\\ 
%\hline 
$v_{\rm LMC, G}$ & $321\pm24$ & \citeauthor{kallivayalil13} \\ 
(Galactocentric velocity) & ($\kms$) &  (\citeyear{kallivayalil13})\\ 
%\hline 
$i$ & $23.4\pm0.5$ & \citeauthor{choi22} \\ 
(inclination) & (deg) & (\citeyear{choi22})\\ 
%\hline 
$\mstar$ & $3\times10^9$ & \citeauthor{vandermarel02} \\ 
(stellar mass) & ($\msun$) & (\citeyear{vandermarel02})\\ 
%\hline 
$M_{\rm HI}$ & $4.4\times10^8$ & \citeauthor{bruns05} \\ 
(\HI\ gas mass) & ($\msun$) & (\citeyear{bruns05})\\ 
%\hline 
$Z$ & 0.5  & \citeauthor{russell92}\\ 
(metallicity) & ($Z_\odot$)  & (\citeyear{russell92})\\ 
%\hline 
SFR & $\sim0.2$ & \citeauthor{harris09}\\ 
(present-day)    & ($\msunyr$) & (\citeyear{harris09})\\ 
%\hline 
$v_{\rm rot, *}$ & $77.5\pm1.3$ & \citeauthor{choi22} \\ 
(stellar rotation$^{[1]}$) & ($\kms$) & (\citeyear{choi22})\\ 
%\hline 
$v_{\rm rot, HI}$ & $\sim70$ & \citeauthor{kim98} \\ 
%(\HI\ rotation$^{[2]}$) & $\sim40~\kms$ & \cite{oh22}\\ 
(\HI\ rotation) & ($\kms$) & (\citeyear{kim98})\\ 
%\hline 
R.A.$^{[2]}$ (J2000) & $80.443$ & \citeauthor{choi22}\\ 
(LMC center) & (deg) & (\citeyear{choi22})\\ 
%\hline 
Decl.$^{[2]}$ (J2000) & $-69.272$ & \citeauthor{choi22}\\ 
(LMC center) & (deg) & (\citeyear{choi22})\\ 
\hline
\end{tabular}
\tablecomments{
\small
[1]: The stellar rotation is fitted for a population of young red supergiants and evolved old red giant branch and asymptotic giant branch stars. 
%[2]: \HI\ rotation curve derived by \cite{oh22} is $\sim30~\kms$ lower than that from \cite{kim98} because the \cite{oh22} value is corrected for transverse, nutation, and procession motions as well as the asymmetric drift in the LMC \HI\ gas. 
[2]: The LMC's kinematic center is derived based on $\sim$10,000 red-giant branch stars, asymptotic giant branch stars, and red supergiant stars. 
% {\red Q for Yumi: \cite{vandermarel02} noted that their dynamical center based on carbon stars is offset from the kinematic center of the \HI\ gas by $1.2\degree\pm0.6\degree$. Is this also true for the LMC RA/DEC that you used in your paper?}.
}
\label{tb:lmc_info}
\end{table}

Gas inflows and outflows have been detected using down-the-barrel observations toward individual massive stars bright in the UV in nearby galaxies \citep[e.g.][]{howk02, danforth02, lehner07, zheng17}. For the LMC, \cite{wakker98} detected \CIV\ absorption with velocities offset from the galaxy's \Halpha\ emission using five stars observed with the Goddard High-Resolution Spectrograph on the Hubble Space Telescope (HST), which they interpreted as evidence for a hot halo around the LMC. \cite{barger16} compared ion absorption toward a pair of an LMC star and a background QSO that are $\sim100$ pc in projected separation; while the star only probes outflows in front of the LMC, the QSO sight line shows nearly symmetrical ion absorption due to outflows from both sides of the galaxy.

\begin{figure*}[t]
    \centering
    \includegraphics[width=\textwidth]{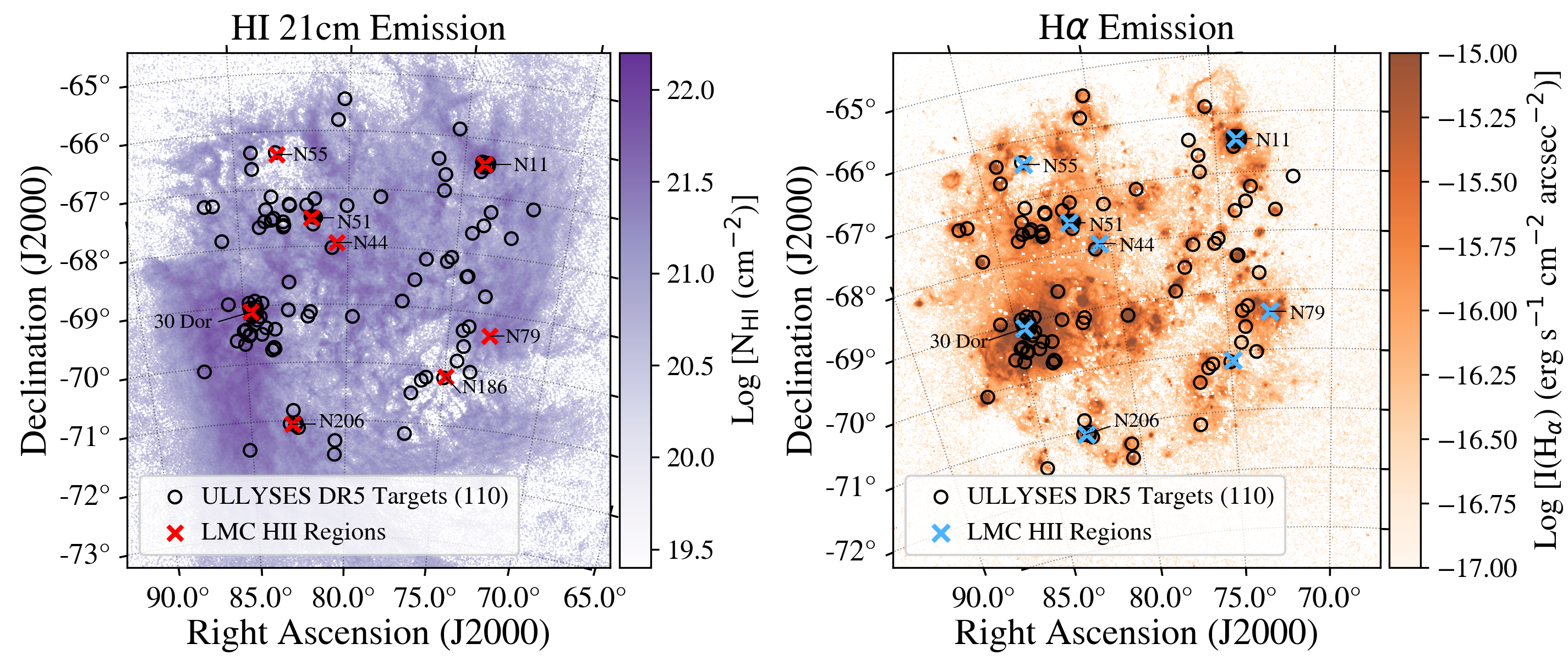}
    \caption{Distribution of 110 ULLYSES DR5 stellar sight lines (black circles) across the LMC. The left panel shows an \HI\ column density map \citep{kim03} and the right panel shows a continuum-subtracted H$\alpha$ intensity map of the LMC \citep{guastad01}. Red and blue crosses in the left and right panels indicate several major \HII\ regions in the LMC, respectively. We introduce the ULLYSES dataset in Section \ref{sec:data_uv}, and the \HI\ and \Halpha\ datasets in Section \ref{sec:data_HI_Halpha}. 
    }
    \label{fig:map_HI_Halpha}
\end{figure*}

Thanks to the HST's Ultraviolet Legacy Library of Young Stars as Essential Standards (ULLYSES\footnote{\label{footnote1}
%ULLYSES is a Director's Discretionary program at Space Telescope Science Institute that provides a comprehensive HST UV spectral library for young stars in various Galactic and low-metallicity extragalactic environments such as the LMC, SMC, NGC3109, and Sextans A. 
https://ullyses.stsci.edu/}) program \citep{Roman-Duval20_ullyses}, we are now able to probe the LMC's outflows on sub-kpc scales using over a hundred UV sight lines (see Figure \ref{fig:map_HI_Halpha}). This manuscript is the first in a series in which we investigate how the interplay between ram pressure and stellar feedback affects the kinematics and ionization structures of outflows and inflows in the LMC (\#HST-AR-16640, PI Zheng\footnote{\label{footnote2}https://www.stsci.edu/cgi-bin/get-proposal-info?id=16640\&observatory=HST}).

\begin{figure*}
    \centering
    \includegraphics[width=\textwidth]{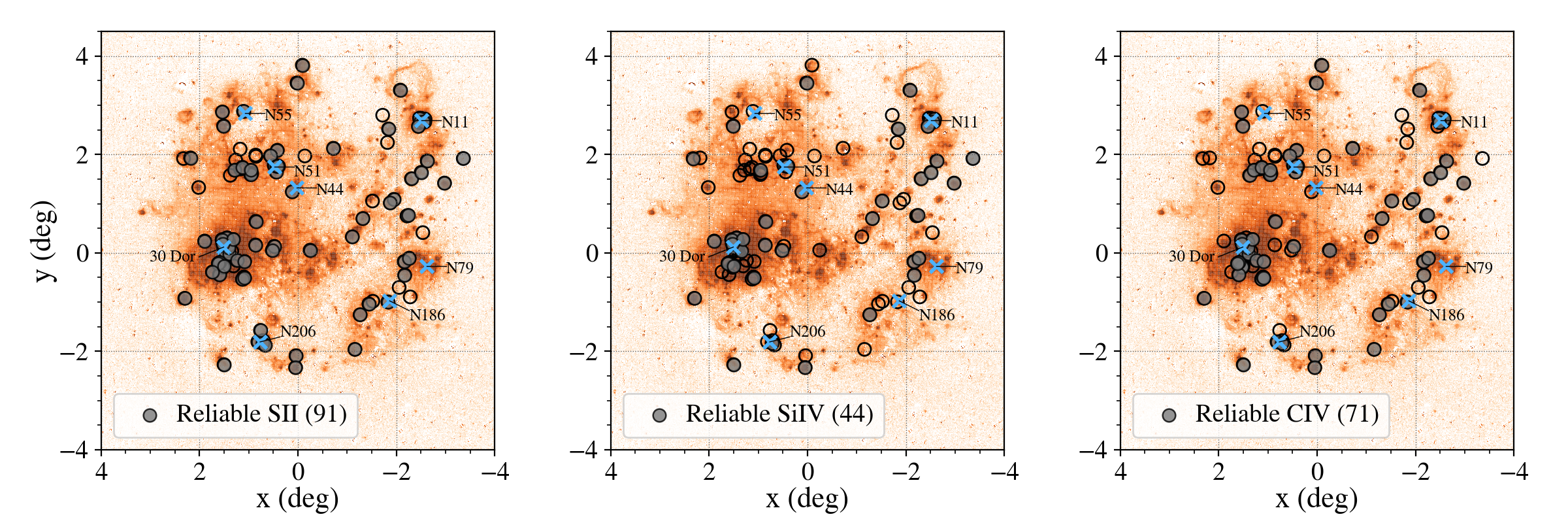}
    \caption{Distribution of the ULLYSES sight lines against the \Halpha\ map in an orthographic projection in the LMC plane, following the method outlined in \cite{choi22}. 
    At the distance of the LMC (50.1 kpc; \citealt{freedman01}), 1 deg $\approx$ 0.9 kpc. Major \HII\ regions are indicated by blue crosses.
    Filled gray circles indicate those sight lines with reliable \SII\ (91/110; left), \SiIV\ (44/110; middle), and \CIV\ (71/109; right) measurements over the LMC's absorption range of $\vhelio=175-375~\kms$. An ion measurement is considered reliable if: (1) its stellar continuum shows a strong P-Cygni profile such that the LMC's interstellar absorption can be reliably identified, and (2) the column density difference between doublet lines is within what is allowed by the apparent optical depth method \citep{Savage91, Savage96}.  
    See Section \ref{sec:data_uv} for further details.}
    \label{fig:map_xy_proj}
\end{figure*}

This paper is organized as follows. In Section \ref{sec:data_uv}, we describe the ULLYSES LMC dataset and relevant spectral analyses. In Section \ref{sec:data_HI_Halpha}, we introduce auxiliary datasets in \Halpha, \HI, and a sample of red supergiant stars to trace the LMC's recent star formation, and neutral and stellar disk kinematics. We show the main results in Section \ref{sec:result}, and compare the LMC's outflows with those detected in starburst galaxies in Section \ref{sec:comparison_wind_literature}. We also compare the observations with outflow simulation predictions in Section \ref{sec:comparison_wind_literature}. We discuss the implications of our work in Section \ref{sec:discuss} and conclude in Section \ref{sec:conclusion}. 

We release our data products, including normalized \SII, \SiIV, and \CIV\ lines and their corresponding best-fit continuum models (when available), as a High Level Science Product called ``LMC-FLOWS" at the Barbara A. Mikulski Archive for Space Telescopes (MAST) via: \dataset[10.17909/hz0m-np43]{\doi{10.17909/hz0m-np43}} \citep{lmc-flows}. Details on the UV data reduction can be found in Section \ref{sec:data_uv}. 

Throughout this paper, the velocity is given in a heliocentric frame, unless otherwise specified. Toward the direction of the LMC, the heliocentric velocity $\vhelio$ and a velocity defined in the Local Standard of Rest (LSR) is generally offset by $\vhelio-\vlsr\sim10~\kms$. We note that outflow velocities are typically measured in two ways in the literature: centroid velocities tracing bulk outflow mass (e.g., \citealt{heckman15}), or maximum velocities tracing terminal velocities of low-density outflowing gas (e.g., \citealt{chisholm15}). In this work, we adopt the first definition to describe the kinematic properties of bulk outflows in the LMC, $\vout$, unless otherwise specified.

\section{Data: UV Absorption}
\label{sec:data_uv}
%%%%%% uv data sample %%%%  

\subsection{ULLYSES DR5 Sample Information}
\label{sec:dr5_info}

We use far UV spectra of 110 massive stars in the LMC that were made public in the ULLYSES's fifth data release (DR5, 2022 June 28, DOI: \dataset[10.17909/t9-jzeh-xy14]{\doi{10.17909/t9-jzeh-xy14}}; \citealt{ullyses_data_doi}). Figure \ref{fig:map_HI_Halpha} shows the distribution of the ULLYSES DR5 stellar sight lines in the LMC against background images of \HI\ 21cm \citep{kim03} and H$\alpha$ maps \citep{guastad01}. 
%The later DR5b includes 17 additional T Tauri stars, but does not affect other data products that were released in DRx5 earlier. 

We are interested in those ULLYSES targets that were observed with the G130M and G160M gratings of the Cosmic Origins Spectrograph (COS), and/or the E140M grating of the Space Telescope Imaging Spectrograph (STIS). The COS data have spectral resolutions of R=12,000--16,000 ($\delta v\approx$19--25 $\kms$) in G130M and 
%cenwave 1291, 1096
R=13,000--20,000 in G160M ($\delta v \approx$15--23 $\kms$; COS Instrument Handbook, \citealt{cos_handbook23}), 
%cenwave 1611
%and R=16,000--20,000 in G185M ($\approx$15--19~$\kms$; COS Instrument Handbook, \citealt{cos_handbook23}). 
%cenwave 1953, 1986. 
and the STIS E140M data have a spectral resolution of R=45,800 ($\delta v \approx 6.6~\kms$; 
%and R=30,000 ($\approx$10 $\kms$) in E230M 
STIS Instrument Handbook, \citealt{stis_handbook}). While the STIS data have a higher spectral resolution, the COS data offer a better sensitivity; this essentially means that STIS was used to observe bright stars while COS was targeted at fainter stars. 

We use the coadded spectra released by the ULLYSES program and refer the reader to \cite{Roman-Duval20_ullyses} and the ULLYSES's data release page$^{\ref{footnote1}}$ for more information. When available, we prioritize data taken with the STIS/E140M grating for the higher spectral resolution. We do not consider data taken with other gratings such as STIS/E140H or COS/G140L, which are less common among the ULLYSES targets; the only exception to this is the \SII\ and \SiIV\ measurements toward star SK-67D83, which has both COS/G130M and STIS/E140H data, we use the STIS/E140H spectrum for its higher resolution (R=114,000 or $\delta v\approx 2.6~\kms$). By design, both the STIS and COS data from the ULLYSES program yield a continuum signal-to-noise ratio of 20--30 per resolution element. 

%The spectral resolution for Echelle Medium-Resolution Mode E140M with $0.2\arcsec\times0.2\arcsec$ is FWHM = 1.3--1.4 in pixels at 1500-1200 \AA (see \href{https://hst-docs.stsci.edu/stisihb/chapter-13-spectroscopic-reference-material/13-6-line-spread-functions/echelle-line-spread-functions}{link}).

% Hereafter, we mainly use the *\_cspec.fits data products that have combined spectra with common instrument and gratings (taking into account different cenwave settings). This means for COS spectra, the spectra were combined for grating of G130M, G160M, and G185M if available. For STIS, the spectra were combined from E140M and E230M for medium resolution, and E140H and E230H for high resolution. (We don't use the g130l-g430l-g750l option). And for FUSE spectra, the spectra were from lwrs or mdrs. 

\subsection{Far UV Line Choices: \SII, \SiIV, and \CIV\ lines}
\label{sec:line_choice}

There are a number of far UV ions that are typically used to study gas flows in nearby galaxies, such as \SiII, \SiIII, \SiIV, \CIV, and \OVI\ \citep[e.g.,][]{wakker98, howk02, lehner07, chisholm15, chisholm16a, barger16, zheng17}. We focus on the \SiIV\ 1393/1402 \AA\ and \CIV\ 1548/1550 \AA\ doublets in the LMC. We choose \SiIV\ and \CIV\ because they are relatively less saturated than \SiII\ and \SiIII\ in the LMC, and thus provide a more accurate characterization of the gas kinematics. \SiIV\ and \CIV\ trace a cool-warm ionized phase with $T\sim10^{4-5}$ K, which is found to contain most of the mass in an outflow in hydrodynamic simulations of feedback-driven outflows \citep[e.g.][]{li17, kim20, rathjen21}.

Although a large fraction of the ULLYSES sight lines also have \OVI\ 1031/1037 \AA\ spectra from the Far Ultraviolet Spectroscopic Explorer (FUSE), we do not use \OVI\ in this work because the \OVI\ 1037 \AA\ line is in a region with multiple contaminants, and the \OVI\ 1031 \AA\ is complicated due to the broad interstellar \OVI\ absorption blended with stellar wind features with unknown continuum shapes. The \OVI\ 1031 \AA\ line is also contaminated by H$_2$ absorption \citep[see][]{howk02}.

The \SiIV\ and \CIV\ doublets are in spectral regions with no other contaminating ISM lines. But, one of the main challenges in analyzing \SiIV\ and \CIV\ is that the widths of the stellar absorption lines are, for some stars in the sample, comparable to the widths of the interstellar absorption lines along the LMC's lines of sight. 
In the next section, we develop an evaluation matrix to select stars with smooth continua that allow accurate stellar continuum modeling over the almost $500~\kms$ range spanned by interstellar absorption from the MW, intervening halo gas, and the LMC.

We also analyze \SII\ 1250/1253 \AA\ lines that trace a less ionized phase of the LMC's ISM.
%with a formation potential of 10.4 eV (from \SI\ to \SII) and an ionization potential of 23.3 eV (from \SII\ to \SIII). 
We do not use the \SII\ 1259 \AA\ line because it is blended with \SiII\ 1260 \AA\ from the MW's ISM. In Figure \ref{fig:map_xy_proj}, we show in filled circles the ULLYSES sight lines with reliable \SII\ (91/110), \SiIV\ (44/110), and \CIV\ (71/109\footnote{There are only 109 stars with \CIV\ coverage; star SK-66D17 was only observed with COS/G130M in the ULLYSES DR5.}) measurements for the LMC's interstellar absorption; we describe how we determine reliable ion measurements in the following sections. The atomic data, including accurate wavelengths and oscillator strengths, are adopted from \cite{morton03}.

%%YZ: For this OVI paragraph, I'm thinking we can leave it for now. I put it there in the first place to briefly indicate that we do plan to analyze OVI even though it won't be in this paper. But maybe that disclaimer isn't necessary. 
%%from KT: OVI 1037 line is in a region with multiple contaminants, OVI 1031 is complicated because OVI absorption tends to be broad and because of the uncertain shape of OVI wind features. There is H2 contamination but that's a smaller issue. 
%Lastly, we note that while ULLYSES does include archival spectra from {\it Far Ultraviolet Spectroscopic Explorer} (FUSE) that probes \OVI\ absorption, we defer the analysis of \OVI\ to future work because the \OVI\ lines require a sophisticated treatment on accurately modeling the stellar \OVI\ P-Cygni profile due to highly ionized winds as well as contamination from molecular hydrogen $H_2$ lines from the MW's and LMC's ISM at similar wavelengths \citep{howk02}. 

\begin{figure*}
    \centering
    \includegraphics[width=\textwidth]{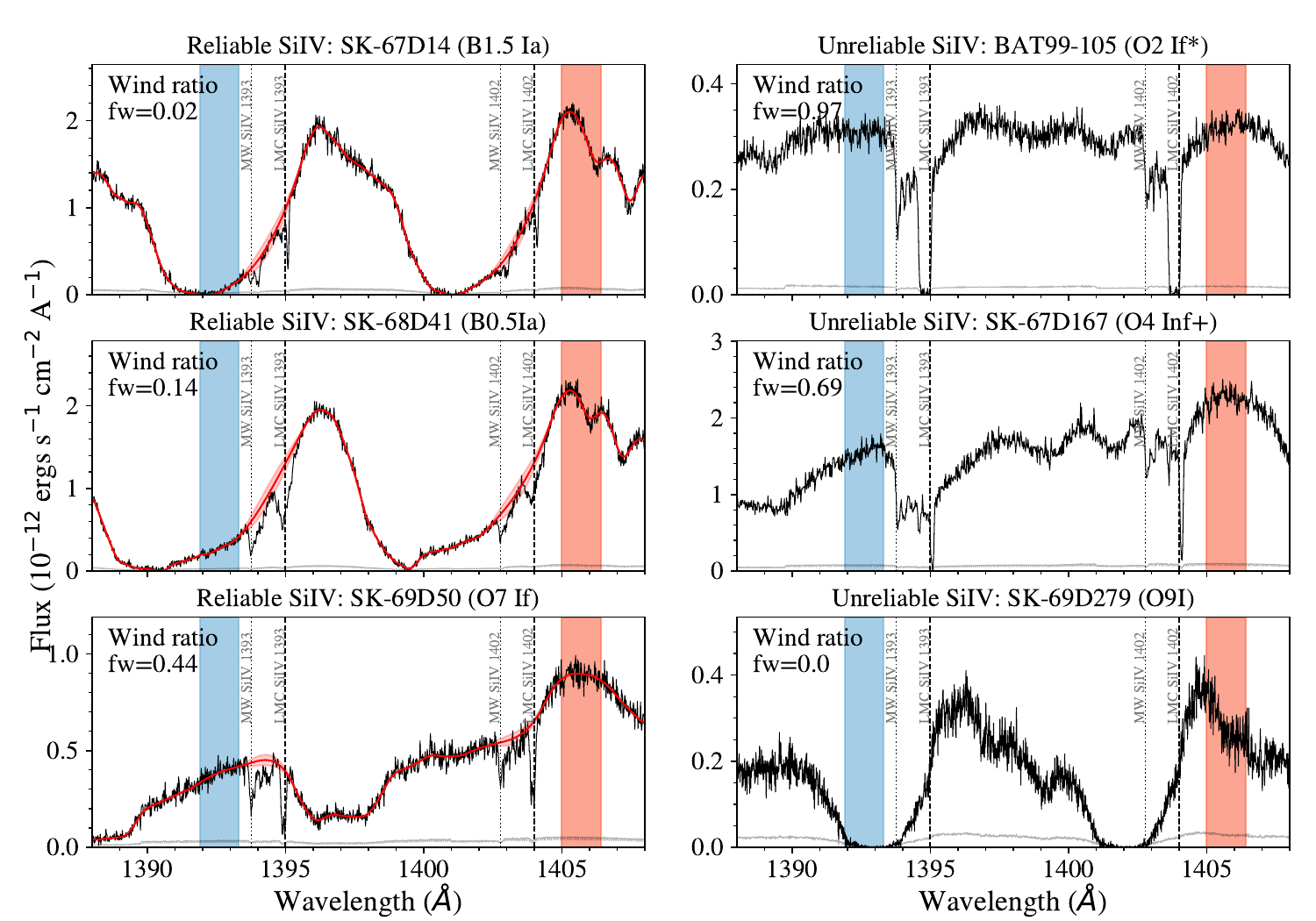}
    \caption{Left: Example spectra with well-developed P-Cygni profiles in \SiIV\ (see Section \ref{sec:wind_ratio}), with fluxes in black curves and errors in gray. The MW and LMC \SiIV\ absorption can be reliably separated from the stellar P-Cygni profiles. We show in red curves the best-fit continuum models, which we discuss in Section \ref{sec:method_contfit}. Right: Example spectra that have either mild or no stellar winds, in which case the \SiIV\ absorption from stellar photospheres blends heavily with the MW and LMC interstellar absorption; we consider these \SiIV\ unreliable. In each panel, the blue and red vertical shades indicate the stellar continuum regions that we use to calculate the wind ratios $f_w$ (Equation \ref{eq:wind_ratio}) to quantify the strengths of stellar winds/P-Cygni profiles.
    }
    \label{fig:goog_bad_spec}
\end{figure*}

\subsection{Selection of Stars with Well-Developed P-Cygni Profiles in \SiIV\ or \CIV}
\label{sec:wind_ratio}

Among the 110 LMC stars in the ULLYSES DR5, there are 57 O-type stars, 28 B-type stars, 15 Wolf-Rayet stars, and 10 with other types such as binaries or luminous blue variables\footnote{Three stars are labelled either as an O type or a Wolf-Rayet star, including LMCe055-1 (WN4/O4), SK-67D22 (O2If*/WN5), and VFTS 482 (O2.5 If*/WN6); for a classification purpose, we group them under the Wolf-Rayet category. Two binaries, HD38029 (WC4+OB) and SK-69D246 (WN5/6h+WN6/7h), are also grouped under the Wolf-Rayet category, in which one or both of the stars are Wolf-Rayet type. The category classification does not significantly affect the results shown in this work.}. The key to extracting reliable \SiIV\ and \CIV\ interstellar absorption lines is to identify OB stellar spectra with well-developed P-Cygni profiles \citep{savage81, howk02}. The P-Cygni profiles often take the form of redshifted emission peaks with optically thick, blueshifted absorption troughs due to absorption and subsequent re-emission of photons in stellar winds \citep{kudritzki00}.

We only consider ULLYSES stars that meet the following criteria: (1) the stars have developed winds/P-Cygni profiles in \SiIV\ or \CIV, (2) the winds have \textsl{high optical depths} such that the fluxes of the absorption troughs are low or approaching zero, and (3) the winds have \textsl{high terminal velocities} such that the blue edges of the troughs are far from line centers. 

Criterion 1 excludes stars with stellar absorption that may have similar widths to the MW and LMC interstellar absorption. Criterion 2 excludes stars with complex spectral shapes, because it is difficult to accurately model pseudo-continua over the interstellar absorption regions for P-Cygni profiles with low optical depths. Lastly, we implement criterion 3, excluding winds with low terminal velocities, because a large fraction of a narrow P-Cygni profile will contain interstellar absorption.

In the left panels of Figure \ref{fig:goog_bad_spec}, we show three examples of \SiIV\ line profiles that meet criteria 1--3. In the right panels, we show another three examples where the stars have either no recognizable \SiIV\ winds (top right), low opacity winds (middle right), or low terminal-velocity winds that cause sharp rising profiles (bottom right). We note that, the \SiIV\ and \CIV\ spectra of the 15 Wolf-Rayet stars are generally dominated by broad stellar features that can be easily distinguished from interstellar absorption, so we use these Wolf-Rayet stars without considering criteria 1--3. % Below, we elaborate on applying criteria 1--3 to the remaining 97 stars. 

To apply criteria 1--3 quantitatively, we develop an automated algorithm based on a general observation that stronger stellar winds with high opacities result in deeper blueshifted absorption troughs where the fluxes approach zero \citep[e.g.][]{hawcroft23}. We define a {\it wind ratio} parameter, $f_w$, which is the depth of a blue absorption trough with respect to the height of a red emission peak. In practice, $f_w$ is computed as the ratio of the median flux over $\vhelio=[-400, -100]~\kms$ in the rest frame of the bluer line (\SiIV\ 1393 or \CIV\ 1548) to the median flux over $\vhelio=[475, 775]~\kms$ in the rest frame of the redder line (\SiIV\ 1402 or \CIV\ 1550). In Figure \ref{fig:goog_bad_spec}, we highlight these two flux regions in blue and red vertical shades, respectively.

The reasoning for the $f_w$ parameterization is as follows. We do not use the ratio of the minimum absorption to the maximum emission fluxes because this is only applicable to spectra that have developed P-Cygni profiles. For those without obvious P-Cygni profiles (e.g., top right panel of Figure \ref{fig:goog_bad_spec}), the locations of the minimum and maximum fluxes are subject to local spectral variations, and in many cases the MW/LMC lines are the strongest absorption features. Secondly, 
%given the proximity of the doublet lines ($\sim9$\AA\ for \SiIV\ and $\sim2$\AA\ for \CIV), 
stellar winds with high terminal velocities will have blended \SiIV\ or \CIV\ profiles instead of distinct peaks and troughs; this is best seen in star SK-69D50 (bottom left in Figure \ref{fig:goog_bad_spec}) where the red peak of \SiIV\ 1393 is absorbed by the blue trough of \SiIV\ 1402. 
% Therefore, we choose to measure the median absorption flux blueward of the bluer lines and the median emission flux redward of the redder lines to account for the doublet blending. 
Lastly, we measure a median absorption flux over $[-400, -100]$ $\kms$ blueward of the bluer lines to avoid the MW's ISM absorption near $\sim0~\kms$; similarly, we measure a median emission flux over $[475, 775]~\kms$ redward of the redder lines to avoid the LMC's interstellar absorption. The $300~\kms$ velocity interval reduces the impact of noise, local spectral variations, and differences in wind terminal velocities.

We examine a set of \SiIV\ and \CIV\ P-Cygni profiles of OB stars from the Potsdam Wolf-Rayet (PoWR) models \citep{Hainich19} at the LMC's metallicity and determine that a threshold at 
\begin{equation}
f_w=\frac{\langle I_{\rm blue}\rangle}{\langle I_{\rm red}\rangle}\leq0.6
\label{eq:wind_ratio}
\end{equation}
can best provide an adequate diagnostic to select stars with well-developed winds. The number of stars that pass the threshold only changes by less than ten when we vary the $f_w$ threshold by $\pm0.1$
%($<6$ for \CIV, and $<9$ for \SiIV)
. For \CIV, Equation \ref{eq:wind_ratio} is analogous to the ``good" or ``best" quality scores set by \cite{hawcroft23} when estimating \CIV\ terminal velocities for 67 OB stars in the ULLYSES LMC dataset, where the minimum flux of the blue absorption trough is roughly less than half of the continuum average. In Section \ref{sec:method_summary}, we will use Equation \ref{eq:wind_ratio} to select stars with reliable interstellar absorption in \SiIV\ and \CIV.

\subsection{Continuum Fitting, Ion Column Densities, and Centroid Velocities}
\label{sec:method_contfit}

We describe our continuum fitting algorithm using the \SiIV\ doublet in Figure \ref{fig:goog_bad_spec} as an example, and note that the same procedure is applied to \SII\ and \CIV. For \SiIV\ and \CIV, the continuum fitting is performed for every star independent of its spectral type or wind ratio $f_w$ (Equation \ref{eq:wind_ratio}).

For each ion doublet, we select a spectral region that covers $\sim$5--10\AA\ blueward of the bluer line and $\sim$5--10\AA\ redward of the redder line\footnote{The exact width of the spectral region does not matter as long as it covers the doublet interstellar absorption and provides enough stellar continuum as a training set for Gaussian Process.}. 
%For \SiIV, this means we select a spectral region from $\sim1383-1388$\AA\ to $\sim1407-1402$\AA, 
We mask a velocity range\footnote{Because COS's line-spread-function (LSF) is broader and less well-defined than STIS's, interstellar absorption lines in COS appear to be $\sim$30--50 $\kms$ broader than those in STIS. For this reason, the velocity mask for each line is determined by visual inspection, and the mask chosen for a COS absorption line is generally $\sim50~\kms$ wider than that of STIS.} of $\vhelio\sim[-100, 380]~\kms$ at the rest frame of each line to cover both the MW and LMC interstellar absorption. To predict stellar continuum over the masked velocity region, a typical approach is to fit low-order Legendre polynomials to absorption-free regions near the lines of interest \citep[e.g.][]{howk02, lehner07, lehner09, barger16, zheng17}. However, as shown in Figure \ref{fig:goog_bad_spec}, stellar continuum is highly variable from star to star and from line to line. 
%; each line requires a considerable amount of effort to manually determine the fitting region, velocity mask, continuum model function, and goodness of the fit for most appropriate outcomes. 
To automate the fitting process and reduce human biases in the continuum placement, we develop a continuum-fitting algorithm based on the concept of Gaussian Process regression \citep{GP-book} and the open-source package \texttt{George}\footnote{https://george.readthedocs.io/en/latest/} \citep{george}.

\begin{figure*}[t]
    \centering
    \includegraphics[width=\textwidth]{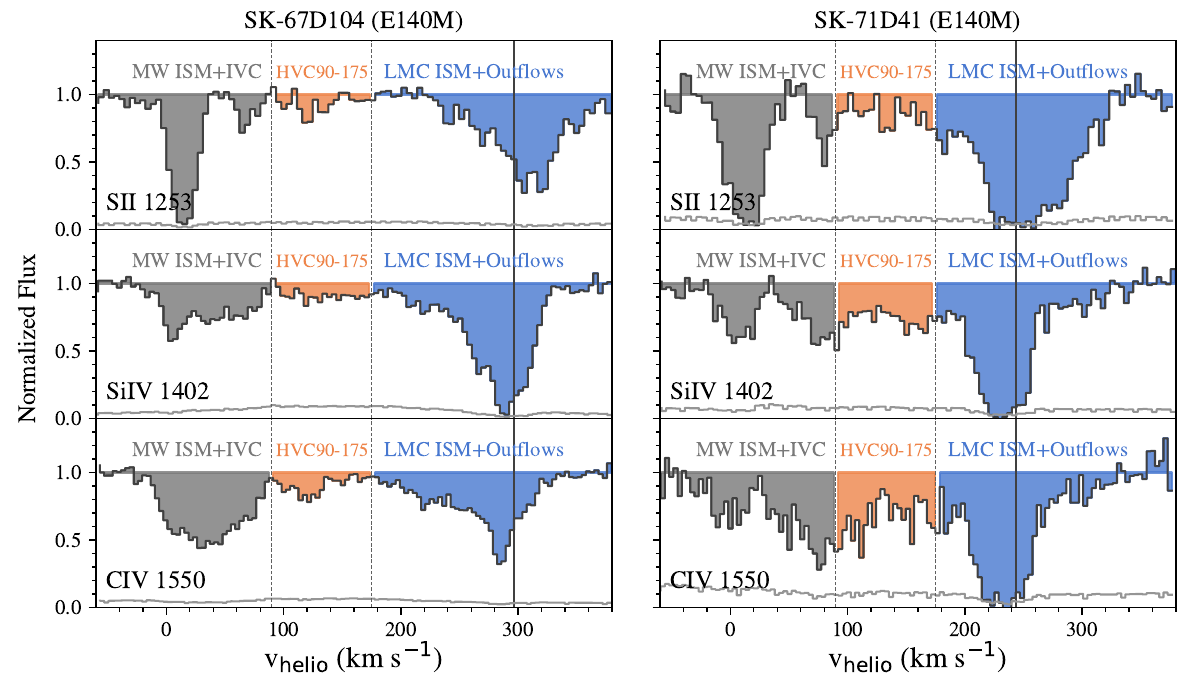}
\caption{Normalized \SII, \SiIV, and \CIV\ spectra for SK-67D104 (left) and SK-71D41 (right). We highlight three velocity components: (1) the MW's ISM and intermediate velocity cloud at $v\lesssim90~\kms$ (gray), (2) a high-velocity cloud at $90\lesssim v \lesssim 175~\kms$ and $d_\odot<13.3$ kpc in the foreground (HVC90-175; orange), and (3) the LMC's ISM and outflows at $175\lesssim v \lesssim 375~\kms$ (blue). The relative spatial locations of the three components are sketched in Figure \ref{fig:lmc_sketch}. The solid vertical lines show the velocities of the LMC's stellar disk at the locations of the stars (see Section \ref{sec:data_HI_Halpha}). 
%The absorption from HVC90-175 and the LMC (ISM+outflows) can be well separated near $v=175~\kms$ along most sight lines (82/91 for \SII, 34/44 for \SiIV, and 49/71 for \CIV).
The left panel shows an example sightline where the absorption from HVC90-175 and the LMC (ISM+outflows) can be well separated near $v=175~\kms$, while the right panel shows an example where the two components are blended. We quantify the degree of blending between HVC90-175 and the LMC absorption near $v=175~\kms$ in Section \ref{sec:method_contfit}.
}
\label{fig:spec_example}
\end{figure*}

Instead of assuming a particular function form (e.g., polynomials), Gaussian Process is a non-parametric process that models the probabilistic distributions of all available model functions. We refer to the unmasked part of the \SiIV\ stellar continuum (free of interstellar absorption) as the training set $X$ and the masked velocity region where we want to predict stellar continuum shape as the test set $Y$. The first step of Gaussian Process is to estimate the probability distribution function of the training set, $\mathcal{N}(\mu_X, \Sigma_X)$, where $\mu_X$ is the sample mean and $\Sigma_X$ is a covariance matrix that describes the correlation of every spectral point $x_i$ with itself and every other spectral point in $X$. We estimate the covariance matrix $\Sigma_X$ by applying a kernel function to model the training set $X$ that takes the form of either an exponential-squared kernel 
%($k(r^2)=\sigma^2$exp(-$\frac{r^2}{2}$)) 
or a Matern 3/2 kernel. 
%($k(r^2)=\sigma^2$(1+$\sqrt{3r^2}$)exp(-$\sqrt{3r^2}$)). 
%can best describe the overall trend or shape of the unmasked stellar continuum. 
We then use the \texttt{scipy.minimize} function to fit for the maximum likelihood parameters for the kernel function and use that to predict the stellar continuum shape over the test set $Y$ (i.e., the masked velocity region).

In the left panels of Figure \ref{fig:goog_bad_spec} where three examples of reliable \SiIV\ interstellar absorption are shown, we plot the best-fit continua as red curves with $1\sigma$ uncertainties. We divide the fluxes by the best-fit continua to normalize each doublet's lines. The uncertainties of the normalized fluxes have combined the original flux errors and the continuum fitting uncertainties through error propagation.

We calculate ion column densities based on the apparent optical depth method (AOD; \citealt{Savage91, Savage96}). The AOD method can also be used to test whether a stellar continuum is placed correctly, which we describe as follows. For an ion line with normalized fluxes of $I_n(v)$\footnote{For some STIS spectra from the ULLYSES DR5, when a line is saturated, the fluxes near the line center appear to be lower than the errors at the same velocities. In this case, we replace those fluxes with the corresponding error values in the AOD calculation.}, its apparent column density as a function of velocity is, 
\begin{equation}
\begin{split}
    N_a(v) & = 3.768\times10^{14}\frac{\tau_a(v)}{f\lambda ({\rm \AA})}~[{\rm cm^{-2}~(\kms})^{-1}] \\ 
    N_a & = \int_{\rm vmin}^{\rm vmax} N_a(v)dv~,
\end{split}
\label{eq:aod_N}
\end{equation}
where $f$ is the oscillator strength, $\lambda$ the rest wavelength in unit of \AA, and $\tau_a(v)$ the apparent optical depth smeared by an instrumental broadening profile.

We adopt a fixed velocity range of $[v_{\rm min}, v_{\rm max}]=[175, 375]~\kms$ to measure the integrated ion column densities of the LMC. The left bound is chosen to avoid contamination from a foreground high-velocity cloud (HVC) at $v\sim90-175~\kms$ within 13.3 kpc from the Sun \citep[see Figure \ref{fig:lmc_sketch};][]{lehner09, richter15, werner15, roman-duval19}. And the right bound is chosen such that the integration range is wide enough to cover the entire LMC absorption. Figure \ref{fig:spec_example} shows two examples where in one case (left panel) the absorption from the foreground HVC and the LMC can be clearly distinguished near $v=175~\kms$, while in the other case (right panel) the two structures blend mildly together near $v=175~\kms$. 

We quantify whether the HVC and the LMC absorption are well separated by measuring the mean absorption flux of an ion line (\SII\ 1253, \SiIV\ 1393, \CIV\ 1548) over a velocity range of $v=175\pm5~\kms$. If the mean flux is more than 80\% of the continuum flux, such as the case in the left panels of Figure \ref{fig:spec_example}, we consider the line to have well separated HVC and LMC absorption. We check all reliable \SII, \SiIV, and \CIV\ normalized lines and confirm that along most sight lines (82/91 for \SII, 34/44 for \SiIV, and 49/71 for \CIV) the HVC and LMC absorption can be well separated at $\sim175~\kms$ -- the spectra shown in the left panels are more common. Thus, setting $v_{\rm min}$ at $175~\kms$ minimizes contamination from the foreground HVC.

The \SII, \SiIV, and \CIV\ doublets all have a doublet ratio of $f_1\lambda_1/f_2\lambda_2=2$, where the subscript 1 is for the stronger line and 2 for the weaker line. This means the integrated column density ratio of an ion doublet is: 
\begin{equation}
    \frac{N_{\rm 2,a}}{N_{\rm 1,a}}=\frac{[\int \tau_{\rm 2,a}(v)dv]/f_2\lambda_2}{[\int \tau_{\rm 1,a} (v)dv]/f_1\lambda_1}=\frac{\int \tau_{\rm 2,a}(v)dv }{\int \tau_{\rm 1,a}(v)dv }\times 2~~.
\label{eq:doublet_ratio}
\end{equation}
%Following the discussion in \cite{Savage91, Savage96} and \cite{jenkins96}, 
In cases where both lines of a doublet are fully resolved without saturation, we expect $N_{\rm 1,a}=N_{\rm 2,a}$ and the column density difference between the doublet lines to be $\Delta \log_{10} N\equiv\log_{10} N_{2,a}-\log_{10} N_{1,a}=0$. On the other hand, when both lines are fully saturated with normalized fluxes near zero, we expect $\int\tau_{\rm 1,a}(v)dv\approx \int \tau_{\rm 2,a}(v)dv$, and the column density difference between the doublet lines to be $N_{\rm 2,a}/N_{\rm 1,a}\approx2$ or  $\Delta \log_{10} N\approx0.3$ dex \citep{Savage91, jenkins96}. Therefore, for lines that are moderately saturated, the column density difference between the doublet lines should be: 
\begin{equation}
    -\sigma_{\rm N}\leq \Delta \log_{10} N \leq +\sigma_{\rm N}+0.3~{\rm dex}~~~,
\label{eq:aod_criteria}
\end{equation}
where $\sigma_{\rm N}$ is the uncertainty tolerance set by the quadratic sum of the uncertainties in $N_{\rm 1,a}$ and $N_{\rm 2,a}$. In Section \ref{sec:method_summary}, we will combine Equation \ref{eq:aod_criteria} with the wind ratio threshold in Equation \ref{eq:wind_ratio} to select stars with reliable interstellar absorption in \SII, \SiIV, and \CIV.

\begin{figure}
      \centering
     \includegraphics[width=\columnwidth]{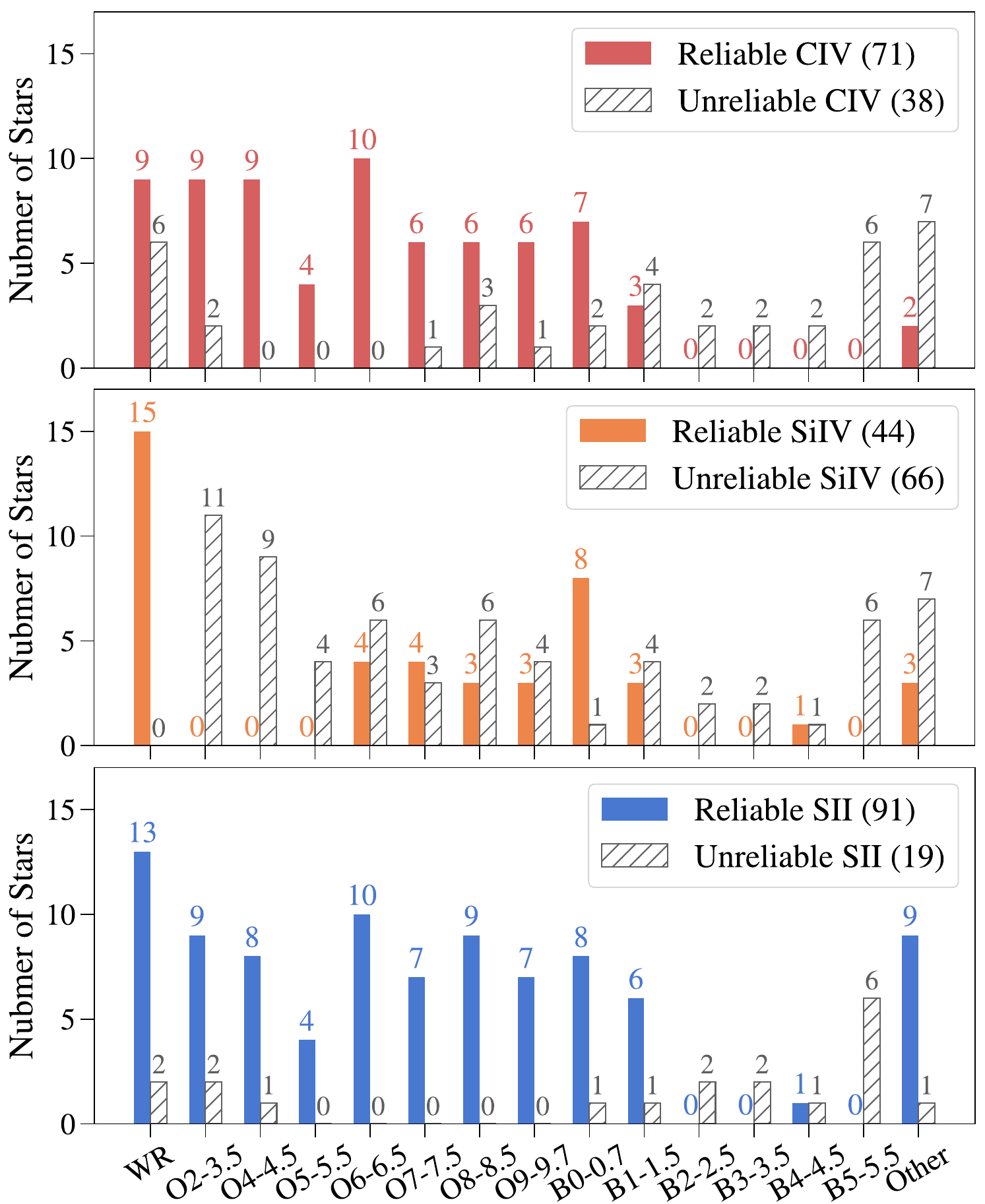}
     \caption{Distribution of stars with reliable \CIV\ (top), \SiIV\ (middle), and \SII\ (bottom panel) LMC interstellar absorption as a function of spectral type. Those sight lines that we adopt in our LMC analysis (i.e., reliable) are shown in blue, red, and orange, respectively, and those that we consider unreliable and thus discarded are shown in gray. See discussion in Section \ref{sec:method_summary}.}
     \label{fig:keep_skip}
 \end{figure}

We compute each line's centroid velocity weighted by the apparent optical depth over the same velocity range. The uncertainty on the centroid velocity is calculated by propagating the errors in the apparent optical depth array over the same velocity range, which are computed based on the continuum-normalized flux errors. Our centroid velocity calculation is similar to the weighted average velocity used in \cite{chisholm15, chisholm16a}, which traces the bulk motion of outflowing gas. While \cite{chisholm15, chisholm16a} show that $v_{90}$, velocity at 90\% of the continuum flux level, can trace low density gas and thus probe the terminal velocity of outflows (of the corresponding transition lines), we do not compute $v_{90}$ because of potential contamination from the foreground HVC at $\sim$90--175 $\kms$ toward some sight lines (see Figure \ref{fig:spec_example}). We discuss the physical properties of this HVC in Section \ref{sec:discuss_hvc}.

\subsection{Summary of Selection Rules and Final Target List}
\label{sec:method_summary}

We adopt two criteria to evaluate whether a best-fit continuum produces reliable \SII, \SiIV, or \CIV\ interstellar absorption lines: (1) whether we expect a star to have a smooth continuum as a result of being a Wolf-Rayet star or having $f_w\leq0.6$ (Equation \ref{eq:wind_ratio}, Section \ref{sec:wind_ratio}), and (2) whether the column density differences of the normalized ion doublet's lines are within theoretical values specified in Equation \ref{eq:aod_criteria} (Section \ref{sec:method_contfit}). Note that criterion 1 is only applied to \SiIV\ and \CIV\ to distinguish narrow ISM lines from broad P-Cygni profiles due to massive stars' stellar winds.

In total, we identify 71 ULLYSES stars (out of 109) with reliable \CIV\ continuum placement over the LMC's absorption range, 44/110 with reliable \SiIV, and 91/110 with reliable \SII. The spatial distribution of these reliable ion measurements across the LMC are shown in Figure \ref{fig:map_xy_proj}. We tabulate each ion's integrated column density and centroid velocity in Table \ref{tb:logN}.

Figure \ref{fig:keep_skip} shows the distribution of selected stars as a function of spectral type. Most stars with reliable interstellar \CIV\ absorption are Wolf-Rayet, O types, or B types earlier than B1, and most stars with reliable interstellar \SiIV\ are Wolf-Rayet, O types later than O6, or B types earlier than B2. There are only 44 stars in \SiIV\ passing our selection rules because most of the early O type stars are with winds that are highly ionized and thus with no significant \SiIV\ P-Cygni profiles; the lack of winds results in \SiIV\ stellar features spanning over similar wavelength widths as the MW and LMC interstellar absorption, such as the top right panel of Figure \ref{fig:goog_bad_spec}. In such cases, our algorithm flags the stars as with unreliable \SiIV. For \SII, the only rule that is used to select reliable interstellar absorption is whether the $\Delta \log_{10} N$ condition in Equation \ref{eq:aod_criteria} is satisfied. So the distribution in the \SII\ panel does not show a particular trend with spectral types, and most (91/110) stars are with reliable \SII\ continuum placement.

\section{Auxiliary Datasets:\\ H$\alpha$, \HI, and Red Supergiant Stars}
\label{sec:data_HI_Halpha}

We supplement the ULLYSES \ulldr\ sample with three additional datasets to estimate the LMC's star formation rate (SFR) surface density ($\sigsfr$; Section \ref{sec:data_halpha}), the total column density and bulk velocity of neutral hydrogen (\HI; Section \ref{sec:data_HI}), and the LMC's stellar disk kinematics (Section \ref{sec:data_rsg}). The derived $\sigsfr$, \HI\ column densities and centroid velocities, and line-of-sight stellar disk velocities are tabulated in Table \ref{tb:logN}.

\begin{figure}[t]
    \includegraphics[width=\columnwidth]{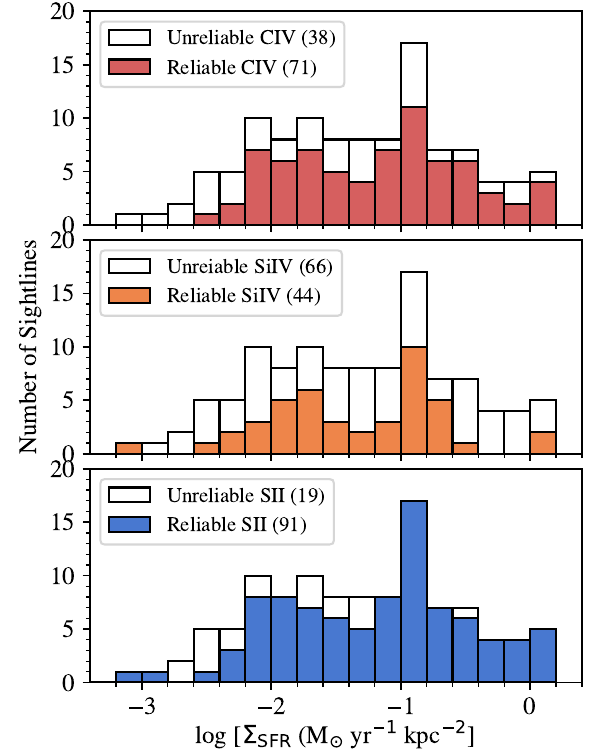}
    \caption{Histogram distributions of the ULLYSES \ulldr\ sight lines as a function of $\sigsfr$ in logrithmic values. Sight lines with reliable \CIV\ (top), \SiIV\ (middle), and \SII\ (bottom) are highlighted in blue, orange and red, respectively.}
    \label{fig:star_hist_sfr}
\end{figure}

\subsection{H$\alpha$ Emission from SHASSA} 
\label{sec:data_halpha}

We obtained a continuum-subtracted H$\alpha$ emission map of the LMC from the Southern H$\alpha$ Sky Survey Atlas (SHASSA; \citealt{guastad01}). The survey has a filter width of 32\AA\ and covers a spatial range of ${\rm Decl.}\sim(-90\degree, +15\degree)$ at a resolution of $\approx0.8\arcmin$ and   
%The SHASSA LMC H$\alpha$ intensity map is in unit of $dR$ (decirayleighs), where $1dR=10^5/4\pi~{\rm photons~cm^{-2}~s^{-1}~sr^{-1}}$. 
with a sensitivity of 2R or $1.2\times10^{-17}~{\rm ergs~cm^{-2}~s^{-1}~arcsec^{-2}}$. The H$\alpha$ map has been corrected for flux contribution from the [\NII] 6549/6585 doublet lines that fall within the filter (see section 4 in \citealt{guastad01}). % Because H$\alpha$ emission originates from recombination of ionized gas into neutral hydrogen atoms in \HII\ regions containing massive OB stars, 
Considering the LMC's metallicity and the SHASSA's filter width, the H$\alpha$ emission traces star-formation activities happening within the last $\sim$7--10 Myr \citep{haydon20}. We calculate the LMC's $\sigsfr$ values as follows. 
%the SFR estimated from H$\alpha$ indicates the lifespan of these stars within the past $\sim10$ Myr. % over the last few Myrs.
% read https://iopscience.iop.org/article/10.1086/323969/pdf

%\textcolor{magenta}{YC added: To measure the current star formation rate from H$\alpha$ luminosity, dust correction is necessary. Due to the lack of a contiguous H$\beta$ emission-line map for the LMC, constructing a Balmer decrement map for direct reddening correction in gas is not possible. \citet{Lah2024} calculated E(B-V) based on the Balmer decrement measurements using IFU observations in 83 discrete fields across the LMC's star-forming disk. Some ULLYSES fields overlap with their observing fields, but a $\sim$1 degree radius was required to find the closest fields for some ULLYSES sightlines, introducing significant reddening variation over that large scale.}

%\textcolor{magenta}{YC added: Instead, we used the red clump's reddening map from \citet{Choi2018} and converted the stellar reddening to nebular gas reddening with the scaling factor of 0.44 \citep{Calzetti2000}. The H$\alpha$ extinction was then computed using the LMC's average extinction curve \citep{Gordon2003}. For the two ULLYSES sightlines outside the \citet{Choi2018} map, SK-65D47 and SK-65D55, we obtained reddening values from the red clump's reddening map by \citet{Skowron2021}, which agrees well with \citet{Choi2018} around the main star-forming disk.}

We first convert the observed H$\alpha$ intensity $I({\rm H\alpha})$ to H$\alpha$ luminosity as $L_{\rm obs}({\rm H\alpha})=4\pi D^2 \theta^2 I({\rm H\alpha})$, where $D=50.1$ kpc \citep{freedman01} and $\theta$ is the angular size of the region of interest in units of arcsec. We factor in the inclination of the LMC and calculate $L_{\rm obs}({\rm H\alpha})$ for every $0.1\times0.1$ kpc$^2$ deprojected area in the LMC; the region size of $0.1\times0.1$ kpc$^2$ is to ensure sufficient sampling of the initial mass function such that the following luminosity to SFR relation holds \citep{kennicutt12}. We adopt 8\% uncertainties in H$\alpha$ fluxes \citep{guastad01} and propagate the errors in the following calculation.

To correct for dust attenuation, we first use a stellar reddening map from red clump stars by \citet{Choi2018} and calculate E(B-V)$_{\rm star}$ values for 108 out of the 110 ULLYSES stellar sight lines. For the remaining two sightlines outside the \citeauthor{Choi2018} map, SK-65D47 and SK-65D55, we obtain E(B-V)$_{\rm star}$ values from \citet{Skowron2021}, which agrees well with \citeauthor{Choi2018} around the main star-forming disk. We then estimate the corresponding nebular gas reddening values as E(B-V)$_{\rm gas}=$E(B-V)$_{\rm star}/0.44$ following the relation from \cite{calzetti97}. 

To evaluate gas reddening uncertainties, we compare our E(B-V)$_{\rm gas}$ values with those available from Balmer decrement measurements using integral field units observations of \HII\ regions by \cite{Lah2024}. For 31 ULLYSES sightlines that are within $<100$pc of \citeauthor{Lah2024}'s \HII\ regions, we find that our E(B-V)$_{\rm gas}$ values agree well with theirs with a median offset of $\sim15\%$. We thus adopt a uniform error of $15\%$ in E(B-V)$_{\rm gas}$ for all our ULLYSES sightlines and propagate the errors in the following calculation. 

Lastly, we compute the H$\alpha$ extinction values $A_{\rm H\alpha}$ using the LMC's average extinction curve \citep{gordon03} and derive the intrinsic H$\alpha$ luminosity as $L_{\rm int}({\rm H\alpha})=L_{\rm obs}({\rm H\alpha})/10^{(-0.4A_{\rm H\alpha})}$. We convert each region's $L_{\rm int}({\rm H\alpha})$ to SFR as $\log_{10} {\rm SFR}=\log_{10} L_{\rm int}({\rm H\alpha}) -41.27$, following the formulation in \cite{kennicutt12} which assumes an initial mass function from \cite{kroupa03}. The SFR surface density $\sigsfr$ is estimated by dividing the SFR value of each region by the corresponding size of $0.1\times0.1$ kpc$^2$, yielding a unit of $\msunyrkpc$.

Figure \ref{fig:star_hist_sfr} shows the distribution of the ULLYSES DR5 sight lines as a function of $\sigsfr$. A majority of the sight lines are in regions with $\sigsfr\sim10^{-2}-1 ~\msunyrkpc$, with a handful of sight lines directly probing either 30 Dor or other \HII\ regions (see Figures \ref{fig:map_HI_Halpha} and \ref{fig:map_xy_proj}). The clustering of the sight lines near major \HII\ regions ultimately affects the range of outflow environments that we will probe, which we discuss in Section \ref{sec:result}. We note that our spectral analyses algorithm do not exacerbate the sampling bias -- Figure \ref{fig:star_hist_sfr} shows that the histogram distributions of reliable \SII, \SiIV\ and \CIV\ measurements are consistent with the distributions of the original ULLYSES dataset.

%%%% 
\subsection{\HI\ 21cm Dataset}
\label{sec:data_HI}

The \HI\ data cube is a combination of an interferometry observation with the Australia Telescope Compact Array (ATCA) at 1$\arcmin$ resolution \citep{kim98}, and a single-dish observation with the Parkes multibeam receiver at 16.9$\arcmin$ \citep{kim03}. The combined data cube has a spatial resolution of 1$\arcmin$, a spectral resolution of $1.6~\kms$, and a flux sensitivity of $\sigma_{\rm T}\sim$2.4 K (or $\sim$15 mJy beam$^{-1}$), and it spans a velocity range from 190 to 386 $\kms$ in the heliocentric frame. 

For each ULLYSES DR5 sight line, we extract median \HI\ fluxes of all spatial pixels within a diameter of 1 beam (1 arcmin) of the sight line. We then integrate the \HI\ spectrum over its entire velocity range to obtain an estimate of the \HI\ column density $N_{\rm HI}$. % as: 
%\begin{equation}
%\begin{split}
%N_{\rm HI} & = 1.823\times10^{18}\int_{\rm vmin}^{\rm vmax} T_b(v) dv \\
%\sigma(N_{\rm HI}) & = 1.823\times10^{18} \sigma_{\rm T} n_v/\sqrt{n_v}dv
%\end{split}
%\label{eq:integ_HI}
%\end{equation}
%where $T_b(v)$ is the \HI\ brightness temperature, $({\rm vmin}, {\rm vmax})=(190, 386)~\kms$ the velocity range of the datacube, $dv=$1.6 $\kms$ the velocity resolution, $\sigma_{\rm T}=$2.4 K the sensitivity level, and $n_v=({\rm vmax}-{\rm vmin})/dv$ the number of velocity channels over the integration range. %The term $n_v-1$ is the degree of freedom assuming no correlation in noises in adjacent pixels. 
The centroid velocity $v_c$ is estimated as the flux-weighted velocity over pixels in the spectrum with fluxes higher than $2\sigma_{\rm T}$. The flux threshold here is to ensure that the centroid velocity of each spectrum reflects the kinematics of the majority of bright (dense) \HI\ gas along a line of sight. We find that this method better traces the center of mass for the \HI\ gas than the velocity estimated at a peak flux, especially in cases where there are multiple \HI\ velocity components toward some LMC regions \citep{kim03, oh22}.

We note that there are five sight lines with negative \HI\ fluxes from the combined data cube that indicate self-absorption: VFTS440, BAT99-105, VFTS-482, SK-65D47, and SK-67D266. % The first three are near 30 Dor, and the latter two are near two \HI\ supergiant shells LMC-4 and LMC-5 on the northern side of the LMC \citep{book08}, respectively. 
For these sight lines, we do not attempt to estimate the total $N_{\rm HI}$ or centroid velocities, and note ``self-abs" in the corresponding entries in Table \ref{tb:logN}. Additionally, there are three sight lines with low \HI\ signals (S/N$<3$); 
%while these sight lines indicate that there is little \HI\ gas along the lines of sight, the spectra are not useful in calculating centroid velocities. 
for these cases, we indicate $3\sigma$ upper limits in $N_{\rm HI}$, but do not obtain $v_c$.
% 09/26 note: we also look into GASKAP data and find some self absorption case also seen in these sightlines. Check yznotes_lmc_ullyses keynote for more details. 

\subsection{Kinematics of the LMC's Young Stellar Disk from Red Supergiant Stars}
\label{sec:data_rsg}

We adopt a kinematic model of the LMC's stellar disk that is based on the line-of-sight heliocentric velocities of 738 red supergiant stars (RSGs) analyzed by \cite{olsen11}. This model includes the effects of the LMC's bulk center-of-mass motion and internal rotation on the observed line-of-sight velocities, as discussed by \cite{vandermarel02}.  The RSGs represent a young ($\lesssim20$ Myr) stellar disk, and their internal rotation curve is found to be consistent with that of the LMC's \HI\ gas \citep{olsen11}. %The line-of-sight velocity dispersion of the RSGs is $\sim8~\kms$ \citep{olsen07}. 

At the locations of the ULLYSES DR5 sight lines, we calculate the line-of-sight velocities predicted by the RSG-based model; these are the velocities that stars would have at the corresponding locations if they resided in the LMC disk plane, shared the LMC's center-of-mass motion, and moved on circular orbits at speeds specified by the fitted rotation curve. We tabulate these model velocities as $v_{\rm RSG}$ in Table \ref{tb:logN}, and use $v_{\rm RSG}$ as the velocity reference of the LMC stellar disk to examine the relative motions of multiphase gas probed by \HI, \SII, \SiIV, and \CIV\ in the following sections.

%%%%%%%%%%%%%%%%%%%%%%%%%%%%%%%%%%%% RESULT %%%%%%%%%%%%%%%%%%%%%%%%%%%%%%%%%%%%

\begin{figure*}[t]
    \centering
    \includegraphics[width=\textwidth]{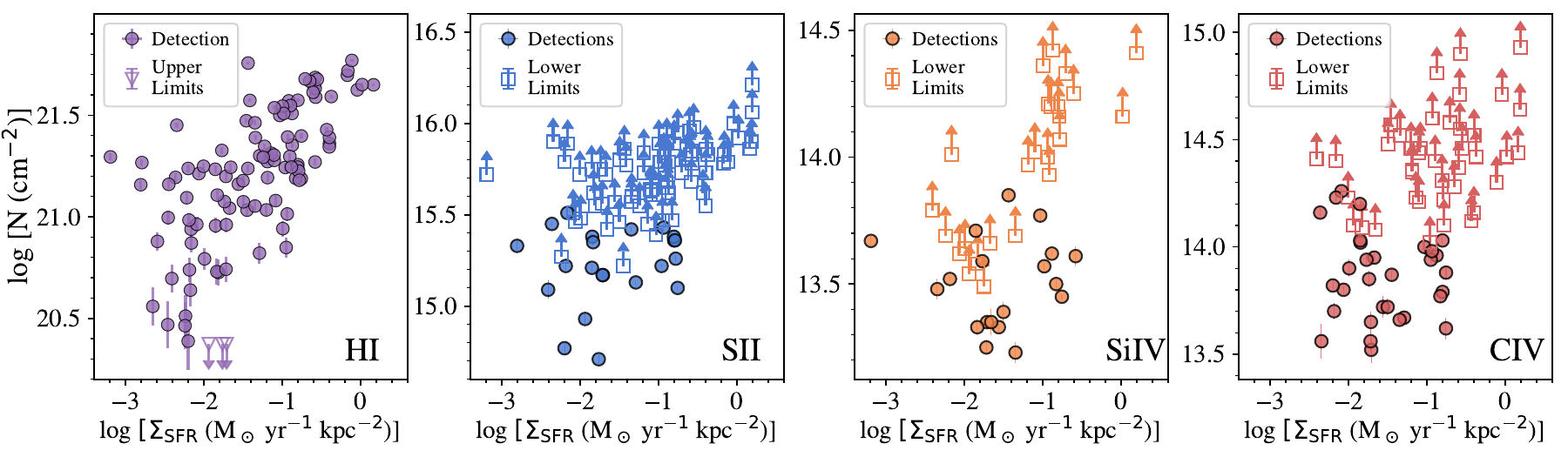}
    \includegraphics[width=\textwidth]{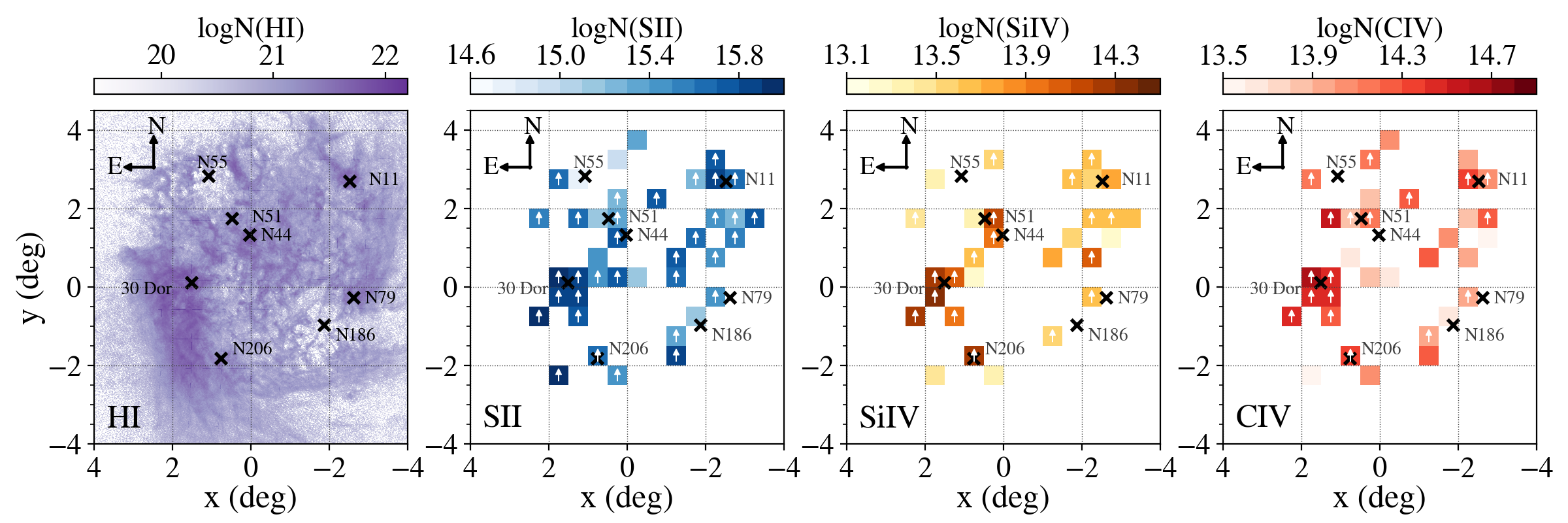}
    \caption{Top: gas column density as a function of $\sigsfr$ in logarithmic values for \HI, \SII, \SiIV, and \CIV, respectively. Solid circles are detections, open squares with upward arrows are saturations with lower limits (only in \SII, \SiIV, and \CIV), and open triangles with downward arrows are non-detections (only in \HI). The error bars on the x-axis ($\sim0.02-0.06$ dex) are smaller than the symbol sizes.
    Bottom: 2D spatial distributions of \HI, \SII, \SiIV, and \CIV\ column densities in the same orthographic projection as shown in Figure \ref{fig:map_xy_proj}. For \SII, \SiIV, and \CIV, the shown $\log N$ values/colors in bins noted with upward arrows should be considered as conservative lower limits. Several major \HII\ regions are highlighted as crosses in each panel. 
    We find high ion column densities correlate with regions with high star-forming activities, such as 30 Dor at $x\sim1.5\degree$ and $y\sim0\degree$. See Section \ref{sec:logN_vs_sfr} for further details.  
    }
    \label{fig:logN_SigSFR_xymap}
\end{figure*}

\section{Results}
\label{sec:result}

We discuss our main results that compare the \SII, \SiIV, and \CIV\ ion properties to other properties of the LMC such as the \HI\ gas content, $\sigsfr$, and stellar kinematics. The measurements of these properties are tabulated in Table \ref{tb:logN}. Specifically, we look into how the ionized gas kinematics is blue-shifted with respect to the \HI\ and stellar disk kinematics, indicating the presence of disk-wide outflows. %Furthermore, we examine how outflow velocity is correlated with gas phases and SFR. 

\begin{figure*}
    \centering
    \includegraphics[width=\textwidth]{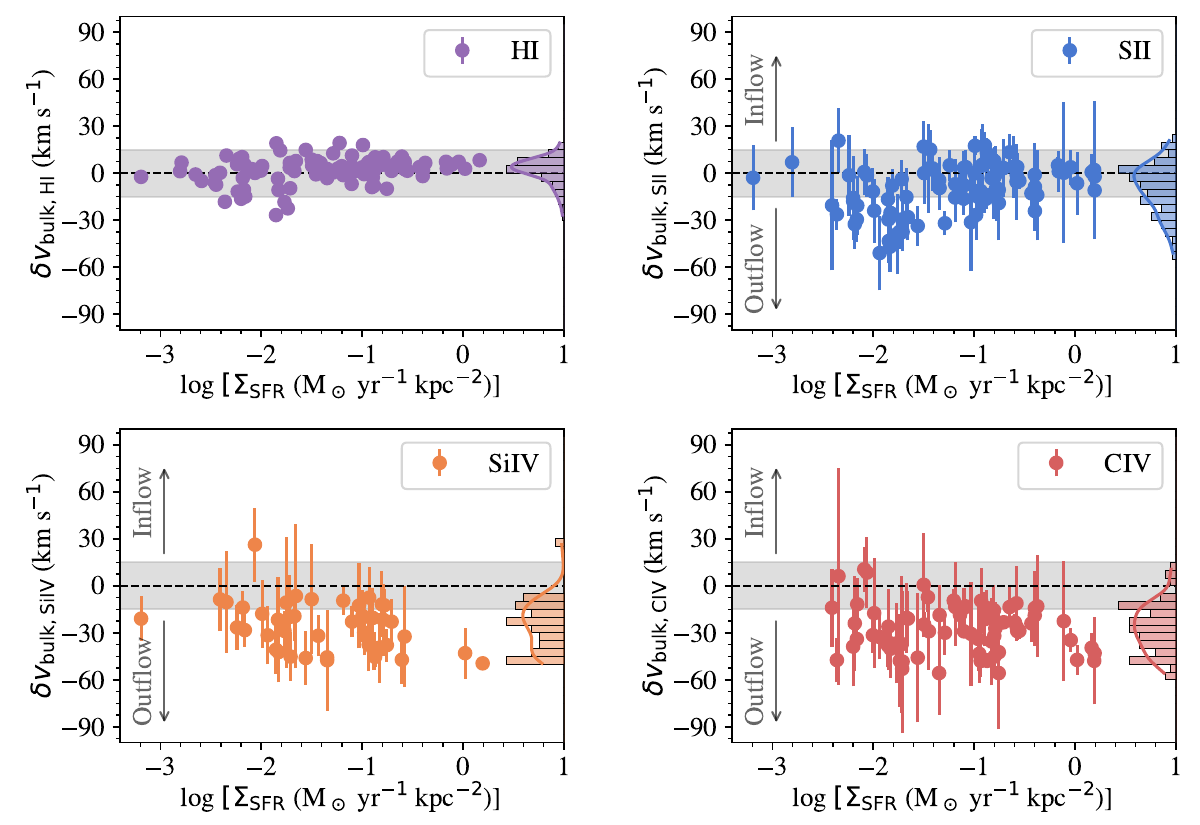}
    \caption{Velocity offsets between the multiphase gas (\HI, \SII, \SiIV, and \CIV) and the LMC's young stellar disk, $\delta v_{\rm bulk}$ (Equation \ref{eq:delta_v}), as a function of $\sigsfr$. The bars on the right sides show histogram distributions of $\delta v_{\rm bulk}$. The top left panel shows that the \HI\ gas largely follows the rotation of the LMC's young stellar disk. For \SII, \SiIV, and \CIV, negative offset velocities indicate outflows; we refer to the negative values at $\delta v_{\rm bulk}<-15~\kms$ as outflow velocities and discuss them in absolute values $|\vout|$. We conservatively do not consider data points within $|\delta v_{\rm bulk}|<15~\kms$ (gray bands) to avoid the LMC's ISM. Generally, we find outflows in \SII, \SiIV, and \CIV\ with bulk velocities of $|\vout|\sim20-60~\kms$, suggesting that the bulk mass of the outflowing gas should be gravitationally bound. At the higher $\sigsfr$ end, we only detect outflows in \SiIV\ and \CIV\ but not in \SII, suggesting that star-forming regions with higher $\sigsfr$ are likely to launch outflows that are more ionized. See Section \ref{sec:voutflow_vs_sfr} for further details. }
    \label{fig:delv_sigsfr}
\end{figure*}

\subsection{Ion Column Density vs. $\sigsfr$}
\label{sec:logN_vs_sfr}

In the top panels of Figure \ref{fig:logN_SigSFR_xymap}, we show the column densities of \HI, \SII, \SiIV, and \CIV\ with respect to $\sigsfr$ toward each line of sight. \HI\ and \SII\ trace neutral and low ionization gas in the LMC's ISM, and \SiIV\ and \CIV\ trace warm ionized gas at $T\sim10^{4-5}$ K. In general, we find that the \HI\ and ionized gas column densities all increase with $\sigsfr$ across the LMC disk.

For \SII, \SiIV, and \CIV, the column density thresholds above which the ion lines become saturated (i.e., lower $N$ limits) are around $\sim10^{15.4}~{\rm cm^{-2}}$, $\sim10^{13.6}~{\rm cm^{-2}}$, and $\sim10^{14.0}~{\rm cm^{-2}}$, respectively. The saturation rates increase toward regions with high $\sigsfr$; in general, all ions are saturated (100\%) at $\sigsfr\gtrsim 10^{-0.5}~\msunyrkpc$. We note that, by design, the ULLYSES sight lines are targeted at massive stars that reside in or near active star-forming regions (see Figure \ref{fig:map_xy_proj}).

In the bottom panels of Figure \ref{fig:logN_SigSFR_xymap}, we show the two-dimensional (2D) distributions of \HI, \SII, \SiIV, and \CIV\ column densities across the LMC disk in the same orthographic projection as in Figure \ref{fig:map_xy_proj}. We grid the \SII, \SiIV, and \CIV\ datasets into $0.5\degree\times0.5\degree$ bins and calculate mean ion column densities of all sight lines within each bin. Most bins have one sight line per bin; and there are $\sim2-8$ sight lines per bin for $\sim10$ bins near major \HII\ regions. For sight lines with lower limit $\log N$ values, we treat the lower limits as detections when calculating the mean and place an upward arrow in the corresponding spatial bin to indicate a lower limit. We have checked that the choice of mean or median values does not change the overall $\log N$ patterns shown in the 2D maps.

We find that regions with high ion column densities coincide with active star-forming sites, such as 30 Dor near $x\sim1.5\degree$ and $y\sim0\degree$. The correlation between ion column densities and $ \sigsfr$ suggests that star formation activities in the past $\sim7-10$ Myr, as traced by \Halpha\ emission, significantly impact the distribution of gas in all phases across the LMC. 

\subsection{Detection of Disk-Wide Outflows in the LMC}
\label{sec:voutflow_vs_sfr}

In Figure \ref{fig:delv_sigsfr}, we show how the bulk velocities of the LMC's multiphase gas offset from its young stellar disk kinematics as a function of $\sigsfr$. We define a velocity offset as 
\begin{equation}
\delta v_{\rm bulk, x}\equiv v_{\rm helio, x} - v_{\rm RSG}~~~, 
\label{eq:delta_v}
\end{equation}
where $v_{\rm helio, x}$ is the centroid velocity of an ion with x being \HI, \SII, \SiIV, or \CIV, and $v_{\rm RSG}$ is the line-of-sight velocity of the LMC's stellar disk as represented by red supergiant stars (see Section \ref{sec:data_rsg}). The values of $v_{\rm helio, x}$ and $v_{\rm RSG}$ for the ULLYSES DR5 sight lines can be found in Table \ref{tb:logN}.

\begin{figure*}[t]
\centering
\includegraphics[width=0.95\textwidth]{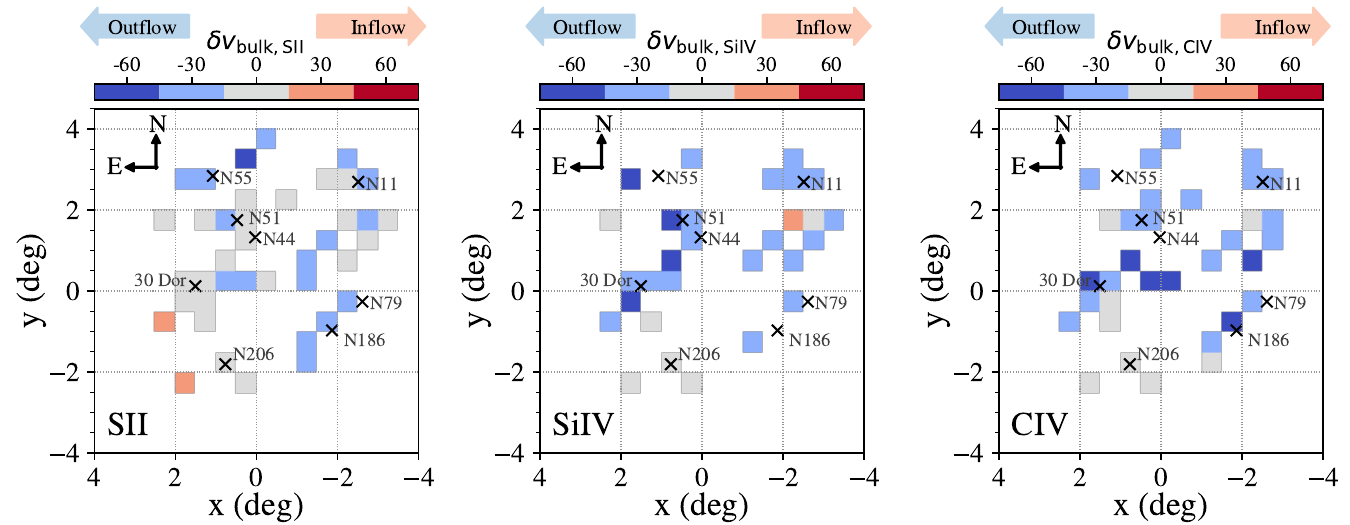}
\caption{2D distribution of $\delta v_{\rm bulk}$ (Equation \ref{eq:delta_v}) for \SII\ (left), \SiIV\ (middle), and \CIV\ (right) in the same orthographic projection as in Figure \ref{fig:map_xy_proj} and in unit of $\kms$. Gray indicates regions with ionized gas kinematics consistent with the LMC's stellar disk (and ISM) within $\pm15~\kms$, blue for bulk outflows and red for bulk inflows. Major \HII\ regions are indicated by crosses. We find disk-wide ionized outflows in \SII, \SiIV\ and \CIV\ with bulk velocities of $|\vout|\sim20-60~\kms$. See Section \ref{sec:voutflow_vs_sfr} for further details. 
\label{fig:delv_xymap}
}
\end{figure*}

For \HI, because the 21cm emission comes from both the front and back sides of the LMC disk, the signs in $\delta v_{\rm bulk, HI}$ cannot be used to diagnose inflows or outflows. The top left panel shows that most of the neutral gas probed by \HI\ has bulk velocities consistent with those of the red supergiant stars within $\sim$20 $\kms$. The similarity between \HI\ kinematics and that of the red supergiant stars suggests that the bulk mass of the \HI\ gas in the LMC is co-rotating with the underlying young stellar disk (see also \citealt{olsen07, olsen11}).

When considering the ionized gas, Figure \ref{fig:delv_sigsfr} shows that the bulk velocities of \SII, \SiIV, and \CIV\ are preferentially offset toward negative values. Given that the stellar sight lines only probe absorption by gas in the foreground of the stars, negative $\delta v_{\rm bulk}$ values in the ion panels indicate outflowing gas from the LMC toward our lines of sight. Hereafter, we refer to those data points with $\delta v_{\rm bulk}<-15~\kms$ as outflows, and discuss the outflow velocities in absolute values as $|\vout|$. 

Figure \ref{fig:delv_sigsfr} shows that the bulk velocities of the ionized outflows are over a range of $|\vout|\sim 20-60~\kms$. Note that the $|\vout|$ values are projected outflow velocities along our lines of sight, and they should be considered as lower limits to the actual outflow velocities in the LMC; in Section \ref{sec:comparison_wind_literature}, we discuss the impact of the LMC's inclination and other factors on the observed outflow velocities. 

We find \SII, \SiIV, and \CIV\ outflows commonly detected over $\sigsfr\sim10^{-2.5}-10^{-0.5}~\msunyrkpc$, and the histogram distributions on the y axes show that the \SiIV\ and \CIV\ outflows are moving faster than the \SII\ outflows by $\sim20-30~\kms$. Toward the higher $\sigsfr$ end, outflows are only detected in \SiIV\ and \CIV\ but not in \SII, which indicates that star-forming regions with high $\sigsfr$ are launching outflows that are likely to be more ionized. We further investigate whether the outflows with high ionization states and in regions with high $\sigsfr$ are preferentially associated with Wolf-Rayet stars or O types earlier than O5, but do not find any significant correlation.

For \SiIV, we find a significant correlation between the outflow velocities and $\sigsfr$ using Kendall's $\tau$ test with $p_{\rm SiIV}=0.04$ (bottom left panel); however, for \SII\ and \CIV, we do not find any significant correlation. In general, Figure \ref{fig:delv_sigsfr} shows large scatters in the outflow velocities with respect to $\sigsfr$. The scatters in the outflow velocities are likely to be caused by multiple factors, such as outflow opening angles, ages and locations (i.e., outflows launched at different times may reach different heights and may not correlate well with present-day SFR traced by H$\alpha$). We further discuss the scatters in Section \ref{sec:comparison_wind_literature}.

Lastly, in Figure \ref{fig:delv_xymap}, we show 2D spatial distributions of $\delta v_{\rm bulk}$ for \SII, \SiIV, and \CIV. The $\delta v_{\rm bulk}$ values are averaged over $0.5\degree\times0.5\degree$ bins to bring out the large-scale kinematic pattern across the LMC. The color bars are arranged such that gray pixels represent regions with ionized gas velocities consistent with the LMC's stellar disk (and ISM) within $\pm15~\kms$, and blue pixels highlight regions with outflows. We find that the \SiIV\ and \CIV\ outflows are commonly found near major \HII\ regions. Some \HII\ regions with high $\sigsfr$, such as 30 Dor, do not have outflows in \SII, which is likely due to outflows being more ionized in these regions, as is also shown in Figure \ref{fig:delv_sigsfr}.

\begin{figure*}[t]
\centering
\includegraphics[width=\textwidth]{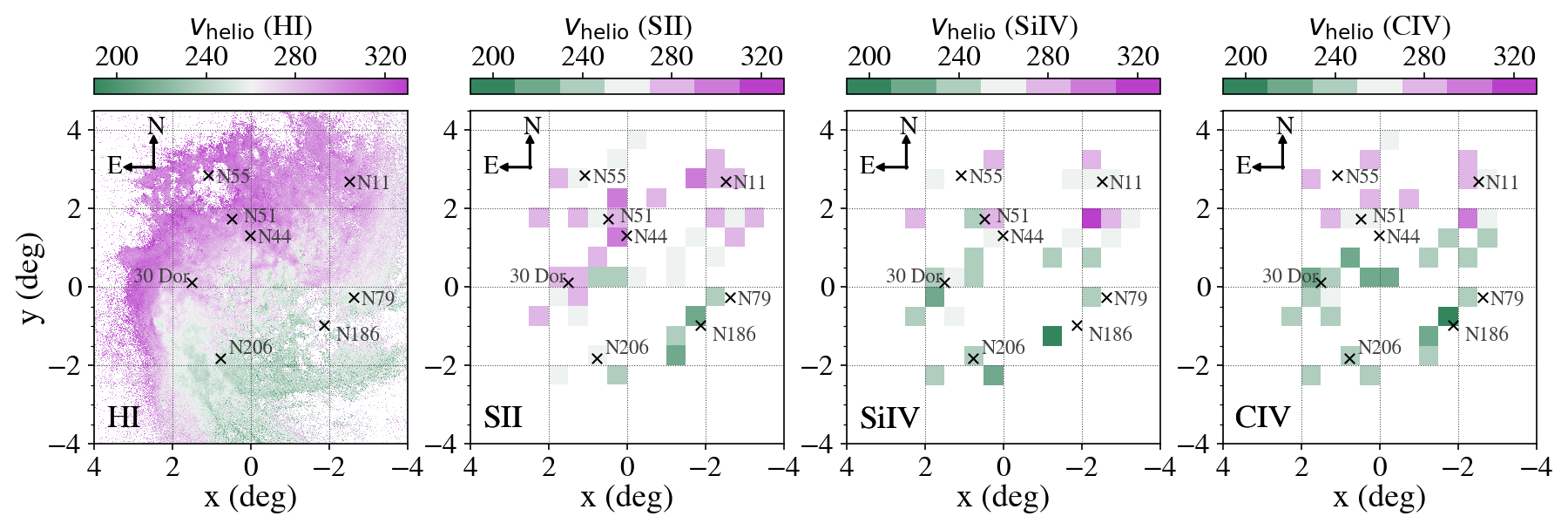}
\caption{2D distributions of \HI, \SII, \SiIV, and \CIV\ bulk velocities in the heliocentric frame and in unit of $\kms$. The systemic velocity of the LMC at the center of its mass (x,y)=(0$\degree$, 0$\degree$) is $v_{\rm sys}=264~\kms$ \citep{choi22}, which is shown as gray color. Major \HII\ regions are indicated by black crosses. All ions show signs of co-rotation with the LMC's disk. See Section \ref{sec:outflow_vrot} for further details. 
\label{fig:vhel_xymap}
}
\end{figure*}

Simulations have shown that a majority of outflow mass is found to be in the cool-warm phase \citep[e.g.][]{li17, kim20, rathjen21}, which is typically traced by the ions that are studied in this work. Given that the observed outflow velocities are within the escape velocity of the LMC near the disk ($v_{\rm esc}\sim90~\kms$; \citealt{barger16}), the bulk mass of the outflowing gas should be gravitationally bound to the LMC. The outflowing gas is thus likely to be part of the LMC's galactic fountain flows and would eventually reverse its course and become inflows toward the LMC at cooler phases, as typically seen in outflow simulations \citep[e.g.][]{kim18, kim20}. However, in Figure \ref{fig:delv_xymap}, the ionized outflows are commonly detected across the LMC disk in all ions, while inflows are not as common. The rare exceptions are two bins in \SII\ near $(x, y)\sim (2\degree, -1.5\degree)$, and one bin in \SiIV\ near $(x, y)\sim(-2\degree$, $2\degree$). We discuss potential causes for these rare inflow detections in Section \ref{sec:inflow}.

\subsection{Outflows Co-rotating with the LMC Disk}
\label{sec:outflow_vrot}

In Figure \ref{fig:vhel_xymap}, we show the 2D distributions of \HI, \SII, \SiIV, and \CIV\ bulk velocities in the heliocentric frame. The \HI\ moment one map (left panel) is based on the 21cm emission data cube from \cite{kim03} as discussed in Section \ref{sec:data_HI}, while the \SII, \SiIV, and \CIV\ are measured along the ULLYSES sight lines and averaged over $0.5\degree\times0.5\degree$ bins in the same way as for the $\log N$ values discussed in Section \ref{sec:logN_vs_sfr}.

The \HI\ panel shows that the northeast half of the LMC disk is moving away from us at a faster speed of $\vhelio\sim280-320~\kms$, while the southwest half is moving slower at $\vhelio\sim200-240~\kms$. Note that the northeast half of the LMC disk is closer to us \citep{vandermarel02}, and the LMC is rotating clockwise. In the \SII, \SiIV, and \CIV\ panels, we find that the ions' bulk velocities exhibit a rotation pattern similar to the one in \HI, indicating that the ionized outflows are co-rotating with the LMC disk.

In all, Figures \ref{fig:logN_SigSFR_xymap}--\ref{fig:vhel_xymap} show a coherent picture that the LMC is currently launching disk-wide, warm-ionized outflows with bulk velocities of $|\vout|\sim20-60~\kms$. The bulk mass of the outflowing gas should be gravitationally bound to the LMC, and is co-rotating with the LMC. Star-forming regions with higher $\sigsfr$ are launching outflows that are likely to be more ionized.

%%%%%%%%%%%%%%%%%%%%%%%%%%%%%%%%%%%% DISCUSSION %%%%%%%%%%%%%%%%%%%%%%%%%%%%%%%%%%%%

\section{A Scaling Relation between Outflow Velocities and Star Formation}
\label{sec:comparison_wind_literature}

\begin{figure*}
    \centering
    \includegraphics[width=0.8\textwidth]{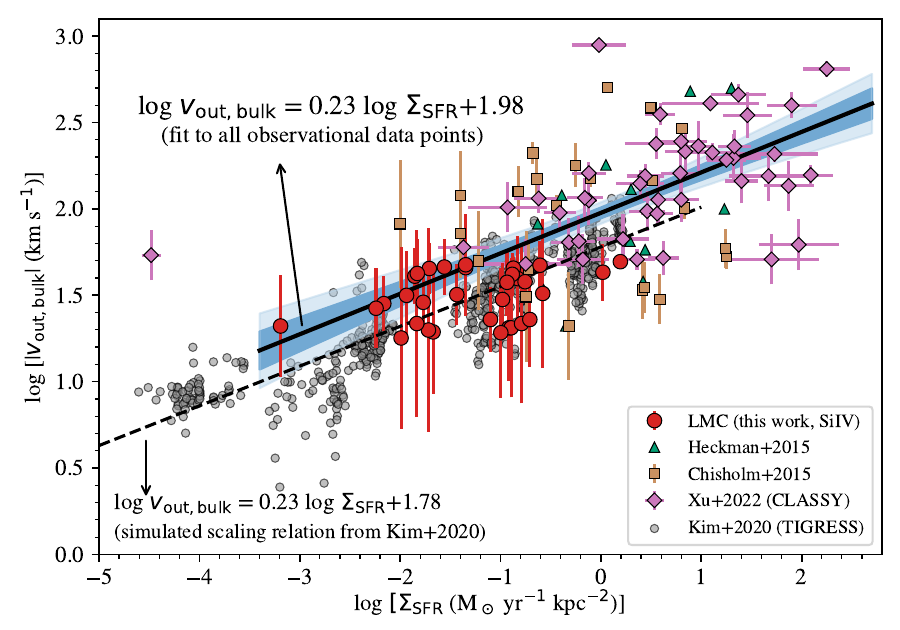}
    \caption{A scaling relation between bulk outflow velocity and $\sigsfr$ in logarithmic values. We compare the LMC outflows (red circles) with three studies, including \citet[][blue triangles]{heckman15}, \citet[][yellow squares]{chisholm15}, and \citet[][magenta diamonds]{xu22_classy3}, on cool outflows in starburst galaxies that used the same or similar instruments (either HST or FUSE) to minimize systemic uncertainties. 
    The solid black line shows our best-fit power-law relation, $\log \vout=0.23\log \sigsfr +1.98$ or $\vout = \frac{95.5}{~\kms}\sigsfr^{0.23}$, for the observational data points, where the dark and light blue shades indicate the 68\% and 95\% confidence intervals. Gray circles show simulation predictions from the TIGRESS-classic simulation suite over a wide range of star-forming conditions \citep{kim20}. The best-fit scaling relation from \cite{kim20} is shown as a black dashed line. See Section \ref{sec:comparison_wind_literature} for further details.}
    \label{fig:vout_sfr_fit}
\end{figure*}

We compare our LMC outflow measurements with both theoretical predictions and existing outflow observations in nearby starbursting galaxies in Figure \ref{fig:vout_sfr_fit}. For theoretical predictions, we focus on simulated data adopted from \cite{kim20}, which study how multiphase outflows develop in a suite of parsec-resolution simulations over a wide range of star-forming conditions with $\sigsfr\sim10^{-4.5}-1~\msunyr$ using the TIGRESS-classic framework \citep{kim17, kim18}. For existing outflow observations, we consider data from \citet[][hereafter, \citetalias{heckman15}]{heckman15}, \citet[][hereafter, \citetalias{chisholm15}]{chisholm15}, and \citet[][hereafter, \citetalias{xu22_classy3}]{xu22_classy3}, which also examine ionized outflows with $T\sim10^{4-5}$ K in commonly accessible ions such as \SiII, \SiIII, \SiIV, and \CIV. 
%\footnote{Note that \cite{heckman15}'s ionized outflow velocities are measured in \NII\ and \CIII, which are detectable by FUSE in the far UV. The formation and ionization potentials of these two ions are similar to other ions that we use in this work.}. 
Because these works measured outflows using either the same or similar instruments (either HST or FUSE spectroscopy\footnote{The HST and FUSE spectra generally have similar velocity resolutions. The HST/COS spectra typically have $\delta v=15-25~\kms$, HST/STIS spectra have $\delta v=6.6~\kms$ in E140M, and FUSE have $\delta v\sim15~\kms$.}), the comparison below avoids significant systemic uncertainties such as differences in spectral resolution and outflow phases. Note that the galaxy samples from these three studies are not mutually exclusive. We elaborate on the key aspects of each work, and adopt the most recent measurements for galaxies that have been used in more than one study.

\citetalias{heckman15} studied warm outflows in 39 starburst galaxies at $z<0.2$ using \SiIII\ (COS) and \NII\ and \CIII\ (FUSE), and their galaxy sample covers a parameter space of $\mstar=10^{7.1-10.9}~\msun$ and SFR=0.016--66 $\msunyr$. 
%For all lines, the outflow velocities are calculated as flux-weighted centroid velocities relative to the systemic velocities of galaxies, consistent with what we have calculated for the LMC. 
%Among the 39 galaxies, 18 were also included in \cite{chisholm15}, and 21 were in \cite{xu22_classy3}. Therefore, we only show 14 unique data points from \cite{heckman15} in Figure \ref{fig:vout_sfr_fit}, with the rest of the galaxy measurements taken from the other two more recent references. 
% They found that the flux-weighted centroid velocities of outflows only correlate weakly with galaxy stellar masses or circular velocities, but strongly with both SFR and SSFR.  
\citetalias{chisholm15} examined outflows traced by \SiII\ in 48 star-forming galaxies 
with $\mstar=10^{7.3-11.4}~\msun$
%(median is 10.25) 
and SFR=0.02--136.8 $\msunyr$ 
%(median is 10.44) 
at $z=0.0007-0.26$. Both studies measured bulk velocities of outflows (center of mass) using weighted average velocities, similar to what we did for the LMC outflows. % Additionally, they estimate the outflow terminal velocity of fast-moving low-density gas as $v_{90}$, which is the velocity at which the absorption flux reaches 90\% of the continuum level of the strongest avaiable \SiII\ lines. 
%They found that both $v_{\rm cen}$ and $v_{90}$ shows significant correlation with SFR, $\mstar$, and circular velocity, while the trend is steeper with $v_{\rm cen}$. 
%An intrinsic scatter of 0.1--0.2 dex are seen in outflow velocity, which they attribute to other galaxy properties such as galaxy inclination, different wind launching mechanism in different star-forming environments as well as surrounding galaxy CGM properties. 
%And they found detection of outflows down to SSFR of 0.01 Msun/yr/kpc2, a factor of 10 lower that what was suggested to be outflow launching thresholds in \cite{heckman02}. One to keep in mind when comparing their outflow measurements with ours: \SiII, which only requires 8.2 eV to produce and 16.3 eV to ionize, trace both neutral and ionized gas, therefore the outflow here might trace the average speed of two phases, especially if they are not co-moving. 

\citetalias{chisholm15} noted that \SiII\ may not be a perfect tracer of warm ionized outflows because it only requires 8.2 eV to produce and 16.3 eV to ionize, which means the ion traces both neutral and ionized outflows. However, a follow-up study by \cite{chisholm16a} using the same sample\footnote{\cite{chisholm16a} studied 37 star-forming galaxies, which is a subset of \cite{chisholm15}'s because of an implementation of $3\sigma$ cut in equivalent widths to only select galaxies with significant outflow detections.} showed that outflows probed by different ions such as \OI, \SiII, \SiIII, and \SiIV\ are most likely to be comoving and cospatial given the similarity in the derived outflow velocities and widths. Therefore, in Figure \ref{fig:vout_sfr_fit}, we compare the LMC's \SiIV\ outflows with \citetalias{chisholm15}'s \SiII\ measurements without additional correction.

\citetalias{xu22_classy3} studied galactic outflows in 45 starburst galaxies with $\mstar\sim10^{6-10}~\msun$ and SFR$\sim0.01-100~\msunyr$ at $0.002<z<0.182$. 
%, as well as 5 additional Lyman break analog galaxies from \cite{heckman15}. 
For each galaxy, they fit Gaussian profiles to ISM absorption, and when available, outflow absorption lines in \OI, \CII, \SiII, \SiIII, and \SiIV. The final outflow velocities are taken as the median values of outflows detected in all available ion lines along the corresponding sight lines. Because the contribution of the ISM absorption has been taken out, by design, their outflow velocities would be slightly faster than those measured for the same galaxies using simply weighted centroid velocities. Therefore, the outflow velocities adopted from \citetalias{xu22_classy3} may be systematically higher than the rest because of their different method. 

Figure \ref{fig:vout_sfr_fit} shows the outflow velocities as a function of $\sigsfr$; the simulated predictions from \cite{kim20} are shown as gray circles, while the observational data are shown as red circles for the LMC, blue triangles for \citetalias{heckman15}, yellow squares for \citetalias{chisholm15}, and magenta diamonds for \citetalias{xu22_classy3}. In a solid black line, we show our linear regression fit, $\log \vout = \beta \log \sigsfr + \alpha + \sigma$, to the observational data points in log-log space using a Python version \footnote{https://github.com/jmeyers314/linmix} of the \texttt{linmix} package detailed in \cite{kelly07}, where $\beta$ is the slope, $\alpha$ the intercept, and $\sigma$ the intrinsic scatter of the data points about the regression. The linear regression is performed over $\sigsfr=10^{-3.4}-10^{2.7}~\msunyrkpc$, excluding the outlier at $\sigsfr\sim10^{-4.5}~\msunyrkpc$ from \citetalias{xu22_classy3}. We find a best-fit scaling relation of: 
\begin{equation}
\begin{split}
    \log (\frac{\vout}{\kms}) = &  
    0.23^{+0.03}_{-0.03}\log(\frac{\sigsfr}{\msunyrkpc}) \\
     & + 1.98^{+0.03}_{-0.03} \pm 0.29~~~, \\ 
\end{split}
\label{eq:vout_sfr_fit}
\end{equation}
where the errors for the coefficients $\alpha$ and $\beta$ are the $84^{\rm th}-50^{\rm th}$ and $16^{\rm th}-50^{\rm th}$ percentile differences from the posterior distributions. During the fitting, because \citetalias{heckman15}'s outflow velocities do not include errors, we assume a uniform error of $e_v=28~\kms$, which is the median outflow velocity error from \citetalias{xu22_classy3}. Additionally, neither \citetalias{chisholm15} nor \citetalias{heckman15} reported errors in their $\log \sigsfr$ values; we assume a uniform error of 18\% of the corresponding $\log\sigsfr$ value, which is the typical median error of \citetalias{xu22_classy3}'s $\log\sigsfr$ measurements. We note that the linear regression result does not change significantly if we use the LMC's \CIV\ outflow measurements instead of \SiIV.

Given that the observational data points from \citetalias{heckman15}, \citetalias{chisholm15}, and \citetalias{xu22_classy3} dominate the middle to higher end of the $\vout$--$\sigsfr$ distribution in Figure \ref{fig:vout_sfr_fit}, it is not surprising that our power-law index of 0.23 is consistent with what have been typically found in the literature ($\vout\propto \sigsfr^{0.1-0.2}$; \citealt{martin05, rupke05, chen10, chisholm15, rupke18, ReichardtChu24}). What is remarkable is that, in spite of spanning five orders of magnitudes in $\sigsfr$, there is a coherent scaling relation between the outflow velocities and star formation activities, with the LMC at the lower end.

Our fit also agrees remarkably well with what is predicted by \cite{kim20}, which find a scaling relation of $\log \vout = 0.23\log\sigsfr+1.78\pm0.14$ for cool outflows over a broad range of star-forming conditions at $\sigsfr\sim10^{-4.5}-1~\msunyrkpc$ (black dashed line in Figure \ref{fig:vout_sfr_fit}). The main discrepancies are the intercept and the intrinsic scatter. For the observations, we find an intercept of $\alpha=1.98$ (or $\vout=95.5~\kms$), while \citeauthor{kim20} predicts a lower intercept of $\alpha=1.78$ (or $\vout=60.3~\kms$). Meanwhile, the observations show an intrinsic scatter of $\sigma=0.29$ dex, which is twice as high as the predicted intrinsic scatter ($\sigma=0.14$ dex). Given that the LMC data points (red circles) in Figure \ref{fig:vout_sfr_fit} generally match well with the simulation predictions (gray circles) over the same $\sigsfr$, we suspect that the discrepancies between our fit and \citeauthor{kim20}'s are mainly driven by the observational data points at the higher $\sigsfr$ end from \citetalias{heckman15}, \citetalias{chisholm15}, and \citetalias{xu22_classy3}.

There are likely to be several reasons for the discrepancies between observations and simulations. Physically, it is possible that star-bursting galaxies from \citetalias{heckman15}, \citetalias{chisholm15}, and \citetalias{xu22_classy3} are driving much faster outflows, which elevate our fit in Figure \ref{fig:vout_sfr_fit}. Furthermore, each data point from \citetalias{heckman15}, \citetalias{chisholm15}, and \citetalias{xu22_classy3} represents a single galaxy or a large fraction of a star-forming disk within an instrument's aperture, which traces spatially averaged galactic scale outflows. These galaxies or regions of galaxies were generally selected for their UV brightness and may be biased towards regions where powerful outflows have already cleared some of the ISM. In contrast, both our LMC measurements and \citeauthor{kim20}'s simulations focus on suc-kpc scale localized outflows that are sensitive to temporal, weak outflow signals from individual star-forming regions. %Therefore, the discrepancies may also reflect the difference between spatially resolved outflows (our data and \citeauthor{kim20}'s data) and aperture-averaged outflows (\citetalias{heckman15}'s, \citetalias{chisholm15}'s, and \citetalias{xu22_classy3}'s). 
Lastly, different methods in calculating outflow velocities among the works may also contribute to the discrepancies.

As for the scatters, as discussed in \citetalias{chisholm15}, there are a number of factors that could cause the scatters such as galaxy inclinations, outflow driving mechanisms (energy or momentum driven), and CGM masses of host galaxies which would impact the propagation of outflows as they leave the disks. Different methods used to derive $\sigsfr$ may also contribute to the scatters. For example, \citetalias{heckman15} and \citetalias{chisholm15} computed their SFR values based on UV and infrared fluxes of galaxies with prescriptions from \cite{kennicutt12}, while \citetalias{xu22_classy3}'s SFR values are based on broadband SED fittings by \cite{berg22}. And, \citetalias{heckman15} and \citetalias{xu22_classy3} calculated their $\sigsfr$ values averaged within half-light radii of their sample galaxies, but \citetalias{chisholm15}'s values are measured within the COS aperture.

The remarkable similarity, but also discrepancies, between observational and simulated outflows warrant further investigation, which is beyond the scope of this work. The key message from Figure \ref{fig:vout_sfr_fit} is that, dwarf galaxies like the LMC are capable of launching outflows not only in star-bursting regions such as 30 Dor, but also in regions with low star formation surface densities. And, despite systemic differences in measurement methods, there is a universal scaling relation between outflow velocities and $\sigsfr$, $\vout\propto\sigsfr^{0.23}$, across a wide range of star-forming conditions at $\sigsfr\sim10^{-4.5}-10^{2}~\msunyrkpc$. 

% Note that previous survey are all have point sources or cover partial of the galaxy disks, while our outflows are directly probing gas in front of the stars. So there may be some difference in the properties here.

\section{Discussion}
\label{sec:discuss}

\subsection{Conservative Estimates on Outflow Mass, \\Outflow Rate, and Mass Loading Factor}
\label{sec:mout_estimate}

We estimate the LMC's bulk outflow mass $\mout$, outflow rate $\dotmout$, and mass loading factor $\eta~(\equiv \dotmout/{\rm SFR})$ by considering the outflows from both sides of the disk. We assume that the back side outflows, though not observable with ULLYSES, are similar to those in the near side. The assumption is based on \cite{barger16}'s star-QSO pair observation that ion absorptions due to outflows from both sides of the LMC show similar absorption depths and velocity spans (see Section \ref{sec:comparison_lmc_literature}). 

The following calculations are based on measurements of \CIV. We choose \CIV\ because there are more ULLYSES sight lines with reliable \CIV\ measurements (71/109) than with \SiIV\ (44/110), as is shown in Figure \ref{fig:map_xy_proj}, which allows us to better estimate the outflow covering fraction across the LMC. Our estimates should be treated as conservative lower-limit values, given that most of the \CIV\ outflows are saturated (see Figure \ref{fig:logN_SigSFR_xymap}) and we assume the maximum \CIV\ ionization fraction of $f_{\rm CIV}=0.3$ possible through either equilibrium collisional ionization or photoionization \citep{gnat07}.

We simplify the LMC outflows as gas moving in a cylindrical volume with a total mass of: 
\begin{equation}
\begin{split}
M_{\rm CIV} & = 2\times m_{\rm C} \times (N_{\rm CIV}\times {\rm cos}~i) \times \pi R_{\rm out}^2 \times c_{\rm f} \\ 
& \gtrsim 2.7\times10^2~\msun (\frac{N_{\rm CIV}}{10^{14}~\rm cm^{-2}})(\frac{R_{\rm out}}{3.5~{\rm kpc}})^2(\frac{c_{\rm f}}{0.4})~~~,
\end{split}
\end{equation}
where the factor of 2 is to take into account the LMC's back side outflows and $m_{\rm C}$ is the mass of a carbon atom. We set $N_{\rm CIV}\approx10^{14}~{\rm cm^{-2}}$, which is the characteristic column density where \CIV\ begins to saturate (see Figure \ref{fig:logN_SigSFR_xymap}), and the factor of ${\rm cos}~i$ is to correct for the increased path length through the outflowing layer due to the LMC's inclination $i=23.4\degree$ (see Table \ref{tb:lmc_info}). $R_{\rm out}\approx3.5$ kpc is the maximum in-plane radius probed by our sight lines that show clear detections of outflows (see Figure \ref{fig:delv_xymap}).

We find a \CIV\ outflow covering fraction of $c_{\rm f}\approx0.4$  (14/33) by counting the number of bins in Figures \ref{fig:logN_SigSFR_xymap} and \ref{fig:delv_xymap} with $\bar{N}_{\rm CIV}\geq10^{14}$ cm$^{-2}$ and $\delta v_{\rm bulk, CIV}\leq-15~\kms$, and then dividing the value by the number of bins with reliable \CIV\ measurements. We choose to calculate $c_{\rm f}$ based on the spatially averaged 2D maps because it avoids over-sampling outflows in high star-forming regions, such as 30 Dor, which are probed by many ULLYSES sight lines simultaneously. Varying the bin sizes of $0.5\degree\times0.5\degree$ to smaller ($0.3\degree\times0.3\degree$) or larger ($1.0\degree\times1.0\degree$) grids only changes the $c_{\rm f}$ value by $\sim0.1$.

We assume that the outflows have the same metallicity as the LMC's ISM ($Z_{\rm out}\approx 0.5~Z_\odot$) and estimate the total mass of the outflows to be: 
\begin{equation}
\begin{split}
    \mout & = 1.4 m_{\rm p}\times \frac{M_{\rm CIV}}{f_{\rm CIV}\times m_{\rm C} \times Z_{\rm out} \times {\rm [C/H]_\odot}} \\ 
    & \gtrsim 8\times10^5~\msun~(\frac{f_{\rm CIV}}{0.3})^{-1}(\frac{Z_{\rm out}}{0.5~\rm Z_\odot})^{-1}~~~,
\end{split}
\end{equation}
where the factor of 1.4 is to account for the helium mass, and [C/H]$_\odot=10^{8.43-12.0}$ is the solar photosphere abundance \citep{asplund09}.

Recent analyses on the LMC's star formation history show that the galaxy is currently experiencing a high star-forming episode that began $\sim$30 Myr ago \citep{mazzi21}. We assume that the LMC began to launch the most recent outflows around the same time and has continued driving outflows until the present day with a time duration of $t_{\rm out}\sim30$ Myr. The LMC's mass outflow rate and the mass loading factor are: 
\begin{equation}
\begin{split}
    \dotmout & =\mout/t_{\rm out}\gtrsim 0.03~\msunyr~ (\frac{t_{\rm out}}{30~\rm Myr})^{-1} \\ 
    \eta & = \dot{M}_{\rm out} / {\rm SFR} \gtrsim 0.15 ~~~,
\end{split} 
\end{equation}
where ${\rm SFR}\approx 0.2~\msunyr$ is the global SFR of the present day LMC adopted from \cite{harris09} and \cite{mazzi21}.

When compared with literature values, we find that our $\mout$ and $\dotmout$ values are consistent with \cite{barger16}'s estimates for the LMC based on their \CIV\ measurements (see their Table 4). Note that our estimates should be treated as strictly conservative values, given that the \CIV\ outflows are mostly saturated in the LMC and we assume the maximum ionization fraction.

Our mass loading factor of $\eta\gtrsim0.15$ is a factor of $\sim10$ lower than those measured by \cite{chisholm17} for galaxies at similar masses as the LMC, which have $\eta\sim0.9-2$; the discrepancy here is likely due to the more active star-forming nature of \citeauthor{chisholm17}'s galaxies with SFR$\sim3.6-26~\msunyr$, which may drive more powerful and efficient outflows. When considering galaxies with similar $\sigsfr$, we find that our $\eta$ value is within what is predicted by \cite{kim20}'s outflow simulations over $\sigsfr\sim10^{-3}-10^{-1}~\msunyrkpc$, but about $\sim1-1.5$ dex lower than their median values (see their Figure 8).

Lastly, considering the time duration of $t_{\rm out}\sim30$ Myr and a mean bulk outflow velocity of $\sim30~\kms$ in \CIV, we can infer that the bulk mass of the LMC outflows is at a height of $z_{\rm out}\sim0.9$ kpc. When compared to the LMC's disk scale height ($\sim0.97$ kpc from Cepheid stars, \citealt{ripepi22}), we find that the bulk outflow mass has not made it out of the LMC's disk. This is consistent with our observation in Figure \ref{fig:vhel_xymap} that the outflowing gas is still under the gravitational influence of the LMC's disk and thus shows kinematic signs of co-rotation with the disk.

%, MRK 1486 ($\mstar=10^{9.3}~\msun$) and KISSR 1578 ($\mstar=10^{9.5}~\msun$), at similar masses to the LMC, which have $\eta\sim1-2$ \citep{chisholm17}. Our mass loading factor is consistent with that measured based on \Halpha\ emission for NGC 4449 ($\mstar=10^{9.3}~\msun$) with $\eta\approx 1.6$ \citep{mcquinn19}. 

\subsection{Shielding of the LMC Outflows by A Potential Bow Shock}
\label{sec:bow_shock}

At a Galactocentric velocity of $321~\kms$ \citep{kallivayalil13}, the LMC is moving in the MW's halo with a Mach number of $\sim2.1$ \citep{setton23}. Using an LMC-specific hydrodynamic simulation, \cite{setton23} showed that the LMC's supersonic movement is likely to generate a bow shock leading the LMC due to ram pressure from the MW, as is illustrated in Figure \ref{fig:lmc_sketch}. The ram pressure impact has been well observed in the LMC's truncated \HI\ disk \citep{salem15}, and it is likely that the ram pressure also shapes the morphology of the LMC's \Halpha\ emission \citep{smart23, setton23}. Although the existence of an LMC bow shock remains to be observationally tested, below we provide indirect evidence of this bow shock by speculating that it may have shielded the LMC outflows from the MW's ram pressure. 

When a satellite galaxy orbits a massive host, its ISM gas is subject to ram pressure stripping from the halo of the massive host. This is commonly seen in dwarf galaxies closer to the MW and M31 \citep{putman21}, as well as in jelly-fish galaxies in large galaxy groups and clusters \citep[e.g.][]{poggianti16}. \cite{zhu23} showed that when a galaxy's disk is at an angle of $45\degree$ against the headwinds due to ram pressure, the gas above the galaxy disk is being swept downstream and flowing mainly parallel to the galaxy disk (see their Figure 13). Had the LMC experienced such strong ram pressure stripping, we would expect the outflowing gas to have been swept in the opposite direction of the LMC's proper motion (east to west), and the outflow column densities are unlikely to correlate with $\sigsfr$.

Our analyses in Figures \ref{fig:logN_SigSFR_xymap}--\ref{fig:vhel_xymap} show that the LMC outflows are well correlated with $\sigsfr$ in ion column densities, and the outflows are co-rotating with the LMC disk. Calculations in Section \ref{sec:mout_estimate} suggest that the bulk mass of the outflowing gas is close to the LMC disk at a height of $\approx0.9$ kpc, which is well within the size of the bow shock predicted in \cite{setton23}'s LMC simulation. Additionally, \cite{barger16} showed that the LMC outflows in the near and back sides show similar absorption strengths and velocity spans, indicating that the outflows on the near side are not significantly suppressed. As there is no significant sign of the outflowing gas being impacted by external forces, such as ram pressure from the MW halo gas, we suspect that the LMC's outflows may have been shielded by a potential bow shock as the LMC orbits the MW supersonically.

\subsection{Comparison with Previous LMC Gas Studies}
\label{sec:comparison_lmc_literature}

In this section, we briefly compare our work with previous studies on the LMC gas over $\vhelio\sim175-375~\kms$. The key message is that the outflows we find are correlated with the most recent star-formation episode of the LMC, and they are gravitationally bound to the LMC with $|\vout|\sim20-60~\kms$, consistent with previous studies using smaller samples of stellar sight lines. These outflows are not connected to the high-velocity cloud at $\vhelio\sim90-175~\kms$ in the foreground of the LMC (see Figure \ref{fig:lmc_sketch});  we discuss the physical properties of the high-velocity cloud and relevant studies in Section \ref{sec:discuss_hvc}.

The LMC gas at $\vhelio\sim175-375~\kms$ is found to be multi-phase as seen in both emission and absorption. Figure \ref{fig:map_HI_Halpha} shows that \HI\ 21cm is found across the galaxy tracing large and small scale neutral gas structures such as outer arms and supergiant shells \citep{kim03, staveley-smith03, nidever08}. This neutral gas mainly follows the rotation of the LMC stellar disk, and does not show signs of outflows (see top left panel of Figure \ref{fig:delv_sigsfr}). \cite{smart23} found that the \Halpha\ emission from the LMC is more extended than the \HI\ by several degrees, and the \Halpha\ gas kinematics is found to weakly trace the \HI\ gas rotation.

\cite{howk02} studied \OVI\ obtained from FUSE toward 12 LMC stellar sight lines over $175\lesssim\vlsr\lesssim375~\kms$ (see also \citealt{danforth02}). The \OVI\ column densities
%measured toward these sight lines 
show a large variation, which is not correlated with underlying structures such as \HI\ superbubbles. They found the \OVI\ centroid velocities to be blueshifted from the LMC's low ion absorption lines (e.g., \FeII) by $\sim-30~\kms$, suggesting the presence of highly ionized outflows among these sight lines, consistent with our findings of outflows in \SiIV\ and \CIV.

\cite{barger16} probed a relatively quiet northwest region of the LMC using a pair of QSO-star sight lines in close projection over $165\lesssim\vlsr\lesssim415~\kms$. Ion spectra %of multiphase gas (e.g., \OI, \SiII, \AlII, \SiIII, \SiIV, \CIV) 
from both the star and the QSO show blueshifted outflows over $\sim165-280~\kms$, while only the QSO spectra show redshifted outflows at $\sim280-415~\kms$ on the back side of the LMC disk. %Because the stellar sightline only probes the nearside outflows while the QSO sightline can access both the near and far sides of the LMC, the absorption difference between the two sightlines uniquely highlights the presence of outflows in the far side in multiphase ions (e.g., \OI, \SiII, \AlII, \SiIII, \SiIV, \CIV). In particular, the outflows in the backside appear to be at more positive velocities ($\vlsr\gtrsim280~\kms$) toward this direction of the galaxy, while the nearside outflows are at $165\lesssim\vlsr\lesssim280~\kms$. 
They found that the outflows in the near and back sides show similar absorption strengths and velocity spans in \OI, \SiII, \AlII, \SiIII, \SiIV, and \CIV. 

Note that the $\sim100~\kms$ outflow speeds quoted by \cite{barger16} were measured toward the edge of the ion absorption, which represents the terminal velocities of the low-density outflowing gas. In contrast, the outflow velocities measured in this work as well as in \cite{howk02} are weighted outflow velocities representing the bulk mass of the outflowing gas. As shown in Figure \ref{fig:spec_example}, we also find outflows with high terminal velocities of $v\sim100~\kms$ toward the ULLYSES sight lines. However, we do not use the terminal outflow velocities in this work to avoid potential contamination due to the high-velocity cloud in the foreground, which we discuss in the next section.

\subsection{The Foreground High-Velocity Cloud at $v\sim90-175~\kms$}
\label{sec:discuss_hvc}

A number of studies have noted the presence of a high-velocity cloud moving at $\vhelio\sim90-175~\kms$ in the foreground of the LMC; hereafter, we refer to this cloud as HVC90-175. As sketched in Figure \ref{fig:lmc_sketch}, \cite{richter15} constrained the distance to HVC90-175 to be within 13.3 kpc from the Sun using an HST/COS spectrum of a hot white dwarf (RX J0439.8−6809; \citealt{werner15}), which means HVC90-175 is located in the inner halo of the MW and at $\delta d \sim40$ kpc from the LMC. 

We show in Section \ref{sec:method_contfit} and Figure \ref{fig:spec_example} that HVC90-175's absorption can be well constrained to be within $\vhelio\lesssim 175~\kms$ and the blending with the LMC absorption is relatively mild. Specifically, for \SiIV\ (\CIV) absorption, we find only 10/44 (22/71) sight lines with non-negligible blending between HVC90-175 and the LMC outflows near $\vhelio=175~\kms$. In these cases, the LMC is likely to launch fast outflows with terminal velocities of $\gtrsim100~\kms$ that are blended with HVC90-175. By focusing on the LMC absorption over $\vhelio=175-375~\kms$ in this work, we minimize contamination from HVC90-175.

The origin of HVC90-175 remains debated. In Appendix \ref{app:discuss_hvc} and Figures \ref{fig:hvc_lsr_gsr_lmcsr}-\ref{fig:hvc_2dmaps}, we show that HVC90-175 is kinematically consistent with being a MW halo cloud at a constant velocity of $\vlsr\sim120~\kms$ (see also \citealt{savage81, deboer90, richter99}). The spread in line of sight velocities towards HVC90-175 can be well accounted for if we assume a temperature of $T\sim10^{4.2}$ K and a non-thermal broadening of $\sigma_{\rm nth}\sim10~\kms$ (see details in Appendix \ref{app:discuss_hvc}), which is typical for MW halo clouds \citep{putman12}. The distance of HVC90-175, $d_\odot<13.3$ kpc, is also consistent with other MW halo clouds \citep{wakker01}.

We note that HVC90-175 has also been suggested to be associated with the LMC as a fast-moving, ancient outflow that was launched by the LMC's previous star-forming episode about $t_{\rm SFR}\sim250-400$ Myr ago \citep[e.g.][]{barger16, ciampa21}. At a distance of $\delta d\sim40$ kpc from the LMC, it would be challenging to launch an outflow that did not decelerate, increase opening angle, or change trajectory. As the LMC moves through the MW halo in a nearly transverse direction (see discussion in Section \ref{sec:bow_shock}), the presumed ancient outflow would have been severely impacted by ram pressure when traveling such a large distance. Therefore, we consider it highly unlikely for HVC90-175 to originate from the LMC as an ancient outflow.

\subsection{Rare Detection of Inflows}
\label{sec:inflow}

We note the rare detection of an inflow in \SiIV\ toward a sight line, SK-67D22, at a velocity of $+26~\kms$ with respect to the LMC's stellar disk, as is shown in the lower left panel of Figure \ref{fig:delv_sigsfr}. The corresponding \CIV\ line shows a similar absorption profile, although its centroid velocity is $+8~\kms$ because of the weighting algorithm that we apply in Section \ref{sec:method_contfit}. SK-67D22 is located at $x\sim-2\degree$ and $y\sim2\degree$ in Figure \ref{fig:delv_xymap}, which is the only bin in the \SiIV\ panel that shows inflow detection. We examine SK-67D22's line spectra (not shown here), and find that the neutral (\HI) and low ionization (\SII) gas is moving at $\vhelio\sim290~\kms$, consistent with the underlying stellar disk ($v_{\rm RSG}\sim285~\kms$). The more ionized \SiIV\ and \CIV\ gas is found over $\vhelio\sim300-340~\kms$, indicating that the ionized gas is inflowing toward the LMC disk at $v_{\rm in}\sim10-50~\kms$ in a relatively quiet region of the galaxy.

Similarly, there are two sight lines with \SII\ inflows at $v_{\rm in}\sim15-20~\kms$ in the southeast corner of the LMC where the star formation is less active (see Figures \ref{fig:map_xy_proj} and \ref{fig:delv_xymap}). In the heliocentric frame, the \SII\ inflows are found at $\vhelio\sim300~\kms$, which coincides with the \HI\ arm E of the LMC as identified by \cite{staveley-smith03}; we further investigate the connection between our \SII\ detection and the arm E in a follow-up paper. 

Overall, the detections of inflows in \SiIV\ and \SII\ suggests that the inflows may still exist in the LMC; they might be much weaker than the outflows along the same sight lines (see Figure \ref{fig:spec_example}), which would be averaged out in our calculation of optical-depth weighted centroid velocities. Additionally, inflows might occur in areas that are relatively less sampled by the ULLYSES sight lines. Another possible explanation for the dominant outflow detection is that the LMC is currently undergoing an active star formation episode \citep{harris09, mazzi21}, which drives disk-wide outflows. Inflows may not occur until $\sim50-60$ Myrs later when outflows turn around, cool down, and rain back down to the disk, as seen in hydrodynamic simulations \citep[e.g.][]{kim18}. We further investigate the occurrence and physical properties of inflows in the LMC in a follow-up paper.

\section{Summary}
\label{sec:conclusion}

Using 110 stellar sight lines from the ULLYSES \ulldr\ \citep{roman-duval21}, we detect prevalent slow-moving, ionized outflows ($|\vout|\sim20-60~\kms$) in \SII, \SiIV\ and \CIV\ across the disk of the LMC. % This work is the first in a series from our HST archival program %(\href{https://www.stsci.edu/cgi-bin/get-proposal-info?id=16640&observatory=HST}{\#HST-AR-16640, PI Zheng})
%(\#HST-AR-16640, PI Zheng$^{\ref{footnote2}}$), in which we examine how the LMC's multiphase gas is impacted by the interplay between its star-forming activities and ram pressure from the MW's CGM. 
Our work provides direct comparison between spatially resolved outflows in a local galaxy (LMC) with aperture-averaged galactic outflows in star-bursting galaxies and simulation predictions. \textit{We demonstrate that there exists a universal scaling relation between outflow velocities and star formation rate surface densities, $|\vout|\propto\sigsfr^{0.23}$, over a wide range of star-forming conditions with $\sigsfr\sim10^{-4.5}-10^2~\msunyrkpc$}. We summarize the main analyses and findings as follows.

We study the LMC neutral and ionized gas over $\vhelio=175-375~\kms$ in \HI, \SII, \SiIV, and \CIV. The velocity range is chosen to encompass the LMC gas in all directions while avoiding contamination from a foreground high-velocity cloud at $d_\odot<13.3$ kpc (Figure \ref{fig:lmc_sketch}). The ion lines are chosen for their relatively less saturated line profiles over the LMC's velocity range. We develop a continuum-fitting algorithm based on the concept of Gaussian Process regression, and select reliable ion spectra with minimal contamination from stellar absorption. Our algorithm results in 91/110 reliable LMC measurements in \SII, 44/110 in \SiIV, and 71/109 in \CIV\ (see Section \ref{sec:data_uv} and Figures \ref{fig:map_xy_proj}--\ref{fig:star_hist_sfr}).

We find that the column densities of the LMC's neutral (\HI) and ionized (\SII, \SiIV, \CIV) gas increase with the star formation rate surface density $\sigsfr$. Most of the \SiIV\ and \CIV\ measurements are heavily saturated with $N(\rm SiIV)\gtrsim10^{13.6}$ cm$^{-2}$ and $N(\rm CIV)\gtrsim10^{14.0}$  cm$^{-2}$, and all ions are saturated at $\sigsfr\gtrsim10^{-0.5}~\msunyrkpc$. As $\sigsfr$ is derived based on \Halpha\ emission that traces the LMC's recent star formation in the past $\sim7-10$ Myrs, the correlation between gas column densities and $\sigsfr$ suggests that the LMC's star-forming activities may have an impact on its multiphase gas over a short timescale (see Section \ref{sec:logN_vs_sfr} and Figure \ref{fig:logN_SigSFR_xymap}).

We compare the centroid velocities of the neutral (\HI) and ionized (\SII, \SiIV, \CIV) gas to the LMC's stellar kinematics in Section \ref{sec:voutflow_vs_sfr} and Figure \ref{fig:delv_sigsfr}, where the centroid velocities indicate the bulk motion of gas where most of the mass is. We find that the velocities of the ionized gas are systemically blue-shifted from the LMC's stellar disk, which indicates prevalent outflows at bulk velocities of $|\vout|\sim20-60~\kms$. While \SiIV\ and \CIV\ outflows are detected ubiquitously in the LMC, \SII\ outflows are only found in regions with relatively low $\sigsfr$ ($\lesssim10^{-0.5}~\msunyrkpc$). This indicates that star-forming regions with high $\sigsfr$ are launching outflows that are likely to be more ionized. 

We release the first 2D UV ion maps of the LMC in Figures \ref{fig:logN_SigSFR_xymap}--\ref{fig:vhel_xymap}, and show that the \SiIV\ and \CIV\ outflows are stronger (in column density) in high star-forming regions such as 30 Dor. Additionally, the \SiIV\ and \CIV\ outflows show signs of co-rotation with the LMC disk. Given that there is no significant sign of impact from external forces such as ram pressure from the MW, we suspect that the outflows are likely to be shielded behind a potential bow shock that is leading the LMC as the galaxy orbits the MW supersonically. The existence and exact location of this potential bow shock remains to be tested observationally (see Sections \ref{sec:outflow_vrot} and \ref{sec:bow_shock}).

We estimate the physical properties of bulk outflows from both sides of the LMC using the \CIV\ measurements, and find strictly conservative lower limits with a total outflow mass of $\mout\gtrsim8\times10^5~\msun$, an outflow rate of $\dotmout\gtrsim 0.03~\msunyr$, and a mass loading factor of $\eta\gtrsim0.15$. When comparing with outflows detected in starburst galaxies from previous observations \citep{heckman15, chisholm15, xu22_classy3}, we find a universal scaling relation of $|\vout|\propto\sigsfr^{0.23}$ (Figure \ref{fig:vout_sfr_fit}). Our measurements also agree remarkably well with what is predicted for cool outflows in the TIGRESS-classic simulation suite \citep{kim20} (see Section \ref{sec:comparison_wind_literature}).

Lastly, we find an intrinsic scatter of 0.29 dex in the $\vout-\sigsfr$ power-law relation for all observational data points combined (Equation \ref{eq:vout_sfr_fit}), which is a factor of $\sim2$ higher than the simulation prediction. As we discuss in Section \ref{sec:comparison_wind_literature}, many factors may contribute to the intrinsic scatter, such as different methods in calculating $\vout$ and $\sigsfr$, potential biases in selecting outflow-dominated galaxies, and intrinsic galaxy properties that have not been accounted for such as inclinations, outflow driving mechanisms, as well as host galaxies' CGM masses. We will continue investigating the LMC outflows (as well as inflows) in comparison with other observational and simulation measurements in follow-up studies.

\begin{acknowledgments}
We thank the anonymous referee for their thorough and helpful feedback. Y.Z. thanks John Chisholm, Grace Telford, Dan Weisz, and Alessandro Savino for discussion at various points during the preparation of this manuscript, thanks David Setton, Gurtina Besla, and Ekta Patel for discussion on the construction of Figure 1, and thanks Chang-Goo Kim for kindly sharing the TIGRESS outflow measurements and providing helpful comments on Section 5.
This work is made possible based on observations obtained with the NASA/ESA Hubble Space Telescope, retrieved from the Mikulski Archive for Space Telescopes (MAST) at the Space Telescope Science Institute (STScI).
Support for Program number HST-AR-16640 was provided by NASA through a grant from STScI, which is operated by the Association of Universities for Research in Astronomy, Incorporated, under NASA contract NAS5-26555. 
E.D.T. was supported by the European Research Council (ERC) under grant agreement no. 101040751. 
Y.F. acknowledges support from NASA award 19-ATP19-0023 and NSF award AST-2007012.
%KITP Acknowledgement: 
%also report manuscript to kitp: 
% https://www.kitp.ucsb.edu/report-publication
Y.Z. and K.T. started collaboration on this topic during a program ``Fundamentals of Gaseous Halos" held in 2021 at Kavli Institute for Theoretical Physics, UC Santa Barbara. This research was supported in part by the National Science Foundation under Grant No. NSF PHY-1748958. 
This work was performed in part at the Aspen Center for Physics, which is supported by National Science Foundation grant PHY-2210452. The LMC's H$\alpha$ image was adopted from the Southern H-Alpha Sky Survey Atlas (SHASSA; \citealt{guastad01}), which is supported by the National Science Foundation. 
\end{acknowledgments}

\vspace{5mm}
\facilities{HST (COS, STIS), Mikulski Archive for Space Telescopes (MAST)}

\software{Astropy \citep{astropy:2013, astropy:2018, astropy:2022}, 
Numpy \citep{numpy}, 
George \citep{george}, 
Matplotlib \citep{matplotlib}}
 
Data Availability: we release our data products, including normalized \SII, \SiIV, and \CIV\ lines and their corresponding best-fit continuum models (when available), as a High Level Science Product called ``LMC-FLOWS" at MAST via: \dataset[10.17909/hz0m-np43]{\doi{10.17909/hz0m-np43}} \citep{lmc-flows}. The ULLYSES DR5 dataset can be found at: \dataset[10.17909/t9-jzeh-xy14]{\doi{10.17909/t9-jzeh-xy14}} \citep{Roman-Duval20_ullyses}.

%guide on dealing with long tables
% https://journals.aas.org/aastexguide/#preamble_deluxetable
%\startlongtable
%\movetableright=-3in
\begin{longrotatetable}
%https://journals.aas.org/aastexguide/
\begin{deluxetable*}{cccccccccccccccc}
\centerwidetable
\tabletypesize{\scriptsize}
\label{tb:logN}
\tablecaption{Column Density and Velocity Measurements of the LMC Gas Along the ULLYSES DR5 Sight Lines}
\tablehead{
\colhead{ID} & \colhead{Star} & \colhead{RA} & \colhead{DEC} & \colhead{$x$} & \colhead{$y$} &\colhead{$\log \sigsfr$} & \colhead{$v_{\rm RSG}$} & \colhead{logN(\HI)} & \colhead{$v_{\rm HI}$} & \colhead{logN(\SII)} & \colhead{$\vcen$(\SII)} & \colhead{logN(\SiIV)} & \colhead{$\vcen$(\SiIV)} & \colhead{logN(\CIV)} & \colhead{$\vcen$(\CIV)}\\
%\colhead{ID}& \colhead{Star} & \colhead{RA} & \colhead{DEC} & \colhead{$x$} & \colhead{$y$} & \colhead{$\sigsfr$} & \colhead{(8)} & \colhead{(9)} & \colhead{(10)} & \colhead{(11)} & \colhead{(12)} & \colhead{(13)} & \colhead{(14)} & \colhead{(15)} & \colhead{(16)} \\
& & \colhead{(deg)} & \colhead{(deg)} & \colhead{(deg)} & \colhead{(deg)} & 
\colhead{($\frac{\msun}{\rm yr~kpc^2}$)} 
& \colhead{($\rm \frac{km}{s}$)} & \colhead{($\rm \frac{1}{cm^2}$)} & \colhead{($\rm \frac{km}{s}$)}  & 
\colhead{($\rm \frac{1}{cm^2}$)} & \colhead{($\rm \frac{km}{s}$)} & \colhead{($\rm \frac{1}{cm^2}$)} & \colhead{($\rm \frac{km}{s}$)} &  \colhead{($\rm \frac{1}{cm^2}$)} & \colhead{($\rm \frac{km}{s}$)} \\
%& & & & & \colhead{kpc$^{-2}$)} & & & & & & & & & \\
 \colhead{(1)}& \colhead{(2)} & \colhead{(3)} & \colhead{(4)} & \colhead{(5)} & \colhead{(6)} & \colhead{(7)} & \colhead{(8)} & \colhead{(9)} & \colhead{(10)} & \colhead{(11)} & \colhead{(12)} & \colhead{(13)} & \colhead{(14)} & \colhead{(15)} & \colhead{(16)} 
 }
\startdata 
1 & SK-68D73 & 80.7491 & -68.0296 & 0.11 & 1.24 & -0.916$\pm$0.052 & 288.5 & 21.57 & 291.2 & $>$15.77 & 294.1$\pm$25.6 & $>$13.93 & 268.0$\pm$14.4 & unreliable & unreliable\\
2 & BAT99-105 & 84.6755 & -69.0987 & 1.51 & 0.12 & 0.194$\pm$0.044 & 271.7 & self-abs & self-abs & $>$16.21 & 260.8$\pm$17.6 & unreliable & unreliable & $>$14.93 & 228.7$\pm$13.4\\
3 & ST92-5-31 & 84.7985 & -69.5104 & 1.52 & -0.29 & -0.597$\pm$0.042 & 265.7 & 21.65 & 273.6 & $>$15.83 & 268.9$\pm$5.8 & unreliable & unreliable & $>$14.37 & 242.4$\pm$5.4\\
4 & SK-67D22 & 74.3644 & -67.6508 & -2.31 & 1.51 & -2.063$\pm$0.055 & 284.7 & 21.21 & 284.9 & $>$15.46 & 283.3$\pm$13.5 & $>$13.62 & 310.9$\pm$23.7 & 13.80$\pm$0.02 & 293.1$\pm$22.7\\
5 & SK-66D172 & 84.2725 & -66.3597 & 1.53 & 2.86 & -1.189$\pm$0.057 & 302.5 & 21.27 & 301.9 & $>$15.73 & 287.0$\pm$15.8 & unreliable & unreliable & $>$14.35 & 291.0$\pm$26.9\\
6 & VFTS72 & 84.3936 & -69.0195 & 1.41 & 0.21 & -0.573$\pm$0.048 & 273.0 & 21.59 & 275.5 & $>$15.78 & 277.0$\pm$8.8 & unreliable & unreliable & $>$14.90 & 244.1$\pm$5.5\\
7 & BI237 & 84.0610 & -67.6553 & 1.37 & 1.58 & -1.458$\pm$0.054 & 292.7 & 21.47 & 292.0 & unreliable & unreliable & unreliable & unreliable & $>$14.57 & 285.4$\pm$9.2\\
8 & SK-67D211 & 83.8079 & -67.5576 & 1.28 & 1.68 & -0.581$\pm$0.054 & 294.2 & 21.27 & 296.9 & $>$15.68 & 288.8$\pm$24.3 & unreliable & unreliable & $>$14.71 & 283.0$\pm$24.0\\
9 & VFTS-482 & 84.6679 & -69.0999 & 1.51 & 0.12 & 0.194$\pm$0.044 & 271.7 & self-abs & self-abs & $>$16.06 & 269.6$\pm$3.5 & $>$14.41 & 222.2$\pm$3.4 & unreliable & unreliable\\
10 & N11-ELS-060 & 74.1756 & -66.4152 & -2.50 & 2.73 & -0.398$\pm$0.056 & 288.3 & 21.36 & 290.9 & $>$15.73 & 264.2$\pm$5.8 & unreliable & unreliable & unreliable & unreliable\\
11 & ST92-5-27 & 84.8065 & -69.5014 & 1.53 & -0.28 & -0.579$\pm$0.042 & 265.8 & 21.63 & 273.2 & $>$15.85 & 264.8$\pm$6.1 & unreliable & unreliable & $>$14.55 & 239.6$\pm$7.6\\
12 & LH114-7 & 85.8042 & -67.8544 & 2.02 & 1.33 & -1.287$\pm$0.055 & 288.9 & 21.30 & 301.6 & unreliable & unreliable & unreliable & unreliable & unreliable & unreliable\\
13 & VFTS-267 & 84.5582 & -69.1299 & 1.46 & 0.09 & 0.164$\pm$0.048 & 271.3 & 21.65 & 279.6 & $>$15.86 & 272.0$\pm$11.7 & unreliable & unreliable & $>$14.44 & 231.7$\pm$9.5\\
14 & VFTS-404 & 84.6410 & -69.1659 & 1.49 & 0.06 & -0.043$\pm$0.045 & 270.7 & 21.62 & 278.0 & $>$16.00 & 274.5$\pm$11.0 & unreliable & unreliable & $>$14.71 & 236.0$\pm$7.5\\
15 & W61-28-23 & 83.7090 & -69.7757 & 1.13 & -0.53 & -1.101$\pm$0.039 & 262.3 & 21.28 & 273.6 & $>$15.63 & 265.8$\pm$10.5 & unreliable & unreliable & $>$14.21 & 248.2$\pm$12.4\\
16 & SK-71D46 & 82.9566 & -71.0606 & 0.82 & -1.80 & -0.875$\pm$0.049 & 243.7 & 21.51 & 245.9 & $>$15.62 & 256.8$\pm$9.4 & unreliable & unreliable & 13.96$\pm$0.02 & 229.5$\pm$11.7\\
17 & SK-67D166 & 82.9342 & -67.6337 & 0.95 & 1.62 & -1.290$\pm$0.054 & 293.7 & 20.82 & 294.2 & 15.13$\pm$0.01 & 261.8$\pm$7.7 & unreliable & unreliable & 13.67$\pm$0.02 & 263.7$\pm$14.1\\
18 & SK-67D105 & 81.5258 & -67.1824 & 0.42 & 2.09 & -1.449$\pm$0.056 & 300.6 & 21.03 & 308.4 & $>$15.22 & 303.7$\pm$28.4 & unreliable & unreliable & 13.87$\pm$0.03 & 271.6$\pm$24.6\\
19 & SK-67D108 & 81.6103 & -67.6223 & 0.44 & 1.65 & -1.141$\pm$0.055 & 294.2 & 21.31 & 296.1 & $>$15.45 & 286.8$\pm$10.6 & unreliable & unreliable & $>$14.23 & 275.6$\pm$8.4\\
20 & HD38029 & 84.2299 & -69.1938 & 1.34 & 0.04 & -0.580$\pm$0.045 & 270.6 & 21.69 & 277.0 & unreliable & unreliable & 13.61$\pm$0.04 & 238.2$\pm$32.2 & unreliable & unreliable\\
21 & SK-67D167 & 82.9663 & -67.6615 & 0.96 & 1.59 & -0.962$\pm$0.054 & 293.3 & 21.24 & 296.8 & 15.22$\pm$0.01 & 278.2$\pm$12.3 & unreliable & unreliable & $>$14.02 & 259.4$\pm$20.6\\
22 & W61-28-5 & 83.6186 & -69.7325 & 1.10 & -0.49 & -1.203$\pm$0.039 & 262.9 & 21.03 & 267.4 & unreliable & unreliable & unreliable & unreliable & $>$14.46 & 253.4$\pm$7.6\\
23 & FARINA-88 & 85.0343 & -69.6548 & 1.59 & -0.44 & -0.552$\pm$0.042 & 263.4 & 21.68 & 265.0 & $>$15.98 & 258.6$\pm$5.0 & unreliable & unreliable & $>$14.46 & 234.6$\pm$5.1\\
24 & LMCE055-1 & 74.2034 & -69.6113 & -2.17 & -0.45 & -2.163$\pm$0.050 & 262.9 & 20.94 & 268.2 & 15.51$\pm$0.02 & 233.4$\pm$14.4 & $>$14.01 & 234.6$\pm$10.5 & $>$14.40 & 228.9$\pm$7.1\\
25 & SK-70D60 & 76.1699 & -70.2596 & -1.44 & -1.04 & -2.200$\pm$0.048 & 255.4 & 20.39 & 242.3 & 14.77$\pm$0.03 & 239.6$\pm$26.8 & unreliable & unreliable & 13.82$\pm$0.03 & 216.6$\pm$25.4\\
26 & SK-65D47 & 80.2280 & -65.4550 & -0.09 & 3.81 & -1.846$\pm$0.063 & 304.5 & self-abs & self-abs & 15.38$\pm$0.02 & 261.3$\pm$14.0 & unreliable & unreliable & 14.02$\pm$0.02 & 268.2$\pm$17.6\\
27 & SK-67D69 & 78.5837 & -67.1342 & -0.72 & 2.13 & -2.004$\pm$0.051 & 301.5 & 21.25 & 303.2 & $>$15.72 & 289.9$\pm$14.8 & unreliable & unreliable & $>$14.23 & 270.0$\pm$11.4\\
28 & VFTS352 & 84.6186 & -69.1886 & 1.48 & 0.03 & -0.169$\pm$0.045 & 270.4 & 21.70 & 277.7 & $>$15.78 & 275.5$\pm$5.0 & unreliable & unreliable & unreliable & unreliable\\
29 & ST92-4-18 & 84.9612 & -69.4076 & 1.59 & -0.19 & -0.800$\pm$0.042 & 267.1 & 21.57 & 272.6 & $>$15.89 & 274.9$\pm$6.5 & unreliable & unreliable & 13.79$\pm$0.02 & 221.9$\pm$13.7\\
30 & N11-ELS-038 & 74.1884 & -66.4197 & -2.50 & 2.72 & -0.400$\pm$0.056 & 288.3 & 21.34 & 286.5 & $>$15.55 & 287.3$\pm$12.7 & unreliable & unreliable & $>$14.16 & 269.4$\pm$13.5\\
31 & PGMW3120 & 74.1951 & -66.4130 & -2.50 & 2.73 & -0.398$\pm$0.056 & 288.4 & 21.39 & 290.4 & $>$15.86 & 279.7$\pm$28.0 & unreliable & unreliable & $>$14.52 & 274.1$\pm$23.9\\
32 & LMCE078-1 & 84.3734 & -69.2478 & 1.39 & -0.02 & -0.649$\pm$0.045 & 269.7 & 21.67 & 277.2 & $>$15.81 & 283.1$\pm$11.9 & unreliable & unreliable & $>$14.28 & 256.1$\pm$12.5\\
33 & SK-65D22 & 75.3462 & -65.8759 & -2.08 & 3.31 & -1.672$\pm$0.054 & 293.5 & 21.04 & 294.9 & $>$15.72 & 278.3$\pm$18.6 & $>$13.66 & 274.1$\pm$16.0 & 13.95$\pm$0.02 & 273.0$\pm$22.4\\
34 & SK-71D19 & 80.5656 & -71.3609 & 0.04 & -2.09 & -2.089$\pm$0.050 & 239.7 & 20.96 & 242.3 & $>$15.51 & 240.5$\pm$14.8 & unreliable & unreliable & 14.26$\pm$0.02 & 250.2$\pm$14.3\\
35 & SK-69D104 & 79.7479 & -69.2152 & -0.25 & 0.06 & -0.757$\pm$0.049 & 271.2 & 21.19 & 276.7 & 15.10$\pm$0.01 & 260.6$\pm$11.9 & unreliable & unreliable & 13.62$\pm$0.05 & 215.5$\pm$36.0\\
36 & VFTS440 & 84.6572 & -69.0892 & 1.50 & 0.13 & 0.189$\pm$0.044 & 271.9 & self-abs & self-abs & $>$15.90 & 274.0$\pm$44.2 & unreliable & unreliable & $>$14.64 & 224.0$\pm$27.7\\
37 & N11-ELS-018 & 74.1710 & -66.4113 & -2.50 & 2.73 & -0.433$\pm$0.056 & 288.3 & 21.43 & 292.2 & $>$15.62 & 275.5$\pm$9.9 & unreliable & unreliable & $>$14.12 & 264.3$\pm$11.9\\
38 & UCAC3-42-30814 & 83.9660 & -69.3886 & 1.24 & -0.15 & -1.084$\pm$0.040 & 267.8 & 21.08 & 269.1 & $>$15.64 & 271.7$\pm$10.7 & unreliable & unreliable & $>$14.46 & 254.6$\pm$4.7\\
39 & SK-67D111 & 81.7003 & -67.4916 & 0.48 & 1.78 & -0.828$\pm$0.056 & 296.1 & 21.19 & 301.7 & $>$15.47 & 293.0$\pm$14.4 & 13.50$\pm$0.03 & 256.6$\pm$22.4 & 13.77$\pm$0.03 & 260.6$\pm$28.0\\
40 & SK-71D50 & 85.1799 & -71.4835 & 1.50 & -2.27 & -2.345$\pm$0.048 & 243.4 & 21.45 & 254.7 & $>$15.90 & 264.2$\pm$20.6 & 13.48$\pm$0.04 & 233.0$\pm$32.4 & 13.56$\pm$0.08 & 249.5$\pm$69.2\\
41 & SK-70D115 & 87.2069 & -70.0661 & 2.30 & -0.92 & -0.895$\pm$0.051 & 259.7 & 21.55 & 257.0 & $>$15.90 & 277.6$\pm$8.4 & $>$14.20 & 239.2$\pm$19.0 & $>$14.40 & 238.4$\pm$17.3\\
42 & BI214 & 83.5258 & -69.4193 & 1.08 & -0.17 & -1.496$\pm$0.040 & 267.6 & 21.07 & 272.4 & $>$15.80 & 267.7$\pm$19.6 & unreliable & unreliable & $>$14.48 & 242.9$\pm$7.9\\
43 & SK-66D19 & 73.9748 & -66.4165 & -2.58 & 2.72 & -1.032$\pm$0.054 & 287.5 & 21.50 & 289.0 & $>$15.39 & 256.3$\pm$31.3 & 13.77$\pm$0.03 & 274.8$\pm$27.3 & 14.00$\pm$0.04 & 256.8$\pm$33.4\\
44 & BI272 & 86.0963 & -67.2414 & 2.18 & 1.93 & -1.812$\pm$0.057 & 295.4 & 20.73 & 310.0 & $>$15.55 & 287.5$\pm$10.3 & unreliable & unreliable & unreliable & unreliable\\
45 & SK-69D50 & 74.3129 & -69.3389 & -2.16 & -0.17 & -1.991$\pm$0.051 & 266.9 & 20.79 & 271.5 & $>$15.48 & 243.0$\pm$20.3 & $>$13.64 & 249.0$\pm$21.7 & 13.90$\pm$0.04 & 249.4$\pm$35.0\\
46 & SK-68D16 & 74.4075 & -68.4100 & -2.22 & 0.75 & -0.949$\pm$0.053 & 278.8 & 20.85 & 284.7 & $>$15.47 & 267.6$\pm$10.0 & unreliable & unreliable & 13.94$\pm$0.02 & 235.4$\pm$18.4\\
47 & SK-67D118 & 81.8888 & -67.2918 & 0.56 & 1.97 & -1.850$\pm$0.056 & 299.0 & 20.96 & 318.1 & 15.21$\pm$0.02 & 269.5$\pm$15.7 & unreliable & unreliable & 14.03$\pm$0.03 & 272.9$\pm$24.1\\
48 & UCAC3-42-33014 & 85.0568 & -69.4264 & 1.62 & -0.22 & -0.873$\pm$0.042 & 266.8 & 21.35 & 277.0 & $>$15.79 & 273.7$\pm$9.2 & $>$14.42 & 221.4$\pm$6.6 & $>$14.81 & 218.9$\pm$4.8\\
49 & SK-68D112 & 82.7835 & -68.6151 & 0.85 & 0.64 & -1.348$\pm$0.052 & 279.5 & 21.46 & 276.4 & 15.42$\pm$0.01 & 270.3$\pm$11.3 & $>$13.69 & 233.8$\pm$8.8 & 13.66$\pm$0.03 & 223.8$\pm$26.6\\
50 & SK-67D191 & 83.3918 & -67.5055 & 1.13 & 1.74 & -1.343$\pm$0.053 & 295.3 & 21.39 & 297.1 & $>$15.60 & 285.9$\pm$14.1 & unreliable & unreliable & $>$14.52 & 276.6$\pm$19.2\\
51 & SK-68D155 & 85.7289 & -68.9485 & 1.90 & 0.24 & -1.436$\pm$0.048 & 273.4 & 21.76 & 280.5 & $>$15.86 & 288.5$\pm$12.5 & 13.85$\pm$0.02 & 241.5$\pm$13.0 & unreliable & unreliable\\
52 & N11-ELS-013 & 74.2536 & -66.4070 & -2.47 & 2.74 & -0.375$\pm$0.056 & 288.6 & 21.59 & 295.2 & $>$15.83 & 274.7$\pm$29.0 & unreliable & unreliable & $>$14.42 & 275.5$\pm$32.5\\
53 & BI173 & 81.7915 & -69.1323 & 0.48 & 0.13 & -1.737$\pm$0.049 & 272.2 & 21.07 & 249.8 & $>$15.75 & 235.0$\pm$11.0 & unreliable & unreliable & 13.85$\pm$0.04 & 224.4$\pm$29.5\\
54 & SK-67D101 & 81.4844 & -67.5080 & 0.40 & 1.76 & -0.801$\pm$0.056 & 295.9 & 21.26 & 301.6 & 15.36$\pm$0.02 & 293.3$\pm$19.1 & unreliable & unreliable & 14.03$\pm$0.02 & 263.9$\pm$19.5\\
55 & SK-67D168 & 82.9672 & -67.5724 & 0.96 & 1.68 & -1.709$\pm$0.054 & 294.6 & $<$20.4 & N/A & 15.17$\pm$0.01 & 257.2$\pm$12.4 & 13.35$\pm$0.02 & 249.4$\pm$15.5 & 13.52$\pm$0.06 & 246.2$\pm$44.8\\
56 & LMCX-4 & 83.2065 & -66.3703 & 1.11 & 2.88 & -1.763$\pm$0.056 & 302.9 & $<$20.4 & N/A & 14.71$\pm$0.03 & 264.2$\pm$25.7 & unreliable & unreliable & unreliable & unreliable\\
57 & BI184 & 82.6278 & -71.0421 & 0.71 & -1.78 & -0.992$\pm$0.047 & 244.0 & 20.94 & 262.0 & $>$15.61 & 261.5$\pm$6.0 & unreliable & unreliable & unreliable & unreliable\\
58 & LH9-34 & 74.1887 & -66.4936 & -2.49 & 2.65 & -0.756$\pm$0.056 & 288.2 & 21.40 & 278.3 & $>$15.95 & 269.2$\pm$23.8 & 13.45$\pm$0.01 & 250.3$\pm$10.5 & 13.88$\pm$0.03 & 245.8$\pm$25.9\\
59 & SK-71D8 & 76.8469 & -71.1983 & -1.16 & -1.96 & -2.159$\pm$0.052 & 243.8 & 20.87 & 228.9 & $>$15.89 & 223.2$\pm$18.6 & unreliable & unreliable & 14.23$\pm$0.02 & 232.0$\pm$16.7\\
60 & VFTS-66 & 84.3879 & -69.0762 & 1.41 & 0.15 & -0.162$\pm$0.048 & 272.2 & 21.72 & 275.3 & $>$15.83 & 273.4$\pm$13.1 & unreliable & unreliable & unreliable & unreliable\\
61 & SK-66D171 & 84.2601 & -66.6436 & 1.51 & 2.58 & -1.559$\pm$0.055 & 300.9 & 21.16 & 315.8 & $>$15.57 & 267.3$\pm$12.7 & 13.33$\pm$0.02 & 254.8$\pm$17.1 & 13.72$\pm$0.05 & 255.0$\pm$41.3\\
62 & SK-69D279 & 85.4361 & -69.5875 & 1.74 & -0.39 & -1.412$\pm$0.043 & 264.2 & 21.57 & 270.7 & $>$15.77 & 267.3$\pm$14.3 & unreliable & unreliable & unreliable & unreliable\\
63 & SK-70D32 & 75.0425 & -70.1860 & -1.83 & -0.99 & -1.714$\pm$0.049 & 256.0 & 20.96 & 246.4 & 15.17$\pm$0.02 & 226.6$\pm$18.8 & unreliable & unreliable & 13.65$\pm$0.05 & 203.0$\pm$41.0\\
64 & SK-66D17 & 73.9812 & -66.4724 & -2.57 & 2.66 & -1.104$\pm$0.054 & 287.4 & 21.30 & 280.6 & $>$15.64 & 271.2$\pm$10.3 & unreliable & unreliable & N/A & N/A\\
65 & SK-68D135 & 84.4548 & -68.9171 & 1.44 & 0.31 & -0.997$\pm$0.051 & 274.5 & 21.55 & 278.3 & $>$15.89 & 274.4$\pm$17.9 & $>$14.36 & 255.3$\pm$16.6 & unreliable & unreliable\\
66 & VFTS87 & 84.4027 & -69.1255 & 1.41 & 0.10 & -0.112$\pm$0.048 & 271.5 & 21.77 & 277.0 & $>$15.79 & 271.8$\pm$44.9 & unreliable & unreliable & $>$14.30 & 248.9$\pm$38.2\\
67 & SK-71D41 & 82.6673 & -71.0936 & 0.72 & -1.83 & -0.931$\pm$0.048 & 243.2 & 21.39 & 246.9 & $>$15.82 & 247.9$\pm$23.5 & $>$14.21 & 235.6$\pm$19.6 & $>$14.60 & 233.6$\pm$20.7\\
68 & SK-67D5 & 72.5789 & -67.6606 & -2.98 & 1.42 & -1.718$\pm$0.057 & 276.4 & 21.20 & 280.0 & $>$15.56 & 272.6$\pm$12.4 & 13.25$\pm$0.03 & 256.5$\pm$27.0 & 13.56$\pm$0.05 & 239.2$\pm$43.7\\
69 & SK-69D220 & 84.1820 & -69.4965 & 1.31 & -0.26 & -1.224$\pm$0.039 & 266.2 & 21.35 & 285.5 & unreliable & unreliable & unreliable & unreliable & unreliable & unreliable\\
70 & SK-68D52 & 76.8351 & -68.5357 & -1.32 & 0.70 & -1.854$\pm$0.051 & 280.8 & 21.23 & 254.1 & $>$15.75 & 264.2$\pm$14.0 & 13.71$\pm$0.02 & 240.0$\pm$15.2 & 14.20$\pm$0.02 & 242.6$\pm$13.9\\
71 & SK-67D107 & 81.5862 & -67.4988 & 0.44 & 1.77 & -0.805$\pm$0.056 & 296.1 & 21.25 & 305.0 & 15.38$\pm$0.02 & 283.7$\pm$16.7 & $>$14.20 & 284.2$\pm$21.0 & $>$14.31 & 280.5$\pm$16.2\\
72 & SK-67D106 & 81.5634 & -67.5000 & 0.43 & 1.77 & -0.791$\pm$0.056 & 296.1 & 21.23 & 306.4 & 15.36$\pm$0.02 & 295.0$\pm$18.6 & $>$14.16 & 274.4$\pm$23.1 & $>$14.22 & 268.9$\pm$16.6\\
73 & HV5622 & 77.3717 & -68.9174 & -1.10 & 0.33 & -0.903$\pm$0.051 & 275.4 & 21.30 & 266.6 & $>$15.68 & 258.9$\pm$8.6 & unreliable & unreliable & unreliable & unreliable\\
74 & N11-ELS-033 & 74.0459 & -66.4734 & -2.55 & 2.66 & -0.883$\pm$0.054 & 287.6 & 21.25 & 279.6 & $>$15.65 & 271.3$\pm$6.7 & 13.62$\pm$0.01 & 245.8$\pm$12.1 & unreliable & unreliable\\
75 & SK-70D79 & 76.6553 & -70.4901 & -1.26 & -1.26 & -1.829$\pm$0.049 & 252.1 & 21.11 & 241.4 & $>$15.62 & 226.8$\pm$19.5 & $>$13.58 & 209.8$\pm$19.4 & $>$14.09 & 211.9$\pm$18.5\\
76 & SK-69D43 & 74.0436 & -69.2606 & -2.26 & -0.11 & -2.183$\pm$0.051 & 267.1 & 20.74 & 264.1 & 15.22$\pm$0.02 & 234.6$\pm$15.8 & 13.52$\pm$0.01 & 253.2$\pm$10.7 & 13.70$\pm$0.04 & 243.1$\pm$30.2\\
77 & SK-68D41 & 76.3630 & -68.1674 & -1.52 & 1.05 & -1.770$\pm$0.050 & 286.0 & 21.33 & 267.6 & unreliable & unreliable & 13.59$\pm$0.02 & 257.2$\pm$12.5 & 13.94$\pm$0.03 & 249.2$\pm$25.5\\
78 & SK-68D140 & 84.7382 & -68.9481 & 1.54 & 0.27 & -0.708$\pm$0.046 & 273.9 & 21.68 & 275.4 & $>$15.73 & 277.2$\pm$18.4 & $>$14.33 & 251.0$\pm$14.5 & $>$14.58 & 250.6$\pm$3.4\\
79 & SK-67D2 & 71.7686 & -67.1148 & -3.36 & 1.92 & -3.191$\pm$0.054 & 275.5 & 21.30 & 273.2 & $>$15.72 & 272.6$\pm$20.8 & 13.67$\pm$0.02 & 254.5$\pm$14.2 & unreliable & unreliable\\
80 & SK-68D23A & 75.2012 & -68.0996 & -1.95 & 1.09 & -2.361$\pm$0.055 & 285.5 & 21.19 & 267.3 & 15.45$\pm$0.01 & 259.3$\pm$9.9 & unreliable & unreliable & 14.16$\pm$0.02 & 238.1$\pm$14.2\\
81 & SK-66D35 & 74.2685 & -66.5774 & -2.45 & 2.57 & -0.981$\pm$0.057 & 288.3 & 21.51 & 287.7 & $>$15.79 & 261.7$\pm$16.1 & 13.57$\pm$0.02 & 258.4$\pm$14.4 & unreliable & unreliable\\
82 & SK-68D129 & 84.1116 & -68.9589 & 1.32 & 0.27 & -1.100$\pm$0.050 & 274.1 & 21.54 & 278.4 & $>$15.61 & 266.4$\pm$13.5 & $>$14.02 & 251.2$\pm$9.9 & $>$14.44 & 245.0$\pm$4.0\\
83 & N206-FS-170 & 82.7622 & -70.8323 & 0.76 & -1.57 & -2.805$\pm$0.046 & 247.1 & 21.16 & 248.4 & 15.33$\pm$0.03 & 254.1$\pm$22.3 & unreliable & unreliable & unreliable & unreliable\\
84 & SK-71D35 & 82.5174 & -71.1323 & 0.67 & -1.87 & -1.242$\pm$0.048 & 242.7 & 21.29 & 245.7 & $>$15.55 & 247.8$\pm$10.0 & unreliable & unreliable & unreliable & unreliable\\
85 & NGC1818-ROB-D1 & 76.1346 & -66.4132 & -1.72 & 2.80 & -2.790$\pm$0.056 & 295.6 & 21.27 & 302.4 & unreliable & unreliable & unreliable & unreliable & unreliable & unreliable\\
86 & SK-67D14 & 73.6329 & -67.2568 & -2.63 & 1.87 & -1.657$\pm$0.059 & 283.3 & 21.24 & 291.1 & $>$15.42 & 255.3$\pm$14.8 & 13.35$\pm$0.05 & 276.8$\pm$46.2 & $>$14.08 & 262.3$\pm$17.5\\
87 & SK-69D52 & 74.4538 & -69.8729 & -2.06 & -0.70 & -2.651$\pm$0.050 & 259.6 & 20.56 & 258.8 & unreliable & unreliable & unreliable & unreliable & unreliable & unreliable\\
88 & SK-68D26 & 75.3844 & -68.1786 & -1.88 & 1.02 & -2.200$\pm$0.054 & 285.5 & 21.24 & 269.2 & $>$15.79 & 267.7$\pm$11.4 & unreliable & unreliable & unreliable & unreliable\\
89 & NGC2004-ELS-26 & 82.6515 & -67.2952 & 0.85 & 1.96 & -2.189$\pm$0.056 & 298.4 & 20.98 & 308.7 & unreliable & unreliable & unreliable & unreliable & unreliable & unreliable\\
90 & SK-70D50 & 75.9412 & -70.1993 & -1.52 & -0.98 & -2.171$\pm$0.051 & 256.2 & 20.64 & 245.7 & unreliable & unreliable & unreliable & unreliable & unreliable & unreliable\\
91 & SK-67D78 & 80.0795 & -67.3016 & -0.14 & 1.97 & -2.455$\pm$0.054 & 299.1 & 21.00 & 297.8 & unreliable & unreliable & unreliable & unreliable & unreliable & unreliable\\
92 & SK-69D140 & 81.9141 & -69.2116 & 0.52 & 0.05 & -1.834$\pm$0.049 & 271.1 & 20.73 & 267.6 & 15.35$\pm$0.02 & 224.2$\pm$15.8 & 13.33$\pm$0.03 & 249.4$\pm$27.1 & unreliable & unreliable\\
93 & SK-70D16 & 73.7390 & -70.0412 & -2.28 & -0.89 & -2.234$\pm$0.051 & 256.2 & 20.51 & 244.5 & unreliable & unreliable & unreliable & unreliable & unreliable & unreliable\\
94 & SK-68D8 & 73.4304 & -68.7148 & -2.54 & 0.41 & -2.592$\pm$0.052 & 271.5 & 20.88 & 266.6 & unreliable & unreliable & unreliable & unreliable & unreliable & unreliable\\
95 & NGC2004-ELS-3 & 82.6684 & -67.2691 & 0.86 & 1.99 & -2.218$\pm$0.056 & 298.5 & 21.09 & 307.8 & unreliable & unreliable & unreliable & unreliable & unreliable & unreliable\\
96 & SK-67D195 & 83.4664 & -67.1339 & 1.17 & 2.11 & -2.460$\pm$0.055 & 298.2 & 20.47 & 292.3 & unreliable & unreliable & unreliable & unreliable & unreliable & unreliable\\
97 & SK-67D197 & 83.4961 & -67.5377 & 1.17 & 1.71 & -1.438$\pm$0.053 & 294.8 & 21.24 & 299.1 & unreliable & unreliable & unreliable & unreliable & unreliable & unreliable\\
98 & SK-66D50 & 75.7868 & -66.9597 & -1.82 & 2.24 & -2.447$\pm$0.055 & 293.6 & 21.16 & 286.3 & unreliable & unreliable & unreliable & unreliable & unreliable & unreliable\\
99 & SK-67D207 & 83.7311 & -67.3519 & 1.27 & 1.89 & -1.717$\pm$0.055 & 296.2 & 20.74 & 302.8 & unreliable & unreliable & unreliable & unreliable & unreliable & unreliable\\
100 & SK-67D20 & 73.8806 & -67.5007 & -2.51 & 1.64 & -2.409$\pm$0.058 & 283.2 & 20.70 & 283.5 & 15.09$\pm$0.05 & 262.7$\pm$41.5 & $>$13.79 & 274.5$\pm$20.2 & $>$14.41 & 269.2$\pm$25.0\\
101 & SK-68D15 & 74.3504 & -68.3992 & -2.24 & 0.76 & -0.936$\pm$0.053 & 278.7 & 21.01 & 285.8 & 15.43$\pm$0.02 & 267.4$\pm$14.1 & $>$14.00 & 241.1$\pm$16.9 & 13.98$\pm$0.01 & 230.7$\pm$10.4\\
102 & SK-66D51 & 75.7871 & -66.6826 & -1.84 & 2.52 & -2.240$\pm$0.055 & 294.1 & 20.46 & 299.8 & $>$15.27 & 292.8$\pm$25.9 & $>$13.69 & 267.5$\pm$14.2 & unreliable & unreliable\\
103 & SK-65D55 & 80.4904 & -65.8167 & 0.02 & 3.45 & -1.937$\pm$0.063 & 304.3 & $<$20.4 & N/A & 14.93$\pm$0.03 & 253.4$\pm$23.8 & $>$13.54 & 272.7$\pm$18.6 & $>$14.10 & 271.6$\pm$14.2\\
104 & SK-71D21 & 80.5939 & -71.5995 & 0.05 & -2.33 & -1.500$\pm$0.049 & 238.3 & 21.18 & 247.0 & $>$15.46 & 255.3$\pm$16.2 & 13.39$\pm$0.04 & 229.7$\pm$35.0 & 13.72$\pm$0.04 & 238.9$\pm$33.3\\
105 & SK-67D104 & 81.5165 & -67.4992 & 0.41 & 1.77 & -0.781$\pm$0.056 & 296.1 & 21.18 & 302.4 & 15.26$\pm$0.02 & 286.9$\pm$21.3 & $>$14.07 & 283.6$\pm$10.4 & $>$14.10 & 270.5$\pm$11.1\\
106 & SK-69D175 & 82.8563 & -69.0940 & 0.86 & 0.16 & -1.347$\pm$0.049 & 272.6 & 21.05 & 275.3 & $>$15.57 & 268.2$\pm$14.5 & 13.23$\pm$0.04 & 225.1$\pm$32.3 & unreliable & unreliable\\
107 & SK-69D191 & 83.5802 & -69.7529 & 1.09 & -0.51 & -1.188$\pm$0.039 & 262.6 & 21.19 & 261.4 & $>$15.84 & 255.6$\pm$16.2 & $>$13.97 & 253.1$\pm$9.2 & $>$14.36 & 249.1$\pm$5.3\\
108 & SK-69D246 & 84.7224 & -69.0336 & 1.53 & 0.18 & 0.020$\pm$0.044 & 272.7 & 21.65 & 275.6 & $>$15.92 & 266.3$\pm$20.2 & $>$14.16 & 229.7$\pm$16.3 & $>$14.42 & 225.4$\pm$9.4\\
109 & HD269927C & 84.7421 & -69.4888 & 1.50 & -0.27 & -0.607$\pm$0.042 & 266.0 & 21.61 & 276.6 & $>$15.96 & 270.2$\pm$19.7 & $>$14.25 & 218.8$\pm$15.4 & unreliable & unreliable\\
110 & SK-67D266 & 86.4664 & -67.2405 & 2.33 & 1.92 & -1.752$\pm$0.059 & 295.4 & self-abs & self-abs & unreliable & unreliable & $>$13.49 & 284.7$\pm$41.7 & unreliable & unreliable\\
\enddata
\tablenotetext{}{Note: \\
Columns (1)--(4): IDs, star names, right ascensions (RA) and declinations (DEC) of the targets from the ULLYSES program (DR5). \\
Columns (5)--(6): $x$ and $y$ coordiantes of each star in the plane of the LMC based on the orthographic projection method outlined in \cite{choi22}. \\
Column (7): Star formation rate surface density measured toward each sight line based on an \Halpha\ emission map from \cite{guastad01}; see Section \ref{sec:data_halpha} for further details. \\
Column (8): Heliocentric velocities of the LMC's young stellar disk at the locations of the corresponding sight lines, measured based on the kinematic model of a population of young supergiant stars (RSGs); see Section \ref{sec:data_rsg} for further details. \\
Columns (9)--(10): Column densities and flux-weighted centroid velocities of the \HI\ gas, measured based on an \HI\ 21cm data cube from \cite{kim03}. We find \HI\ self absorption in five sight lines, BAT99-105 (ID 2),
VFTS-482 (ID 9), SK-65D47 (ID 26), VFTS440 (ID 36), and SK-67D266 (ID 110), which we note as ``self-abs" in the corresponding entries. Additionally, we do not find significant \HI\ detection along three sight lines, SK-67D168 (ID 55), LMCX-4 (ID 56), and SK-65D55 (ID 103), for which we indicate $3\sigma$ upper limits in $\log$ N(\HI) based on the data cube's sensitivity level; see Section \ref{sec:data_HI} for further details. \\
Columns (11)--(16): Column densities and centroid velocities of \SII, \SiIV, and \CIV\ integrated from $\vhelio=175$ to 375 $\kms$ based on the AOD method (see Section \ref{sec:method_contfit} for details). A measurement is labelled as ``unreliable" when our algorithm does not find a reliable continuum placement over the MW+LMC interstellar absorption velocity range; see Section \ref{sec:method_summary} for further details. For each ion doublet, we use the weaker lines (\SII\ 1250, \SiIV\ 1402, \CIV\ 1550) when the stronger lines are saturated. The only exception is \SiIV\ measured toward SK-68D73, which only has STIS/E140H coverage extending up to \SiIV\ 1393 \AA\ but not 1402 \AA; in this case, we use the measurement from \SiIV\ 1393\AA. When both lines of an ion doublet are non-saturated, we take the mean values of the doublet lines and the errors are combined in quadrature. %\\ 
% [a]: When spectral types are with more than 8 letters in string length, we note them here to keep the table width within the page limit. 
% (3) ST92-5-31: O2-3(n)f*p; 
% (4) SK-67D22: O2 If*/WN5; 
% (5) SK-66D172: O2III(f*)+OB; 
% (6) VFTS72: O2 V-III(n)((f*)); 
% (7) BI237: O2 V((f*)); 
% (8) SK-67D211: O2III(f*); 
% (9) VFTS-482: O2.5 If*/WN6; 
% (10) N11-ELS-060: O3 V((f*)); 
% (11) ST92-5-27: O3 V((f)); 
% (12) LH114-7: O3 III(f)*; 
% (13) VFTS-267: O3 III-I(n)f*; 
% (14) VFTS-404: O3.5 V(n)((fc)); 
% (15) W61-28-23: O3.5 III(f *); 
% (18) SK-67D105: O4f + O6 V; 
% (22) W61-28-5: O4 V((f+)); 
% (23) FARINA-88: O4 III(f); 
% (27) SK-67D69: O4 III(f); 
% (28) VFTS352: O4.5V(n)((fc)):z:+O5.5V(n)((fc)):z:; 
% (30) N11-ELS-038: O5 III(f+); 
% (31) PGMW3120: O5.5 V ((f*)); 
% (36) VFTS440: O6-6.5II(f); 
% (37) N11-ELS-018: O6 II(f\^+ ); 
% (42) BI214: O6.5(n)(f)p; 
% (44) BI272: O7:III-II:; 
% (48) UCAC3-42-33014: O7.5(f)np; 
% (49) SK-68D112: O7.5(n)(f)p; 
% (54) SK-67D101: O8II((f)); 
% (60) VFTS-66: O9V + B0.2V; 
% (65) SK-68D135: ON9.7 Ia+; 
% (66) VFTS87: O9.7Ib-II; 
% (78) SK-68D140: B0.7 Ib-Iab Nwk; 
% (108) SK-69D246: WN5/6h+WN6/7h. 
}
\end{deluxetable*}
\end{longrotatetable}

\appendix
\restartappendixnumbering
\section{The Foreground High-Velocity Cloud at $v\sim90-175~\kms$}
\label{app:discuss_hvc}

Section \ref{sec:discuss_hvc} shows that HVC90-175's absorption can be constrained to be at $\vhelio\lesssim175~\kms$ and it is located at $d_{\odot}<13.3$ kpc \citep{richter15, werner15}. The cloud is observed in \HI\ \citep{deboer90, staveley-smith03}, \Halpha\ emission \citep{ciampa21}, molecular hydrogen \citep{richter99, tchernyshyov22}, and UV absorption tracing ionized gas \citep[e.g.][]{savage81, lehner07, lehner09, barger16, roman-duval19}. Here we further discuss HVC90-175 based on UV measurements from the literature and newly obtained in this work. We show that the kinematics of HVC90-175 is consistent with being a MW halo cloud at $\vlsr\sim120~\kms$ with a temperature of $T\sim10^{4.2}$ K and a non-thermal broadening of $\sigma_{\rm nth}\sim10~\kms$. As noted in Section \ref{sec:intro}, in the general direction of the LMC, the $\vhelio$ and $\vlsr$ velocities are offset by $\vhelio - \vlsr\sim10~\kms$.

\begin{figure*}[t]
\centering
\includegraphics[width=0.95\textwidth]{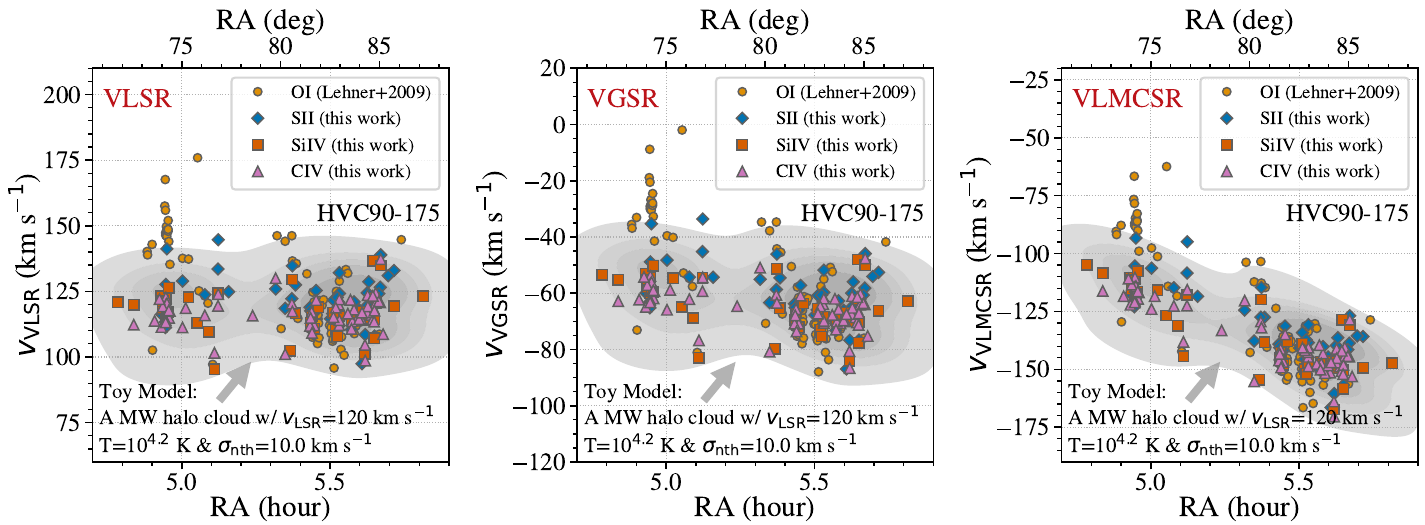}
\caption{Multiphase ion velocities of HVC90-175 as a function of R.A. in the Local Standard of Rest (LSR; left), Galactic Standard of Red (GSR; middle), and LMC Standard of Rest (LMCSR; right) frames (see Equation \ref{eq:lsr_gsr_lmcsr}). HVC90-175 is detected in low ions such as \OI\ (yellow circles; \citealt{lehner09}, see also \citealt{roman-duval19}) as well as in \SII, \SiIV, and \CIV\ (this work). HVC90-175's gas kinematics is consistent with being a MW halo cloud at $\vlsr=120~\kms$ with a temperature of $T=10^{4.2}$ K and a non-thermal broadening of $\sigma_{\rm nth}\sim10~\kms$. The distance to HVC90-175 is constrained to be $d_\odot<13.3$ kpc \citep{richter15, werner15}. 
\label{fig:hvc_lsr_gsr_lmcsr}
}
\end{figure*}

\begin{figure*}[t]
\centering
\includegraphics[width=0.95\textwidth]{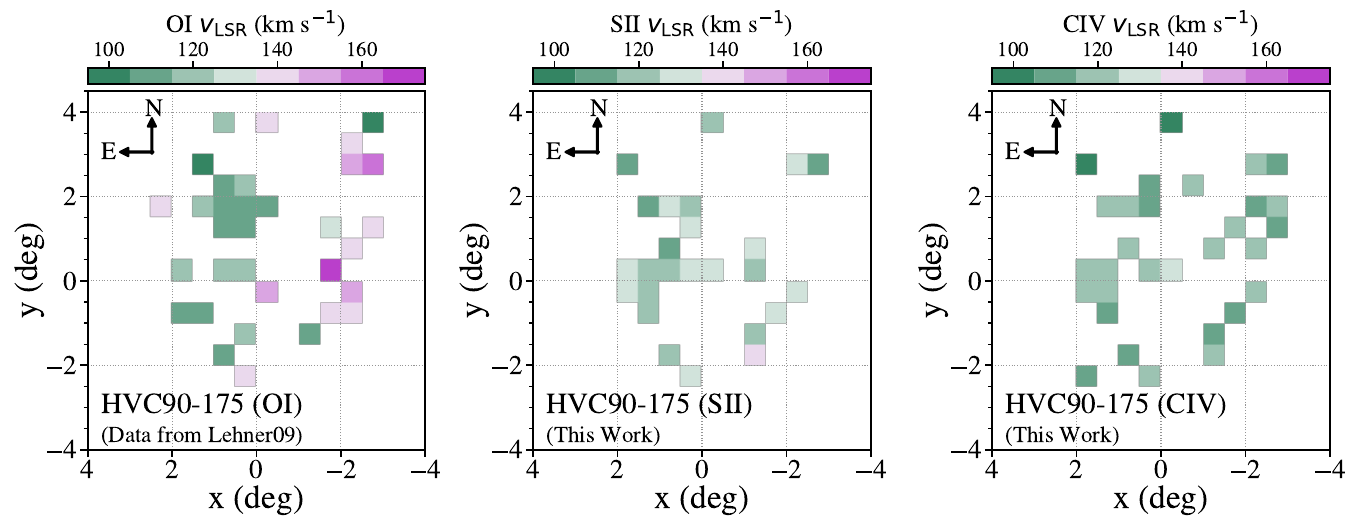}
\caption{2D velocity distributions of HVC90-175 in the same orthographic projection as in Figure \ref{fig:vhel_xymap}. In the left panel, we find that the \OI\ gas, as measured by \cite{lehner09}, is moving faster at $\vlsr\sim150-175~\kms$ toward the west. Meanwhile, \SII\ (middle), \SiIV\ (not shown), and \CIV\ (right) show nearly constant $\vlsr$ at $\sim 120~\kms$ across the surface of the cloud. 
\label{fig:hvc_2dmaps}
}
\end{figure*}

In Figure \ref{fig:hvc_lsr_gsr_lmcsr}, we show HVC90-175's velocity distributions in \OI, \SII, \SiIV, and \CIV\ as a function of R.A.. The \SII, \SiIV, and \CIV\ are weighted centroid velocities integrated over a velocity range of $\vhelio=90-175~\kms$ using the same ULLYSES sight lines as discussed in Section \ref{sec:method_contfit}; here we prioritize using the stronger lines of \SII\ 1253, \SiIV\ 1393, and \CIV\ 1548 unless the absorption is saturated. The \OI\ data points are adopted from \cite{lehner09}, who studied HVC90-175 toward 139 FUSE sight lines in the direction of the LMC.

The ion velocities are shown in three reference frames: the Local Standard of Rest (LSR; left panel), the Galactic Standard of Rest (GSR; middle panel), and the LMC Standard of Rest (LMCSR; right panel). The velocity conversions among these frames are based on the equations given by \cite{lehner09} in their section 3.2: 
\begin{equation}
    \begin{split}
        v_{\rm LMCSR} & =v_{\rm GSR}+86{\rm cos\ell}{\rm cos}b+268{\rm sin\ell}{\rm cos}b-252{\rm sin}b\\
        v_{\rm GSR} & = \vlsr+220{\rm sin}\ell{\rm cos}b~~~.\\
    \end{split}
    \label{eq:lsr_gsr_lmcsr}
\end{equation}
We note that the \OI\ distribution in the right panel (VLMCSR) reproduces \citeauthor{lehner09}'s figure 5, in which they first noted that HVC90-175 shows an apparent velocity gradient with R.A. in the LMCSR frame (see also \citealt{roman-duval19}).

When we examine the velocity distributions with R.A. in all three reference frames, we find that HVC90-175's velocity gradient can only be mildly observed in the LSR (left panel) and the GSR frames (middle panel). In Figure \ref{fig:hvc_2dmaps}, we show the 2D $\vlsr$ distribution of the HVC90-175's gas across the surface of the LMC in the same orthographic projection as in Figure \ref{fig:vhel_xymap}. For \SII\ and \CIV, there is no apparent trend in $\vlsr$; and in \OI, the southwest half (bottom right corner in the left panel) is moving faster, which is in the opposite direction of the LMC's rotation.

To better understand HVC90-175's ion velocities, we model a hypothetical MW halo cloud lying in front of the LMC at a constant velocity of $\vlsr=120~\kms$ and show the modelled velocity distributions as gray contours in Figure \ref{fig:hvc_lsr_gsr_lmcsr}. The velocities of the modeled halo cloud are calculated for the same ULLYSES sight lines used in this work, and the contours are generated using the \texttt{kdeplot} function from the \texttt{seaborn} package.

The spreads in the velocities are calculated by assuming a temperature of $T=10^{4.2}$ K for our \SII\ measurements and a non-thermal broadening of $\sigma_{\rm nth}=10~\kms$, which corresponds to a total velocity dispersion of $\sigma_v \equiv \sqrt{(k_{\rm B}T/m_{\rm S}) + \sigma_{\rm nth}^2}= 10.2~\kms$. The choice of $\sigma_{\rm nth}=10~\kms$ is based on what is typically measured for non-thermal broadening in the ionized CGM gas \citep{chen23}. We note that the match between the modeled and the observed data in Figure \ref{fig:hvc_lsr_gsr_lmcsr} would be further improved if we assumed a higher temperature ($T\gtrsim10^{4.2}$ K), adopted measurements from the lighter ions such as \OI, \SiIV\ or \CIV, or more contribution from non-thermal broadening.

The right panel of Figure \ref{fig:hvc_lsr_gsr_lmcsr} shows that the modeled halo cloud, which has a constant $\vlsr=120~\kms$, exhibits a similar velocity gradient with R.A. as the observed ion data. We thus demonstrate that the velocity gradient with R.A. in HVC90-175 is mainly caused by velocity transformation between rest frames. The kinematic signature of HVC90-175 is consistent with that of a foreground cloud moving at $\vlsr\sim120~\kms$ with $T\sim10^{4.2}$ K and a non-thermal broadening of $\sigma_{\rm nth}\sim10~\kms$, which are physical properties commonly found in known high-velocity clouds in the MW and ionized CGM gas in low-redshift galaxies \citep{wakker01, putman12, chen23}. 

Additionally, previous studies have measured the metallicity of HVC90-175 to be at [\OI/\HI] $=-0.51^{+0.12}_{-0.16}$ dex \citep{lehner09} and an ion ratio of $N_{\rm OVI}$/$N_{\rm CIV}\sim1-10$ \citep{lehner07}. HVC90-175's metallicity is consistent with those of high-velocity clouds in the MW such as Complex C ($Z\sim0.1-0.3~\zsun$; \citealt{shull11}), and its ion ratio is also consistent with those measured in the MW's ionized gas in various directions ($N_{\rm OVI}$/$N_{\rm CIV}\sim1-7$; \citealt{sembach03}, see their table 11).

Lastly, we note that there is a small sample of \OI\ absorbers at R.A.$\sim5$ hour or $\sim75\degree$ that cannot be accounted for by our MW halo cloud model in Figure \ref{fig:hvc_lsr_gsr_lmcsr}. The \OI\ gas at this R.A. is moving at $\vlsr\sim150-175~\kms$ and located at the west half of the LMC disk, which are shown as pink-purple pixels in Figure \ref{fig:hvc_2dmaps}. Coincidently, the LMC's neutral and ionized gas in this area is also moving at similar velocity of $\vlsr\sim200~\kms$, as shown in Figure \ref{fig:vhel_xymap}. It is possible that the excess \OI\ absorbers with $\vlsr\sim150-175~\kms$ near R.A.$\sim$5 hour are either an extension of the LMC's gas with low $\vlsr$ or blended with the LMC at similar velocities.

\newpage 
\bibliographystyle{aasjournal}
\bibliography{main}

\begin{thebibliography}{}
\expandafter\ifx\csname natexlab\endcsname\relax\def\natexlab#1{#1}\fi
\providecommand{\url}[1]{\href{#1}{#1}}
\providecommand{\dodoi}[1]{doi:~\href{http://doi.org/#1}{\nolinkurl{#1}}}
\providecommand{\doeprint}[1]{\href{http://ascl.net/#1}{\nolinkurl{http://ascl.net/#1}}}
\providecommand{\doarXiv}[1]{\href{https://arxiv.org/abs/#1}{\nolinkurl{https://arxiv.org/abs/#1}}}

\bibitem[{{Ambikasaran} {et~al.}(2015){Ambikasaran}, {Foreman-Mackey},
  {Greengard}, {Hogg}, \& {O'Neil}}]{george}
{Ambikasaran}, S., {Foreman-Mackey}, D., {Greengard}, L., {Hogg}, D.~W., \&
  {O'Neil}, M. 2015, IEEE Transactions on Pattern Analysis and Machine
  Intelligence, 38, 252, \dodoi{10.1109/TPAMI.2015.2448083}

\bibitem[{{Ambrocio-Cruz} {et~al.}(2016){Ambrocio-Cruz}, {Le Coarer}, {Rosado},
  {Russeil}, {Amram}, {Laval}, {Epinat}, {Ram{\'\i}rez}, {Odonne}, \&
  {Goldes}}]{ambrocio-cruz16}
{Ambrocio-Cruz}, P., {Le Coarer}, E., {Rosado}, M., {et~al.} 2016, \mnras, 457,
  2048, \dodoi{10.1093/mnras/stw054}

\bibitem[{{Andersson} {et~al.}(2023){Andersson}, {Agertz}, {Renaud}, \&
  {Teyssier}}]{andersson23}
{Andersson}, E.~P., {Agertz}, O., {Renaud}, F., \& {Teyssier}, R. 2023, \mnras,
  521, 2196, \dodoi{10.1093/mnras/stad692}

\bibitem[{{Asplund} {et~al.}(2009){Asplund}, {Grevesse}, {Sauval}, \&
  {Scott}}]{asplund09}
{Asplund}, M., {Grevesse}, N., {Sauval}, A.~J., \& {Scott}, P. 2009, \araa, 47,
  481, \dodoi{10.1146/annurev.astro.46.060407.145222}

\bibitem[{{Astropy Collaboration} {et~al.}(2013){Astropy Collaboration},
  {Robitaille}, {Tollerud}, {Greenfield}, {Droettboom}, {Bray}, {Aldcroft},
  {Davis}, {Ginsburg}, {Price-Whelan}, {Kerzendorf}, {Conley}, {Crighton},
  {Barbary}, {Muna}, {Ferguson}, {Grollier}, {Parikh}, {Nair}, {Unther},
  {Deil}, {Woillez}, {Conseil}, {Kramer}, {Turner}, {Singer}, {Fox}, {Weaver},
  {Zabalza}, {Edwards}, {Azalee Bostroem}, {Burke}, {Casey}, {Crawford},
  {Dencheva}, {Ely}, {Jenness}, {Labrie}, {Lim}, {Pierfederici}, {Pontzen},
  {Ptak}, {Refsdal}, {Servillat}, \& {Streicher}}]{astropy:2013}
{Astropy Collaboration}, {Robitaille}, T.~P., {Tollerud}, E.~J., {et~al.} 2013,
  \aap, 558, A33, \dodoi{10.1051/0004-6361/201322068}

\bibitem[{{Astropy Collaboration} {et~al.}(2018){Astropy Collaboration},
  {Price-Whelan}, {Sip{\H{o}}cz}, {G{\"u}nther}, {Lim}, {Crawford}, {Conseil},
  {Shupe}, {Craig}, {Dencheva}, {Ginsburg}, {Vand erPlas}, {Bradley},
  {P{\'e}rez-Su{\'a}rez}, {de Val-Borro}, {Aldcroft}, {Cruz}, {Robitaille},
  {Tollerud}, {Ardelean}, {Babej}, {Bach}, {Bachetti}, {Bakanov}, {Bamford},
  {Barentsen}, {Barmby}, {Baumbach}, {Berry}, {Biscani}, {Boquien}, {Bostroem},
  {Bouma}, {Brammer}, {Bray}, {Breytenbach}, {Buddelmeijer}, {Burke},
  {Calderone}, {Cano Rodr{\'\i}guez}, {Cara}, {Cardoso}, {Cheedella}, {Copin},
  {Corrales}, {Crichton}, {D'Avella}, {Deil}, {Depagne}, {Dietrich}, {Donath},
  {Droettboom}, {Earl}, {Erben}, {Fabbro}, {Ferreira}, {Finethy}, {Fox},
  {Garrison}, {Gibbons}, {Goldstein}, {Gommers}, {Greco}, {Greenfield},
  {Groener}, {Grollier}, {Hagen}, {Hirst}, {Homeier}, {Horton}, {Hosseinzadeh},
  {Hu}, {Hunkeler}, {Ivezi{\'c}}, {Jain}, {Jenness}, {Kanarek}, {Kendrew},
  {Kern}, {Kerzendorf}, {Khvalko}, {King}, {Kirkby}, {Kulkarni}, {Kumar},
  {Lee}, {Lenz}, {Littlefair}, {Ma}, {Macleod}, {Mastropietro}, {McCully},
  {Montagnac}, {Morris}, {Mueller}, {Mumford}, {Muna}, {Murphy}, {Nelson},
  {Nguyen}, {Ninan}, {N{\"o}the}, {Ogaz}, {Oh}, {Parejko}, {Parley}, {Pascual},
  {Patil}, {Patil}, {Plunkett}, {Prochaska}, {Rastogi}, {Reddy Janga},
  {Sabater}, {Sakurikar}, {Seifert}, {Sherbert}, {Sherwood-Taylor}, {Shih},
  {Sick}, {Silbiger}, {Singanamalla}, {Singer}, {Sladen}, {Sooley},
  {Sornarajah}, {Streicher}, {Teuben}, {Thomas}, {Tremblay}, {Turner},
  {Terr{\'o}n}, {van Kerkwijk}, {de la Vega}, {Watkins}, {Weaver}, {Whitmore},
  {Woillez}, {Zabalza}, \& {Astropy Contributors}}]{astropy:2018}
{Astropy Collaboration}, {Price-Whelan}, A.~M., {Sip{\H{o}}cz}, B.~M., {et~al.}
  2018, \aj, 156, 123, \dodoi{10.3847/1538-3881/aabc4f}

\bibitem[{{Astropy Collaboration} {et~al.}(2022){Astropy Collaboration},
  {Price-Whelan}, {Lim}, {Earl}, {Starkman}, {Bradley}, {Shupe}, {Patil},
  {Corrales}, {Brasseur}, {N{"o}the}, {Donath}, {Tollerud}, {Morris},
  {Ginsburg}, {Vaher}, {Weaver}, {Tocknell}, {Jamieson}, {van Kerkwijk},
  {Robitaille}, {Merry}, {Bachetti}, {G{"u}nther}, {Aldcroft},
  {Alvarado-Montes}, {Archibald}, {B{'o}di}, {Bapat}, {Barentsen}, {Baz{'a}n},
  {Biswas}, {Boquien}, {Burke}, {Cara}, {Cara}, {Conroy}, {Conseil}, {Craig},
  {Cross}, {Cruz}, {D'Eugenio}, {Dencheva}, {Devillepoix}, {Dietrich},
  {Eigenbrot}, {Erben}, {Ferreira}, {Foreman-Mackey}, {Fox}, {Freij}, {Garg},
  {Geda}, {Glattly}, {Gondhalekar}, {Gordon}, {Grant}, {Greenfield}, {Groener},
  {Guest}, {Gurovich}, {Handberg}, {Hart}, {Hatfield-Dodds}, {Homeier},
  {Hosseinzadeh}, {Jenness}, {Jones}, {Joseph}, {Kalmbach}, {Karamehmetoglu},
  {Ka{l}uszy{'n}ski}, {Kelley}, {Kern}, {Kerzendorf}, {Koch}, {Kulumani},
  {Lee}, {Ly}, {Ma}, {MacBride}, {Maljaars}, {Muna}, {Murphy}, {Norman},
  {O'Steen}, {Oman}, {Pacifici}, {Pascual}, {Pascual-Granado}, {Patil},
  {Perren}, {Pickering}, {Rastogi}, {Roulston}, {Ryan}, {Rykoff}, {Sabater},
  {Sakurikar}, {Salgado}, {Sanghi}, {Saunders}, {Savchenko}, {Schwardt},
  {Seifert-Eckert}, {Shih}, {Jain}, {Shukla}, {Sick}, {Simpson},
  {Singanamalla}, {Singer}, {Singhal}, {Sinha}, {Sip{H{o}}cz}, {Spitler},
  {Stansby}, {Streicher}, {{{S}}umak}, {Swinbank}, {Taranu}, {Tewary},
  {Tremblay}, {Val-Borro}, {Van Kooten}, {Vasovi{'c}}, {Verma}, {de Miranda
  Cardoso}, {Williams}, {Wilson}, {Winkel}, {Wood-Vasey}, {Xue}, {Yoachim},
  {Zhang}, {Zonca}, \& {Astropy Project Contributors}}]{astropy:2022}
{Astropy Collaboration}, {Price-Whelan}, A.~M., {Lim}, P.~L., {et~al.} 2022,
  \apj, 935, 167, \dodoi{10.3847/1538-4357/ac7c74}

\bibitem[{{Barger} {et~al.}(2016){Barger}, {Lehner}, \& {Howk}}]{barger16}
{Barger}, K.~A., {Lehner}, N., \& {Howk}, J.~C. 2016, \apj, 817, 91,
  \dodoi{10.3847/0004-637X/817/2/91}

\bibitem[{{Berg} {et~al.}(2022){Berg}, {James}, {King}, {McDonald}, {Chen},
  {Chisholm}, {Heckman}, {Martin}, {Stark}, {Aloisi}, {Amor{\'\i}n},
  {Arellano-C{\'o}rdova}, {Bayliss}, {Bordoloi}, {Brinchmann}, {Charlot},
  {Chevallard}, {Clark}, {Erb}, {Feltre}, {Gronke}, {Hayes}, {Henry},
  {Hernandez}, {Jaskot}, {Jones}, {Kewley}, {Kumari}, {Leitherer}, {Llerena},
  {Maseda}, {Mingozzi}, {Nanayakkara}, {Ouchi}, {Plat}, {Pogge},
  {Ravindranath}, {Rigby}, {Sanders}, {Scarlata}, {Senchyna}, {Skillman},
  {Steidel}, {Strom}, {Sugahara}, {Wilkins}, {Wofford}, {Xu}, \& {Classy
  Team}}]{berg22}
{Berg}, D.~A., {James}, B.~L., {King}, T., {et~al.} 2022, \apjs, 261, 31,
  \dodoi{10.3847/1538-4365/ac6c03}

\bibitem[{{Besla} {et~al.}(2007){Besla}, {Kallivayalil}, {Hernquist},
  {Robertson}, {Cox}, {van der Marel}, \& {Alcock}}]{besla07}
{Besla}, G., {Kallivayalil}, N., {Hernquist}, L., {et~al.} 2007, \apj, 668,
  949, \dodoi{10.1086/521385}

\bibitem[{{Br{\"u}ns} {et~al.}(2005){Br{\"u}ns}, {Kerp}, {Staveley-Smith},
  {Mebold}, {Putman}, {Haynes}, {Kalberla}, {Muller}, \& {Filipovic}}]{bruns05}
{Br{\"u}ns}, C., {Kerp}, J., {Staveley-Smith}, L., {et~al.} 2005, \aap, 432,
  45, \dodoi{10.1051/0004-6361:20040321}

\bibitem[{{Calzetti}(1997)}]{calzetti97}
{Calzetti}, D. 1997, in American Institute of Physics Conference Series, Vol.
  408, The ultraviolet universe at low and High redshift, ed. W.~H. {Waller}
  (AIP), 403--412, \dodoi{10.1063/1.53764}

\bibitem[{{Chen} {et~al.}(2023){Chen}, {Qu}, {Rauch}, {Chen}, {Zahedy},
  {Johnson}, {Schaye}, {Rudie}, {Boettcher}, {Cantalupo},
  {Faucher-Gigu{\`e}re}, {Greene}, {Lopez}, \& {Simcoe}}]{chen23}
{Chen}, H.-W., {Qu}, Z., {Rauch}, M., {et~al.} 2023, \apjl, 955, L25,
  \dodoi{10.3847/2041-8213/acf85b}

\bibitem[{{Chen} {et~al.}(2010){Chen}, {Tremonti}, {Heckman}, {Kauffmann},
  {Weiner}, {Brinchmann}, \& {Wang}}]{chen10}
{Chen}, Y.-M., {Tremonti}, C.~A., {Heckman}, T.~M., {et~al.} 2010, \aj, 140,
  445, \dodoi{10.1088/0004-6256/140/2/445}

\bibitem[{{Chisholm} {et~al.}(2017){Chisholm}, {Tremonti}, {Leitherer}, \&
  {Chen}}]{chisholm17}
{Chisholm}, J., {Tremonti}, C.~A., {Leitherer}, C., \& {Chen}, Y. 2017, \mnras,
  469, 4831, \dodoi{10.1093/mnras/stx1164}

\bibitem[{{Chisholm} {et~al.}(2016){Chisholm}, {Tremonti}, {Leitherer}, {Chen},
  \& {Wofford}}]{chisholm16a}
{Chisholm}, J., {Tremonti}, C.~A., {Leitherer}, C., {Chen}, Y., \& {Wofford},
  A. 2016, \mnras, 457, 3133, \dodoi{10.1093/mnras/stw178}

\bibitem[{{Chisholm} {et~al.}(2015){Chisholm}, {Tremonti}, {Leitherer}, {Chen},
  {Wofford}, \& {Lundgren}}]{chisholm15}
{Chisholm}, J., {Tremonti}, C.~A., {Leitherer}, C., {et~al.} 2015, \apj, 811,
  149, \dodoi{10.1088/0004-637X/811/2/149}

\bibitem[{{Choi} {et~al.}(2022){Choi}, {Olsen}, {Besla}, {van der Marel},
  {Zivick}, {Kallivayalil}, \& {Nidever}}]{choi22}
{Choi}, Y., {Olsen}, K. A.~G., {Besla}, G., {et~al.} 2022, \apj, 927, 153,
  \dodoi{10.3847/1538-4357/ac4e90}

\bibitem[{{Choi} {et~al.}(2018){Choi}, {Nidever}, {Olsen}, {Blum}, {Besla},
  {Zaritsky}, {van der Marel}, {Bell}, {Gallart}, {Cioni}, {Johnson}, {Vivas},
  {Saha}, {de Boer}, {No{\"e}l}, {Monachesi}, {Massana}, {Conn},
  {Martinez-Delgado}, {Mu{\~n}oz}, \& {Stringfellow}}]{Choi2018}
{Choi}, Y., {Nidever}, D.~L., {Olsen}, K., {et~al.} 2018, \apj, 866, 90,
  \dodoi{10.3847/1538-4357/aae083}

\bibitem[{{Ciampa} {et~al.}(2021){Ciampa}, {Barger}, {Lehner}, {Horn},
  {Hernandez}, {Haffner}, {Smart}, {Bustard}, {Barber}, \& {Boot}}]{ciampa21}
{Ciampa}, D.~A., {Barger}, K.~A., {Lehner}, N., {et~al.} 2021, \apj, 908, 62,
  \dodoi{10.3847/1538-4357/abd320}

\bibitem[{{Danforth} {et~al.}(2002){Danforth}, {Howk}, {Fullerton}, {Blair}, \&
  {Sembach}}]{danforth02}
{Danforth}, C.~W., {Howk}, J.~C., {Fullerton}, A.~W., {Blair}, W.~P., \&
  {Sembach}, K.~R. 2002, \apjs, 139, 81, \dodoi{10.1086/338239}

\bibitem[{{Davis} {et~al.}(2023){Davis}, {Tremonti}, {Swiggum}, {Moustakas},
  {Diamond-Stanic}, {Coil}, {Geach}, {Hickox}, {Perrotta}, {Petter}, {Rudnick},
  {Rupke}, {Sell}, \& {Whalen}}]{davis23}
{Davis}, J.~D., {Tremonti}, C.~A., {Swiggum}, C.~N., {et~al.} 2023, \apj, 951,
  105, \dodoi{10.3847/1538-4357/accbbf}

\bibitem[{{de Boer} {et~al.}(1990){de Boer}, {Morras}, \& {Bajaja}}]{deboer90}
{de Boer}, K.~S., {Morras}, R., \& {Bajaja}, E. 1990, \aap, 233, 523

\bibitem[{{Erb} {et~al.}(2012){Erb}, {Quider}, {Henry}, \& {Martin}}]{erb12}
{Erb}, D.~K., {Quider}, A.~M., {Henry}, A.~L., \& {Martin}, C.~L. 2012, \apj,
  759, 26, \dodoi{10.1088/0004-637X/759/1/26}

\bibitem[{{Freedman} {et~al.}(2001){Freedman}, {Madore}, {Gibson}, {Ferrarese},
  {Kelson}, {Sakai}, {Mould}, {Kennicutt}, {Ford}, {Graham}, {Huchra},
  {Hughes}, {Illingworth}, {Macri}, \& {Stetson}}]{freedman01}
{Freedman}, W.~L., {Madore}, B.~F., {Gibson}, B.~K., {et~al.} 2001, \apj, 553,
  47, \dodoi{10.1086/320638}

\bibitem[{{Gaustad} {et~al.}(2001){Gaustad}, {McCullough}, {Rosing}, \& {Van
  Buren}}]{guastad01}
{Gaustad}, J.~E., {McCullough}, P.~R., {Rosing}, W., \& {Van Buren}, D. 2001,
  \pasp, 113, 1326, \dodoi{10.1086/323969}

\bibitem[{{Gnat} \& {Sternberg}(2007)}]{gnat07}
{Gnat}, O., \& {Sternberg}, A. 2007, \apjs, 168, 213, \dodoi{10.1086/509786}

\bibitem[{{Gordon} {et~al.}(2003){Gordon}, {Clayton}, {Misselt}, {Landolt}, \&
  {Wolff}}]{gordon03}
{Gordon}, K.~D., {Clayton}, G.~C., {Misselt}, K.~A., {Landolt}, A.~U., \&
  {Wolff}, M.~J. 2003, \apj, 594, 279, \dodoi{10.1086/376774}

\bibitem[{{Hainich} {et~al.}(2019){Hainich}, {Ramachandran}, {Shenar},
  {Sander}, {Todt}, {Gruner}, {Oskinova}, \& {Hamann}}]{Hainich19}
{Hainich}, R., {Ramachandran}, V., {Shenar}, T., {et~al.} 2019, \aap, 621, A85,
  \dodoi{10.1051/0004-6361/201833787}

\bibitem[{Harris {et~al.}(2020)Harris, Millman, van~der Walt, Gommers,
  Virtanen, Cournapeau, Wieser, Taylor, Berg, Smith, Kern, Picus, Hoyer, van
  Kerkwijk, Brett, Haldane, del R{'{\i}}o, Wiebe, Peterson,
  G{'{e}}rard-Marchant, Sheppard, Reddy, Weckesser, Abbasi, Gohlke, \&
  Oliphant}]{numpy}
Harris, C.~R., Millman, K.~J., van~der Walt, S.~J., {et~al.} 2020, Nature, 585,
  357, \dodoi{10.1038/s41586-020-2649-2}

\bibitem[{{Harris} \& {Zaritsky}(2009)}]{harris09}
{Harris}, J., \& {Zaritsky}, D. 2009, \aj, 138, 1243,
  \dodoi{10.1088/0004-6256/138/5/1243}

\bibitem[{{Hawcroft} {et~al.}(2023){Hawcroft}, {Sana}, {Mahy}, {Sundqvist}, {de
  Koter}, {Crowther}, {Bestenlehner}, {Brands}, {David-Uraz}, {Decin}, {Erba},
  {Garcia}, {Hamann}, {Herrero}, {Ignace}, {Kee}, {Kub{\'a}tov{\'a}},
  {Lefever}, {Moffat}, {Najarro}, {Oskinova}, {Pauli}, {Prinja}, {Puls},
  {Sander}, {Shenar}, {St-Louis}, {ud-Doula}, \& {Vink}}]{hawcroft23}
{Hawcroft}, C., {Sana}, H., {Mahy}, L., {et~al.} 2023, arXiv e-prints,
  arXiv:2303.12165, \dodoi{10.48550/arXiv.2303.12165}

\bibitem[{{Haydon} {et~al.}(2020){Haydon}, {Kruijssen}, {Chevance}, {Hygate},
  {Krumholz}, {Schruba}, \& {Longmore}}]{haydon20}
{Haydon}, D.~T., {Kruijssen}, J.~M.~D., {Chevance}, M., {et~al.} 2020, \mnras,
  498, 235, \dodoi{10.1093/mnras/staa2430}

\bibitem[{{Heckman} {et~al.}(2015){Heckman}, {Alexandroff}, {Borthakur},
  {Overzier}, \& {Leitherer}}]{heckman15}
{Heckman}, T.~M., {Alexandroff}, R.~M., {Borthakur}, S., {Overzier}, R., \&
  {Leitherer}, C. 2015, \apj, 809, 147, \dodoi{10.1088/0004-637X/809/2/147}

\bibitem[{{Heckman} {et~al.}(2000){Heckman}, {Lehnert}, {Strickland}, \&
  {Armus}}]{heckman00}
{Heckman}, T.~M., {Lehnert}, M.~D., {Strickland}, D.~K., \& {Armus}, L. 2000,
  \apjs, 129, 493, \dodoi{10.1086/313421}

\bibitem[{{Hopkins} {et~al.}(2014){Hopkins}, {Kere{\v{s}}}, {O{\~n}orbe},
  {Faucher-Gigu{\`e}re}, {Quataert}, {Murray}, \& {Bullock}}]{hopkins14}
{Hopkins}, P.~F., {Kere{\v{s}}}, D., {O{\~n}orbe}, J., {et~al.} 2014, \mnras,
  445, 581, \dodoi{10.1093/mnras/stu1738}

\bibitem[{{Howk} {et~al.}(2002){Howk}, {Sembach}, {Savage}, {Massa},
  {Friedman}, \& {Fullerton}}]{howk02}
{Howk}, J.~C., {Sembach}, K.~R., {Savage}, B.~D., {et~al.} 2002, \apj, 569,
  214, \dodoi{10.1086/339322}

\bibitem[{Hunter(2007)}]{matplotlib}
Hunter, J.~D. 2007, Computing in Science \& Engineering, 9, 90,
  \dodoi{10.1109/MCSE.2007.55}

\bibitem[{{Jenkins}(1996)}]{jenkins96}
{Jenkins}, E.~B. 1996, \apj, 471, 292, \dodoi{10.1086/177969}

\bibitem[{{Kallivayalil} {et~al.}(2013){Kallivayalil}, {van der Marel},
  {Besla}, {Anderson}, \& {Alcock}}]{kallivayalil13}
{Kallivayalil}, N., {van der Marel}, R.~P., {Besla}, G., {Anderson}, J., \&
  {Alcock}, C. 2013, \apj, 764, 161, \dodoi{10.1088/0004-637X/764/2/161}

\bibitem[{{Kelly}(2007)}]{kelly07}
{Kelly}, B.~C. 2007, \apj, 665, 1489, \dodoi{10.1086/519947}

\bibitem[{{Kennicutt} \& {Evans}(2012)}]{kennicutt12}
{Kennicutt}, R.~C., \& {Evans}, N.~J. 2012, \araa, 50, 531,
  \dodoi{10.1146/annurev-astro-081811-125610}

\bibitem[{{Kim} \& {Ostriker}(2017)}]{kim17}
{Kim}, C.-G., \& {Ostriker}, E.~C. 2017, \apj, 846, 133,
  \dodoi{10.3847/1538-4357/aa8599}

\bibitem[{{Kim} \& {Ostriker}(2018)}]{kim18}
---. 2018, \apj, 853, 173, \dodoi{10.3847/1538-4357/aaa5ff}

\bibitem[{{Kim} {et~al.}(2020){Kim}, {Ostriker}, {Somerville}, {Bryan},
  {Fielding}, {Forbes}, {Hayward}, {Hernquist}, \& {Pandya}}]{kim20}
{Kim}, C.-G., {Ostriker}, E.~C., {Somerville}, R.~S., {et~al.} 2020, \apj, 900,
  61, \dodoi{10.3847/1538-4357/aba962}

\bibitem[{{Kim} {et~al.}(1998){Kim}, {Staveley-Smith}, {Dopita}, {Freeman},
  {Sault}, {Kesteven}, \& {McConnell}}]{kim98}
{Kim}, S., {Staveley-Smith}, L., {Dopita}, M.~A., {et~al.} 1998, \apj, 503,
  674, \dodoi{10.1086/306030}

\bibitem[{{Kim} {et~al.}(2003){Kim}, {Staveley-Smith}, {Dopita}, {Sault},
  {Freeman}, {Lee}, \& {Chu}}]{kim03}
---. 2003, \apjs, 148, 473, \dodoi{10.1086/376980}

\bibitem[{{Kroupa} \& {Weidner}(2003)}]{kroupa03}
{Kroupa}, P., \& {Weidner}, C. 2003, \apj, 598, 1076, \dodoi{10.1086/379105}

\bibitem[{{Kudritzki} \& {Puls}(2000)}]{kudritzki00}
{Kudritzki}, R.-P., \& {Puls}, J. 2000, \araa, 38, 613,
  \dodoi{10.1146/annurev.astro.38.1.613}

\bibitem[{{Lah} {et~al.}(2024){Lah}, {Colless}, {D'Eugenio}, {Groves}, \&
  {Gelfand}}]{Lah2024}
{Lah}, P., {Colless}, M., {D'Eugenio}, F., {Groves}, B., \& {Gelfand}, J.~D.
  2024, \mnras, 529, 2611, \dodoi{10.1093/mnras/stae671}

\bibitem[{{Lehner} \& {Howk}(2007)}]{lehner07}
{Lehner}, N., \& {Howk}, J.~C. 2007, \mnras, 377, 687,
  \dodoi{10.1111/j.1365-2966.2007.11631.x}

\bibitem[{{Lehner} {et~al.}(2009){Lehner}, {Staveley-Smith}, \&
  {Howk}}]{lehner09}
{Lehner}, N., {Staveley-Smith}, L., \& {Howk}, J.~C. 2009, \apj, 702, 940,
  \dodoi{10.1088/0004-637X/702/2/940}

\bibitem[{{Li} {et~al.}(2017){Li}, {Bryan}, \& {Ostriker}}]{li17}
{Li}, M., {Bryan}, G.~L., \& {Ostriker}, J.~P. 2017, \apj, 841, 101,
  \dodoi{10.3847/1538-4357/aa7263}

\bibitem[{{Martin}(2005)}]{martin05}
{Martin}, C.~L. 2005, \apj, 621, 227, \dodoi{10.1086/427277}

\bibitem[{{Mazzi} {et~al.}(2021){Mazzi}, {Girardi}, {Zaggia}, {Pastorelli},
  {Rubele}, {Bressan}, {Cioni}, {Clementini}, {Cusano}, {Rocha}, {Gullieuszik},
  {Kerber}, {Marigo}, {Ripepi}, {Bekki}, {Bell}, {de Grijs}, {Groenewegen},
  {Ivanov}, {Oliveira}, {Sun}, \& {van Loon}}]{mazzi21}
{Mazzi}, A., {Girardi}, L., {Zaggia}, S., {et~al.} 2021, \mnras, 508, 245,
  \dodoi{10.1093/mnras/stab2399}

\bibitem[{{McKee} \& {Ostriker}(1977)}]{mckee77}
{McKee}, C.~F., \& {Ostriker}, J.~P. 1977, \apj, 218, 148,
  \dodoi{10.1086/155667}

\bibitem[{{McLeod} {et~al.}(2019){McLeod}, {Dale}, {Evans}, {Ginsburg},
  {Kruijssen}, {Pellegrini}, {Ramsay}, \& {Testi}}]{mcleod19}
{McLeod}, A.~F., {Dale}, J.~E., {Evans}, C.~J., {et~al.} 2019, \mnras, 486,
  5263, \dodoi{10.1093/mnras/sty2696}

\bibitem[{{Medallon} \& {Welty}(2023)}]{stis_handbook}
{Medallon}, S., \& {Welty}, D. 2023, in STIS Instrument Handbook for Cycle 31
  v. 22, Vol.~22, 22

\bibitem[{{Morton}(2003)}]{morton03}
{Morton}, D.~C. 2003, \apjs, 149, 205, \dodoi{10.1086/377639}

\bibitem[{{Nidever} {et~al.}(2008){Nidever}, {Majewski}, \& {Butler
  Burton}}]{nidever08}
{Nidever}, D.~L., {Majewski}, S.~R., \& {Butler Burton}, W. 2008, \apj, 679,
  432, \dodoi{10.1086/587042}

\bibitem[{{Oh} {et~al.}(2022){Oh}, {Kim}, {For}, \& {Staveley-Smith}}]{oh22}
{Oh}, S.-H., {Kim}, S., {For}, B.-Q., \& {Staveley-Smith}, L. 2022, \apj, 928,
  177, \dodoi{10.3847/1538-4357/ac5905}

\bibitem[{{Olsen} \& {Massey}(2007)}]{olsen07}
{Olsen}, K. A.~G., \& {Massey}, P. 2007, \apjl, 656, L61,
  \dodoi{10.1086/512484}

\bibitem[{{Olsen} {et~al.}(2011){Olsen}, {Zaritsky}, {Blum}, {Boyer}, \&
  {Gordon}}]{olsen11}
{Olsen}, K. A.~G., {Zaritsky}, D., {Blum}, R.~D., {Boyer}, M.~L., \& {Gordon},
  K.~D. 2011, \apj, 737, 29, \dodoi{10.1088/0004-637X/737/1/29}

\bibitem[{{Peeples} {et~al.}(2019){Peeples}, {Corlies}, {Tumlinson}, {O'Shea},
  {Lehner}, {O'Meara}, {Howk}, {Earl}, {Smith}, {Wise}, \&
  {Hummels}}]{peeples19}
{Peeples}, M.~S., {Corlies}, L., {Tumlinson}, J., {et~al.} 2019, \apj, 873,
  129, \dodoi{10.3847/1538-4357/ab0654}

\bibitem[{{Poggianti} {et~al.}(2016){Poggianti}, {Fasano}, {Omizzolo},
  {Gullieuszik}, {Bettoni}, {Moretti}, {Paccagnella}, {Jaff{\'e}}, {Vulcani},
  {Fritz}, {Couch}, \& {D'Onofrio}}]{poggianti16}
{Poggianti}, B.~M., {Fasano}, G., {Omizzolo}, A., {et~al.} 2016, \aj, 151, 78,
  \dodoi{10.3847/0004-6256/151/3/78}

\bibitem[{{Putman} {et~al.}(2012){Putman}, {Peek}, \& {Joung}}]{putman12}
{Putman}, M.~E., {Peek}, J.~E.~G., \& {Joung}, M.~R. 2012, \araa, 50, 491,
  \dodoi{10.1146/annurev-astro-081811-125612}

\bibitem[{{Putman} {et~al.}(2021){Putman}, {Zheng}, {Price-Whelan}, {Grcevich},
  {Johnson}, {Tollerud}, \& {Peek}}]{putman21}
{Putman}, M.~E., {Zheng}, Y., {Price-Whelan}, A.~M., {et~al.} 2021, \apj, 913,
  53, \dodoi{10.3847/1538-4357/abe391}

\bibitem[{{Rasmussen} \& {Williams}(2006)}]{GP-book}
{Rasmussen}, C.~E., \& {Williams}, C. K.~I. 2006, {Gaussian Processes for
  Machine Learning}

\bibitem[{{Rathjen} {et~al.}(2021){Rathjen}, {Naab}, {Girichidis}, {Walch},
  {W{\"u}nsch}, {Dinnbier}, {Seifried}, {Klessen}, \& {Glover}}]{rathjen21}
{Rathjen}, T.-E., {Naab}, T., {Girichidis}, P., {et~al.} 2021, \mnras, 504,
  1039, \dodoi{10.1093/mnras/stab900}

\bibitem[{{Reichardt Chu} {et~al.}(2022){Reichardt Chu}, {Fisher}, {Nielsen},
  {Chisholm}, {Girard}, {Kacprzak}, {Bolatto}, {Herrera-Camus}, {Sandstrom},
  {Li}, {Rickards Vaught}, \& {McPherson}}]{chu22}
{Reichardt Chu}, B., {Fisher}, D.~B., {Nielsen}, N.~M., {et~al.} 2022, \mnras,
  511, 5782, \dodoi{10.1093/mnras/stac420}

\bibitem[{{Reichardt Chu} {et~al.}(2024){Reichardt Chu}, {Fisher}, {Chisholm},
  {Berg}, {Bolatto}, {Cameron}, {Fielding}, {Herrera-Camus}, {Kacprzak}, {Li},
  {McLeod}, {McPherson}, {Nielsen}, {Rickards Vaught}, {Ridolfo}, \&
  {Sandstrom}}]{ReichardtChu24}
{Reichardt Chu}, B., {Fisher}, D.~B., {Chisholm}, J., {et~al.} 2024, arXiv
  e-prints, arXiv:2402.17830, \dodoi{10.48550/arXiv.2402.17830}

\bibitem[{{Richter} {et~al.}(2015){Richter}, {de Boer}, {Werner}, \&
  {Rauch}}]{richter15}
{Richter}, P., {de Boer}, K.~S., {Werner}, K., \& {Rauch}, T. 2015, \aap, 584,
  L6, \dodoi{10.1051/0004-6361/201527451}

\bibitem[{{Richter} {et~al.}(1999){Richter}, {de Boer}, {Widmann},
  {Kappelmann}, {Gringel}, {Grewing}, \& {Barnstedt}}]{richter99}
{Richter}, P., {de Boer}, K.~S., {Widmann}, H., {et~al.} 1999, \nat, 402, 386,
  \dodoi{10.1038/46492}

\bibitem[{{Ripepi} {et~al.}(2022){Ripepi}, {Chemin}, {Molinaro}, {Cioni},
  {Bekki}, {Clementini}, {de Grijs}, {De Somma}, {El Youssoufi}, {Girardi},
  {Groenewegen}, {Ivanov}, {Marconi}, {McMillan}, \& {van Loon}}]{ripepi22}
{Ripepi}, V., {Chemin}, L., {Molinaro}, R., {et~al.} 2022, \mnras, 512, 563,
  \dodoi{10.1093/mnras/stac595}

\bibitem[{Roman-Duval(2020)}]{ullyses_data_doi}
Roman-Duval, J. 2020, Hubble UV Legacy Library of Young Stars as Essential
  Standards ("ULLYSES"),  STScI/MAST, \dodoi{10.17909/T9-JZEH-XY14}

\bibitem[{{Roman-Duval} {et~al.}(2019{\natexlab{a}}){Roman-Duval}, {Jenkins},
  {Williams}, {Tchernyshyov}, {Gordon}, {Meixner}, {Hagen}, {Peek},
  {Sandstrom}, {Werk}, \& {Yanchulova Merica-Jones}}]{roman-duval19}
{Roman-Duval}, J., {Jenkins}, E.~B., {Williams}, B., {et~al.}
  2019{\natexlab{a}}, \apj, 871, 151, \dodoi{10.3847/1538-4357/aaf8bb}

\bibitem[{{Roman-Duval} {et~al.}(2019{\natexlab{b}}){Roman-Duval}, {Jenkins},
  {Williams}, {Tchernyshyov}, {Gordon}, {Meixner}, {Hagen}, {Peek},
  {Sandstrom}, {Werk}, \& {Yanchulova Merica-Jones}}]{roman-duval21}
---. 2019{\natexlab{b}}, \apj, 871, 151, \dodoi{10.3847/1538-4357/aaf8bb}

\bibitem[{{Roman-Duval} {et~al.}(2020){Roman-Duval}, {Proffitt}, {Taylor},
  {Monroe}, {Fischer}, {Fischer}, {Fullerton}, {Aloisi}, {Britt}, {Busko},
  {Carlberg}, {De Rosa}, {Jedrzejewski}, {Lockwood}, {Frazer}, {Hernandez},
  {James}, {Oliveira}, {Plesha}, {Riedel}, {Riley}, {Sahnow}, {Sankrit},
  {Shaw}, {Smith}, {Sohn}, {Som}, {Ubeda}, \& {Welty}}]{Roman-Duval20_ullyses}
{Roman-Duval}, J., {Proffitt}, C.~R., {Taylor}, J.~M., {et~al.} 2020, Research
  Notes of the American Astronomical Society, 4, 205,
  \dodoi{10.3847/2515-5172/abca2f}

\bibitem[{{Rubin} {et~al.}(2014){Rubin}, {Prochaska}, {Koo}, {Phillips},
  {Martin}, \& {Winstrom}}]{rubin14}
{Rubin}, K. H.~R., {Prochaska}, J.~X., {Koo}, D.~C., {et~al.} 2014, \apj, 794,
  156, \dodoi{10.1088/0004-637X/794/2/156}

\bibitem[{{Rupke} {et~al.}(2005){Rupke}, {Veilleux}, \& {Sanders}}]{rupke05}
{Rupke}, D.~S., {Veilleux}, S., \& {Sanders}, D.~B. 2005, \apjs, 160, 115,
  \dodoi{10.1086/432889}

\bibitem[{{Rupke}(2018)}]{rupke18}
{Rupke}, D. S.~N. 2018, Galaxies, 6, 138, \dodoi{10.3390/galaxies6040138}

\bibitem[{{Russell} \& {Dopita}(1992)}]{russell92}
{Russell}, S.~C., \& {Dopita}, M.~A. 1992, \apj, 384, 508,
  \dodoi{10.1086/170893}

\bibitem[{{Salem} {et~al.}(2015){Salem}, {Besla}, {Bryan}, {Putman}, {van der
  Marel}, \& {Tonnesen}}]{salem15}
{Salem}, M., {Besla}, G., {Bryan}, G., {et~al.} 2015, \apj, 815, 77,
  \dodoi{10.1088/0004-637X/815/1/77}

\bibitem[{{Savage} \& {de Boer}(1981)}]{savage81}
{Savage}, B.~D., \& {de Boer}, K.~S. 1981, \apj, 243, 460,
  \dodoi{10.1086/158613}

\bibitem[{{Savage} \& {Sembach}(1991)}]{Savage91}
{Savage}, B.~D., \& {Sembach}, K.~R. 1991, \apj, 379, 245,
  \dodoi{10.1086/170498}

\bibitem[{{Savage} \& {Sembach}(1996)}]{Savage96}
---. 1996, \araa, 34, 279, \dodoi{10.1146/annurev.astro.34.1.279}

\bibitem[{{Schaye} {et~al.}(2015){Schaye}, {Crain}, {Bower}, {Furlong},
  {Schaller}, {Theuns}, {Dalla Vecchia}, {Frenk}, {McCarthy}, {Helly},
  {Jenkins}, {Rosas-Guevara}, {White}, {Baes}, {Booth}, {Camps}, {Navarro},
  {Qu}, {Rahmati}, {Sawala}, {Thomas}, \& {Trayford}}]{schaye15}
{Schaye}, J., {Crain}, R.~A., {Bower}, R.~G., {et~al.} 2015, \mnras, 446, 521,
  \dodoi{10.1093/mnras/stu2058}

\bibitem[{{Sembach} {et~al.}(2003){Sembach}, {Wakker}, {Savage}, {Richter},
  {Meade}, {Shull}, {Jenkins}, {Sonneborn}, \& {Moos}}]{sembach03}
{Sembach}, K.~R., {Wakker}, B.~P., {Savage}, B.~D., {et~al.} 2003, \apjs, 146,
  165, \dodoi{10.1086/346231}

\bibitem[{{Setton} {et~al.}(2023){Setton}, {Besla}, {Patel}, {Hummels},
  {Zheng}, \& {Schneider}}]{setton23}
{Setton}, D.~J., {Besla}, G., {Patel}, E., {et~al.} 2023, arXiv e-prints,
  arXiv:2308.10963, \dodoi{10.48550/arXiv.2308.10963}

\bibitem[{{Shull} {et~al.}(2011){Shull}, {Stevans}, {Danforth}, {Penton},
  {Lockman}, \& {Arav}}]{shull11}
{Shull}, J.~M., {Stevans}, M., {Danforth}, C., {et~al.} 2011, \apj, 739, 105,
  \dodoi{10.1088/0004-637X/739/2/105}

\bibitem[{{Sirressi} {et~al.}(2024){Sirressi}, {Adamo}, {Hayes},
  {Rivera-Thorsen}, {Aloisi}, {Bik}, {Calzetti}, {Chisholm}, {Fox},
  {Fumagalli}, {Grasha}, {Hernandez}, {Messa}, {Osborne}, {{\"O}stlin},
  {Sabbi}, {Schinnerer}, {Smith}, {Usher}, \& {Wofford}}]{sirressi24}
{Sirressi}, M., {Adamo}, A., {Hayes}, M., {et~al.} 2024, \aj, 167, 166,
  \dodoi{10.3847/1538-3881/ad29f9}

\bibitem[{{Skowron} {et~al.}(2021){Skowron}, {Skowron}, {Udalski},
  {Szyma{\'n}ski}, {Soszy{\'n}ski}, {Wyrzykowski}, {Ulaczyk}, {Poleski},
  {Koz{\l}owski}, {Pietrukowicz}, {Mr{\'o}z}, {Rybicki}, {Iwanek}, {Wrona}, \&
  {Gromadzki}}]{Skowron2021}
{Skowron}, D.~M., {Skowron}, J., {Udalski}, A., {et~al.} 2021, \apjs, 252, 23,
  \dodoi{10.3847/1538-4365/abcb81}

\bibitem[{{Smart} {et~al.}(2023){Smart}, {Haffner}, {Barger}, {Ciampa}, {Hill},
  {Krishnarao}, \& {Madsen}}]{smart23}
{Smart}, B.~M., {Haffner}, L.~M., {Barger}, K.~A., {et~al.} 2023, \apj, 948,
  118, \dodoi{10.3847/1538-4357/acc06e}

\bibitem[{{Soderblom}(2023)}]{cos_handbook23}
{Soderblom}, D.~R. 2023, in COS Instrument Handbook v. 15.0, Vol.~15, 15

\bibitem[{{Staveley-Smith} {et~al.}(2003){Staveley-Smith}, {Kim}, {Calabretta},
  {Haynes}, \& {Kesteven}}]{staveley-smith03}
{Staveley-Smith}, L., {Kim}, S., {Calabretta}, M.~R., {Haynes}, R.~F., \&
  {Kesteven}, M.~J. 2003, \mnras, 339, 87,
  \dodoi{10.1046/j.1365-8711.2003.06146.x}

\bibitem[{{Tan} \& {Fielding}(2023)}]{tan23}
{Tan}, B., \& {Fielding}, D.~B. 2023, arXiv e-prints, arXiv:2305.14424,
  \dodoi{10.48550/arXiv.2305.14424}

\bibitem[{{Tchernyshyov}(2022)}]{tchernyshyov22}
{Tchernyshyov}, K. 2022, \apj, 931, 78, \dodoi{10.3847/1538-4357/ac68e0}

\bibitem[{{van der Marel} {et~al.}(2002){van der Marel}, {Alves}, {Hardy}, \&
  {Suntzeff}}]{vandermarel02}
{van der Marel}, R.~P., {Alves}, D.~R., {Hardy}, E., \& {Suntzeff}, N.~B. 2002,
  \aj, 124, 2639, \dodoi{10.1086/343775}

\bibitem[{{Veilleux} {et~al.}(2020){Veilleux}, {Maiolino}, {Bolatto}, \&
  {Aalto}}]{veilleux20}
{Veilleux}, S., {Maiolino}, R., {Bolatto}, A.~D., \& {Aalto}, S. 2020, \aapr,
  28, 2, \dodoi{10.1007/s00159-019-0121-9}

\bibitem[{{Vogelsberger} {et~al.}(2014){Vogelsberger}, {Genel}, {Springel},
  {Torrey}, {Sijacki}, {Xu}, {Snyder}, {Bird}, {Nelson}, \&
  {Hernquist}}]{vogelsberger14}
{Vogelsberger}, M., {Genel}, S., {Springel}, V., {et~al.} 2014, \nat, 509, 177,
  \dodoi{10.1038/nature13316}

\bibitem[{{Wakker} {et~al.}(1998){Wakker}, {Howk}, {Chu}, {Bomans}, \&
  {Points}}]{wakker98}
{Wakker}, B., {Howk}, J.~C., {Chu}, Y.-H., {Bomans}, D., \& {Points}, S.~D.
  1998, \apjl, 499, L87, \dodoi{10.1086/311334}

\bibitem[{{Wakker}(2001)}]{wakker01}
{Wakker}, B.~P. 2001, \apjs, 136, 463, \dodoi{10.1086/321783}

\bibitem[{{Weiner} {et~al.}(2009){Weiner}, {Coil}, {Prochaska}, {Newman},
  {Cooper}, {Bundy}, {Conselice}, {Dutton}, {Faber}, {Koo}, {Lotz}, {Rieke}, \&
  {Rubin}}]{weiner09}
{Weiner}, B.~J., {Coil}, A.~L., {Prochaska}, J.~X., {et~al.} 2009, \apj, 692,
  187, \dodoi{10.1088/0004-637X/692/1/187}

\bibitem[{{Werner} \& {Rauch}(2015)}]{werner15}
{Werner}, K., \& {Rauch}, T. 2015, \aap, 584, A19,
  \dodoi{10.1051/0004-6361/201527261}

\bibitem[{{Xu} {et~al.}(2022){Xu}, {Heckman}, {Henry}, {Berg}, {Chisholm},
  {James}, {Martin}, {Stark}, {Aloisi}, {Amor{\'\i}n}, {Arellano-C{\'o}rdova},
  {Bordoloi}, {Charlot}, {Chen}, {Hayes}, {Mingozzi}, {Sugahara}, {Kewley},
  {Ouchi}, {Scarlata}, \& {Steidel}}]{xu22_classy3}
{Xu}, X., {Heckman}, T., {Henry}, A., {et~al.} 2022, \apj, 933, 222,
  \dodoi{10.3847/1538-4357/ac6d56}

\bibitem[{{Zheng} {et~al.}(2017){Zheng}, {Peek}, {Werk}, \& {Putman}}]{zheng17}
{Zheng}, Y., {Peek}, J.~E.~G., {Werk}, J.~K., \& {Putman}, M.~E. 2017, \apj,
  834, 179, \dodoi{10.3847/1538-4357/834/2/179}

\bibitem[{{Zheng, Yong}(2024)}]{lmc-flows}
{Zheng, Yong}. 2024, Quantifying Diskwide Ionized Outflows in the LMC with
  ULLYSES ("LMC-FLOWS"),  STScI/MAST, \dodoi{10.17909/HZ0M-NP43}

\bibitem[{{Zhu} {et~al.}(2023){Zhu}, {Tonnesen}, \& {Bryan}}]{zhu23}
{Zhu}, J., {Tonnesen}, S., \& {Bryan}, G.~L. 2023, arXiv e-prints,
  arXiv:2309.07037, \dodoi{10.48550/arXiv.2309.07037}

\end{thebibliography}

\end{CJK*}
\end{document}